\documentclass[fleqn, usenatbib]{mnras}
\usepackage{newtxtext, newtxmath} 	
\usepackage[T1]{fontenc}
\usepackage{graphicx}
\usepackage{amsmath}
\usepackage{mathtools}
\usepackage{multicol}
\usepackage{bm}
\usepackage{pdflscape}
\usepackage{natbib}
\usepackage[section]{placeins}
\usepackage{lipsum}
\usepackage{etoolbox}
\usepackage{tabularx}
\usepackage{xcolor}
\usepackage[toc, page]{appendix}
\usepackage{subfiles}
\usepackage{geometry}
\usepackage{CJKutf8}
\hypersetup{ 
	colorlinks		= true,
	urlcolor 		= blue,
	linkcolor 		= blue,
	citecolor 		= blue
}

\newcommand{\mc}{\textsc{emcee}}
\newcommand{\vice}{\textsc{VICE}}

\newcommand{\feh}{\ensuremath{\text{[Fe/H]}}}
\newcommand{\afe}{\ensuremath{\text{[$\alpha$/Fe]}}}

\newcommand{\ddfrac}[2]{\frac{\displaystyle{#1}}{\displaystyle{#2}}}
\newcommand{\refp}[1]{(\ref{#1})}
\newcommand{\yacc}{\ensuremath{y_\alpha^\text{CC}}}
\newcommand{\yfecc}{\ensuremath{y_\text{Fe}^\text{CC}}}
\newcommand{\yfeia}{\ensuremath{y_\text{Fe}^\text{Ia}}}
\newcommand{\script}[1]{\ensuremath{\mathcal{#1}}}
\newcommand{\scinote}[2]{\ensuremath{#1\times10^{#2}}}
\newcommand{\erf}{\ensuremath{\text{erf}}}
\newcommand{\python}{\textsc{python}}
\newcommand{\msun}{\ensuremath{\text{M}_\odot}}
\newcommand{\gaia}{\textit{Gaia}}

\providecommand{\noopsort}[1]{}

\title[Dwarf Galaxy Archaeology]{Dwarf galaxy archaeology from chemical
abundances and star formation histories}

\author[J.W. Johnson et al.]{James W. Johnson,$^{1, 2}$\thanks{
	Contact e-mail:~\href{mailto:johnson.7419@osu.edu}{johnson.7419@osu.edu}}
	Charlie Conroy,$^{3}$
	Benjamin D. Johnson,$^{3}$
	Annika H. G. Peter,$^{1, 2, 4}$
	\newauthor
	Phillip A. Cargile,$^{3}$
	Ana Bonaca,$^{5}$
	Rohan P. Naidu,$^{3, 6}$\thanks{NASA Hubble Fellow}
	Turner Woody,$^{3}$
	Yuan-Sen Ting
	(\begin{CJK*}{UTF8}{gbsn}丁源森\ignorespacesafterend\end{CJK*}),$^{7, 8}$
	\newauthor
	Jiwon Jesse Han,$^{3}$
	and Joshua S. Speagle
	(\begin{CJK*}{UTF8}{gbsn}沈佳士\ignorespacesafterend\end{CJK*})$^{9, 10, 11,
	12}$
	\\
	$^{1}$Department of Astronomy, The Ohio State University,
	140 W. 18th Ave., Columbus, OH, 43210, USA
	\\
	$^{2}$Center for Cosmology and Astroparticle Physics (CCAPP),
	The Ohio State University, 191 W. Woodruff Ave., Columbus, OH, 43210, USA
	\\
	$^{3}$Center for Astrophysics | Harvard \& Smithsonian, 60 Garden Street,
	Cambridge, MA, 02138, USA
	\\
	$^{4}$Department of Physics, The Ohio State University, 191 W. Woodruff
	Ave., Columbus, OH, 43210, USA
	\\
	$^{5}$The Observatories of the Carnegie Institution for Science, 813 Santa
	Barbara St., Pasadena, CA, 91101, USA
	\\
	$^{6}$Kavli Institute for Astrophysics and Space Research, Massachusetts
	Institute of Technology, 70 Vassar St., Cambridge, MA, 02139, USA
	\\
	$^{7}$Research School of Astronomy \& Astrophysics, Australian National
	University, Cotter Rd., Weston, ACT 2611, Australia
	\\
	$^{8}$School of Computing, Australian National University, Acton, ACT
	2601, Australia
	\\
	$^{9}$Department of Statistical Sciences, University of Toronto, 9th Floor,
	Ontario Power Building, 700 University Ave, Toronto, ON M5G 1Z5, Canada
	\\
	$^{10}$David A. Dunlap Department of Astronomy \& Astrophysics, University
	of Toronto, 50 St George Street, Toronto ON M5S 3H4, Canada
	\\
	$^{11}$Dunlap Institute for Astronomy \& Astrophysics, University of
	Toronto, 50 St George Street, Toronto, ON M5S 3H4, Canada
	\\
	$^{12}$Data Sciences Institute, University of Toronto, 17th Floor, Ontario
	Power Building, 700 University Ave, Toronto, ON M5G 1Z5, Canada
}

\date{Accepted XXX; Received YYY; in original form ZZZ}
\pubyear{2022}

\begin{document}
\label{firstpage}
\pagerange{\pageref{firstpage}--\pageref{lastpage}}
\maketitle

\begin{abstract}
We model the stellar abundances and ages of two disrupted dwarf galaxies in the
Milky Way stellar halo:~\gaia-Sausage Enceladus (GSE) and Wukong/LMS-1.
Using a statistically robust likelihood function, we fit one-zone models of
galactic chemical evolution with exponential infall histories to both systems,
deriving e-folding timescales of~$\tau_\text{in} = 1.01 \pm 0.13$ Gyr for GSE
and~$\tau_\text{in} = 3.08^{+3.19}_{-1.16}$ Gyr for Wukong/LMS-1.
GSE formed stars for~$\tau_\text{tot} = 5.40^{+0.32}_{-0.31}$ Gyr, sustaining
star formation for~$\sim$$1.5 - 2$ Gyr after its first infall into the Milky
Way~$\sim$10 Gyr ago.
Our fit suggests that star formation lasted for
$\tau_\text{tot} = 3.36^{+0.55}_{-0.47}$ Gyr in Wukong/LMS-1, though our sample
does not contain any age measurements.
The differences in evolutionary parameters between the two are qualitatively
consistent with trends with stellar mass~$M_\star$ predicted by simulations and
semi-analytic models of galaxy formation.
Our fitting method is based only on poisson sampling from an evolutionary
track and requires no binning of the data.
We demonstrate its accuracy by testing against mock data, showing that
it accurately recovers the input model across a broad range of sample sizes
($20 \leq N \leq 2000$) and measurement uncertainties
($0.01 \leq \sigma_\afe,~\sigma_\feh \leq 0.5$;
$0.02 \leq \sigma_{\log_{10}(\text{age})} \leq 1$).
Our inferred values of the outflow mass-loading factor reasonably match
$\eta \propto M_\star^{-1/3}$ as predicted by galactic wind models.
Due to the generic nature of our derivation, this likelihood function should
be applicable to one-zone models of any parametrization and easily extensible
to other astrophysical models which predict tracks in some observed space.
\end{abstract}


\section{Introduction}
\label{sec:intro}

Dwarf galaxies provide a unique window into galaxy formation and evolution.
In the local universe, dwarfs can be studied in detail using resolved stellar
populations across a wide range of mass, morphology and star formation history
(SFH).
Field dwarfs have more drawn-out SFHs than more massive galaxies like the Milky
Way and Andromeda~\citep[e.g.,][]{Behroozi2019, GarrisonKimmel2019}, while
satellites often have their star formation ``quenched'' by ram pressure
stripping from the hot halo of their host
\citep*[see discussion in, e.g.,][]{Steyrleithner2020} if they are not
disintegrated by the tidal forces of the host.
As a result, disrupted dwarf galaxies assembled much of their stellar mass at
high redshift, but their resolved stellar populations encode a wealth of
information on their progenitor's evolutionary history.
\par
Photometrically, one can constrain the SFH by fitting the observed
color-magnitude diagram (CMD) with a composite set of theoretical isochrones
\citep[e.g.,][]{Dolphin2002, Weisz2014b}.
The CMD also offers constraints on the metallicity distribution function
(MDF;~\citealp*[e.g.,][]{Lianou2011}).
In some cases, the MDF can also be constrained with narrow-band imaging
\citep{Fu2022}, especially when combined with machine learning algorithms
trained on spectroscopic measurements as in~\citet{Whitten2011}.
Depending on the limiting magnitude of the survey and the evolutionary stages
of the accessible stars, it may or may not be feasible to estimate ages on a
star-by-star basis.
When these measurements are made spectroscopically, however, multi-element
abundance information becomes available, and age estimates become more precise
by pinning down various stellar parameters such as effective temperatures and
surface gravities.
\par
Chemical abundances in a dwarf galaxy can also offer independent constraints
on the evolutionary histories of dwarf galaxies, including the earliest epochs
of star formation.
Stars are born with the same composition as their natal molecular clouds --
spectroscopic abundance measurements in open clusters have demonstrated that
FGK main-sequence and red giant stars exhibit chemical homogeneities
within~$\sim$$0.02 - 0.03$ dex~\citep{DeSilva2006, Bovy2016, Liu2016b,
Casamiquela2020} while inhomogeneities at the~$\sim$$0.1 - 0.2$ dex level can
be attributed to diffusion~\citep{BertelliMotta2018, Liu2019, Souto2019} or
planet formation~\citep{Melendez2009, Liu2016a, Spina2018}.
A star's detailed metal content is therefore a snapshot of the galactic
environment that it formed from.
This connection is the basis of galactic chemical evolution (GCE), which
bridges the gap between nuclear physics and astrophysics by combining galactic
processes such as star formation with nuclear reaction networks to estimate the
production rates of various nuclear species by stars and derive their
abundances in the intertsellar medium (ISM).
GCE models that accurately describe the observed abundances of resolved stars
in intact and disrupted dwarf galaxies can offer constraints on their star
formation and accretion histories, the efficiency of outflows, and the origin
of the observed abundance pattern.
\par
In this paper, we systematically assess the information that can be extracted
from the abundances and ages of stars in dwarf galaxies when modelling the
data in this framework.
The simplest and most well-studied GCE models are called ``one-zone'' models,
reviews of which can be found in works such as~\citet{Tinsley1980},
\citet{Pagel2009} and \citet{Matteucci2012, Matteucci2021}.
One-zone models are computationally cheap, and with reasonable approximations,
even allow analytic solutions to the evolution of the abundances for simple
SFHs~\citep*[e.g.,][]{Weinberg2017}.
This low expense expedites the application of statistical likelihood estimates
to infer best-fit parameters for some set of assumptions regarding a galaxy's
evolutionary history.
There are both simple and complex examples in the literature of how one might
go about these calculations.
For example,~\citet{Kirby2011} measure and fit the MDFs of eight Milky Way
dwarf satellite galaxies with the goal of determining which evolved according
to ``leaky-box,'' ``pre-enriched'' or ``extra-gas'' analytic models.
\citet{delosReyes2022} used abundances for a wide range of elements to
constrain the evolutionary history of the Sculptor dwarf Spheroidal.
To derive best-fit parameters for the two-infall model of the Milky Way disc
\citep[e.g.,][]{Chiappini1997},~\citet{Spitoni2020, Spitoni2021} use Markov
chain Monte Carlo (MCMC) methods and base their likelihood function off of the
minimum distance between each star and the evolutionary track in
the~\afe-\feh\footnote{
	We follow the conventional definition in which
	[X/Y]~$\equiv \log_{10}(N_\text{X} / N_\text{Y}) -
	\log_{10}(N_{\text{X},\odot} / N_{\text{Y},\odot})$
	is the logarithmic difference in the abundance ratio of the nuclear species
	X and Y between some star and the sun.
} plane.
\citet{Hasselquist2021} used similar methods to derive evolutionary parameters
for the Milky Way's most massive satellites with the~\textsc{FlexCE}
\citep{Andrews2017} and the~\citet{Lian2018, Lian2020} chemical evolution
codes.
\par
While these studies have employed various methods to estimate the relative
likelihood of different parameter choices, to our knowledge there is no
demonstration of the statistical validity of these methods in the literature.
The distribution of stars in abundance space is generally non-uniform, and the
probability of randomly selecting a star from a given epoch of some galaxy's
evolution scales with the star formation rate (SFR) at that time (modulo the
selection function of the survey).
Describing the enrichment history of a galaxy as a one-zone model casts the
observed stellar abundances as a stochastic sample from the predicted
evolutionary track, a process which proceeds mathematically according to an
\textit{inhomogeneous poisson point process} (IPPP; see, e.g.,
\citealt{Press2007}).
To this end, we apply the principles of an IPPP to an arbitrary model-predicted
track in some observed space.
We demonstrate that this combination results in the derivation of a single
likelihood function which is required to ensure the accuracy of best-fit
parameters.
Our derivation does not assume that the track was predicted by a GCE model,
and it should therefore be easily extensible to other astrophysical models
which predict evolutionary tracks in some observed space, such as stellar
streams in kinematic space or isochrones on CMDs.
We however limit our discussion in this paper to our use case of one-zone GCE
models.
\par
After discussing the one-zone model framework in~\S~\ref{sec:onezone} and
our fitting method in~\S~\ref{sec:fitting}, we establish the accuracy of this
likelihood function by means of tests against mock data in~\S~\ref{sec:mocks},
simultaneously exploring how the precision of inferred parameters is affected
by sample size, measurement uncertainties and the portion of the sample that
has age information.
These methods are able to reconstruct the SFHs of dwarf galaxies because the
GCE framework allows one to convert the number of stars versus metallicity into
the number of stars versus time.
Abundance ratios such as~\afe~quantify the relative importance of type Ia
supernova (SN Ia) enrichment, and constraints on its associated delay-time
distribution (DTD) set an overall timescale.
In~\S~\ref{sec:h3}, we demonstrate our method in action by modelling two
disrupted dwarf galaxies in the Milky Way halo.
One has received a considerable amount of attention in the literature: the
\gaia-Sausage Enceladus (GSE;~\citealp{Belokurov2018, Helmi2018}), and the
other, discovered more recently, is a less deeply studied system: Wukong
\citep{Naidu2020, Naidu2022}, independently discovered as LMS-1
by~\citet{Yuan2020}.

\section{Galactic Chemical Evolution}
\label{sec:onezone}

One-zone GCE models connect the star formation and accretion histories of
galaxies to the enrichment rates in the ISM through prescriptions for
nucleosynthetic yields, outflows, and star formation efficiency (SFE) within
a simple mathematical framework.
Their fundamental assumption is that newly produced metals mix instantaneously
throughout the star-forming gas reservoir.
In detail, this assumption is valid as long as the mixing timescale is short
compared to the depletion timescale (i.e., the average time a fluid element
remains in the ISM before getting incorporated into new stars or ejected in an
outflow).
Based on the observations of~\citet{Leroy2008},~\citet{Weinberg2017} calculate
that characteristic depletion times can range from~$\sim$500 Myr up to~$\sim$10
Gyr for conditions in typical star forming disc galaxies.
In the dwarf galaxy regime, the length scales are short, star formation is slow
\citep[e.g.,][]{Hudson2015}, and the ISM velocities are turbulent
\citep{Dutta2009, Stilp2013, Schleicher2016}.
With this combination, instantaneous mixing should be a good approximation,
though we are unaware of any studies which address this observationally.
As long as the approximation is valid, then there should exist an evolutionary
track in chemical space (e.g., the~\afe-\feh~plane) about which the intrinsic
scatter is negligible compared to the measurement uncertainty.
This empirical test should be feasible on a galaxy-by-galaxy basis.
\par
With the goal of assessing the information content of one-zone GCE models
applied to dwarf galaxies, we emphasize that the accuracy of the methods we
outline in this paper are contingent on the validity of the instantaneous
mixing approximation.
This assumption reduces GCE to a system of coupled integro-differential
equations, which we solve using the publicly available~\textsc{Versatile
Integrator for Chemical Evolution} (\vice\footnote{
	\url{https://pypi.org/project/vice}
};~\citealp{Johnson2020}).
We provide an overview of the model framework below and refer to
\citet{Johnson2020} and the~\vice~science documentation\footnote{
	\url{https://vice-astro.readthedocs.io/en/latest/science_documentation}
} for further details.
\par
At a given moment in time, gas is added to the ISM via inflows and recycled
stellar envelopes and is removed from the ISM by star formation and outflows,
if present.
The sum of these terms gives rise to the following differential equation
describing the evolution of the gas supply:
\begin{equation}
\dot{M}_\text{g} = \dot{M}_\text{in} - \dot{M}_\star - \dot{M}_\text{out}
+ \dot{M}_\text{r},
\label{eq:mdotgas}
\end{equation}
where~$\dot{M}_\text{in}$ is the infall rate,~$\dot{M}_\star$ is the SFR,
$\dot{M}_\text{out}$ is the outflow rate, and~$\dot{M}_\text{r}$ describes
the return of stellar envelopes from previous generations of stars.
\par
\vice~implements the same characterization of outflows as the~\textsc{FlexCE}
\citep{Andrews2017} and~\textsc{OMEGA}~\citep{Cote2017} chemical evolution
codes in which a ``mass-loading factor''~$\eta$ describes a linear relationship
between the outflow rate itself and the SFR:
\begin{equation}
\eta \equiv \frac{\dot{M}_\text{out}}{\dot{M}_\star}.
\label{eq:massloading}
\end{equation}
This parametrization is appropriate for models in which massive stars are the
dominant source of energy for outflow-driving winds.
Empirically, the strength of outflows (i.e., the value of~$\eta$) is strongly
degenerate with the absolute scale of nucleosynthetic yields.
We discuss this further below and quantify the strength of the degeneracy in
more detail in Appendix~\ref{sec:degeneracy}.
\par
The SFR and the mass of the ISM are related by the SFE timescale~$\tau_\star$,
defined as the ratio of the two:
\begin{equation}
\tau_\star \equiv \frac{M_\text{g}}{\dot{M}_\star}.
\label{eq:taustar}
\end{equation}
The inverse~$\tau_\star^{-1}$ is the SFE itself, quantifying the
\textit{fractional} rate at which some ISM fluid element is forming stars.
Some authors refer to~$\tau_\star$ as the ``depletion time''
\citep[e.g.,][]{Tacconi2018} because it describes the e-folding decay timescale
of the ISM mass due to star formation if no additional gas is added.
Our nomenclature follows~\citet{Weinberg2017}, who demonstrate that depletion
times in GCE models can shorten significantly in the presence of outflows.
\par
The recycling rate~$\dot{M}_\text{r}$ is a complicated function which depends
on the stellar initial mass function~\citep[IMF; e.g.,][]{Salpeter1955,
Miller1979, Kroupa2001, Chabrier2003}, the initial-final remnant mass relation
\citep[e.g.,][]{Kalirai2008}, and the mass-lifetime relation\footnote{
	We assume a~\citet{Kroupa2001} IMF and the~\citet{Larson1974} mass-lifetime
	relation throughout this paper.
	These choices do not significantly impact our conclusions as~$\eta$
	and~$\tau_\star$ play a much more significant role in establish the
	evolutionary histories of our GCE models.
	Our fitting method is nonetheless easily extensible to models which relax
	these assumptions.
}~\citep*[e.g.,][]{Larson1974, Maeder1989, Hurley2000}, all of which must then
be convolved with the SFH.
However, the detailed rate of return of stellar envelopes has only a
second-order effect on the gas-phase evolutionary track in the~\afe-\feh~plane.
The first-order details are instead determined by the SFE timescale~$\tau_\star$
and the mass-loading factor~$\eta$ (see discussion in~\citealt{Weinberg2017}).
In the absence of sudden events such as a burst of star formation, the detailed
form of the SFH actually has minimal impact of the shape of the model track
\citep{Weinberg2017, Johnson2020}.
That information is instead encoded in the stellar MDFs (i.e., the density of
stars along the track).
\par
In the present paper, we focus on the enrichment of the so-called ``alpha''
(e.g., O, Ne, Mg) and ``iron-peak'' elements (e.g., Cr, Fe, Ni, Zn), with the
distribution of stars in the~\afe-\feh~plane being our primary observational
diagnostic to distinguish between GCE models.
Massive stars and their core collapse SNe (CCSNe) are the dominant enrichment
source of alpha elements in the universe, while iron-peak elements are produced
in significant amounts by both massive stars and SNe Ia~\citep[e.g.,][]{Johnson2019}.
In detail, some alpha and iron-peak elements also have contributions from slow
neutron capture nucleosynthesis, an enrichment pathway responsible for much of
the abundances of yet heavier nuclei (specifically Sr and up).
Because the neutron capture yields of alpha and iron-peak elements are
small compared to their SN yields, we do not discuss this process further.
Our fitting method is nonetheless easily extensible to GCE models which do,
provided that the data contain such measurements.
\par
Due to the steep nature of the stellar mass-lifetime relation
\citep[e.g.,][]{Larson1974, Maeder1989, Hurley2000}, massive stars, their winds,
and their SNe enrich the ISM on~$\sim$few Myr timescales.
As long as these lifetimes is shorter than the relevant timescales for a
galaxy's evolution and the present-day stellar mass is sufficiently high such
that stochastic sampling of the IMF does not significantly impact the yields,
then it is adequate to approximate this nucleosynthetic material as some
population-averaged yield ejected instantaneously following a single stellar
population's formation.
This implies a linear relationship between the CCSN enrichment rate and the
SFR:
\begin{equation}
\dot{M}_\text{x}^\text{CC} = y_\text{x}^\text{CC} \dot{M}_\star,
\end{equation}
where~$y_\text{x}^\text{CC}$ is the IMF-averaged fractional net yield from
massive stars of some element x.
That is, for a fiducial value of~$y_\text{x}^\text{CC} = 0.01$, 100~\msun~of
star formation would produce 1~\msun~of~\textit{newly produced} element x (the
return of previously produced metals is implemented as a separate term
in~\vice; see~\citealt{Johnson2020} or the~\vice~science documentation for
details).
\par
Unlike CCSNe, SNe Ia occur on a significantly extended DTD.
The details of the DTD are a topic of active inquiry~\citep[e.g.,][]{Greggio2005,
Strolger2020, Freundlich2021}, and at least a portion of the uncertainty can be
traced to uncertainties in both galactic and cosmic SFHs.
Comparisons of the cosmic SFH~\citep[e.g.,][]{Hopkins2006, Madau2014, Davies2016,
Madau2017, Driver2018} with volumetric SN Ia rates as a function of redshift
indicate that the cosmic DTD is broadly consistent with a uniform~$\tau^{-1}$
power-law (\citealp{Maoz2012a};~\citealp*{Maoz2012b};~\citealp{Graur2013,
Graur2014}).
Following~\citet{Weinberg2017}, we take a~$\tau^{-1.1}$ power-law DTD with a
minimum delay time of~$t_\text{D} = 150$ Myr, though in principle this
delay can be as short as~$t_\text{D} \approx 40$ Myr due to the lifetimes
of the most massive white dwarf progenitors.
For any selected DTD~$R_\text{Ia}(\tau)$, the SN Ia enrichment rate can be
expressed as an integral over the SFH weighted by the DTD:
\begin{equation}
\dot{M}_\text{x}^\text{Ia} = y_\text{x}^\text{Ia}\ddfrac{
	\int_0^{T - t_\text{D}} \dot{M}_\star(t) R_\text{Ia}(T - t) dt
}{
	\int_0^\infty R_\text{Ia}(t) dt
}.
\end{equation}
In general, the mass of some element x in the ISM is also affected by outflows,
recycling and star formation.
The total enrichment rate can be computed by simply adding up all of the source
terms and subtracting the sink terms:
\begin{equation}
\dot{M}_\text{x} = \dot{M}_\text{x}^\text{CC} + \dot{M}_\text{x}^\text{Ia} -
Z_\text{x}\dot{M}_\star - Z_\text{x}\dot{M}_\text{out} + \dot{M}_{\text{x,r}},
\label{eq:enrichment}
\end{equation}
where~$Z_x = M_\text{x} / M_\text{ISM}$ is the abundance by mass of the nuclear
species x in the ISM.
This equation as written assumes that the outflowing material is of the same
composition as the ISM, but in principle, the various nuclear species of
interest may be some factor above or below the ISM abundance.
In the present paper we assume all accreting material to be zero metallicity
gas; when this assumption is relaxed, an additional term
$Z_\text{x,in}\dot{M}_\text{in}$ appears in this equation.
\par
As mentioned above, the strength of outflows is degenerate with the absolute
scale of nucleosynthetic yields.
This ``yield-outflow degeneracy'' is remarkably strong, and it arises because
yields and outflows are the dominant source and sink terms in equation
\refp{eq:enrichment} above.
As a consequence, high-yield and high-outflow models generally have a
low-yield and low-outflow counterpart that predicts a similar enrichment
history.
In order to break this degeneracy, only a single parameter setting the absolute
scale is required.
To this end, we set the alpha element yield from massive stars to be exactly
$\yacc = 0.01$ and let our Fe yields be free parameters.
Appropriate for O, this value is loosely motivated by nucleosynthesis theory in
that massive star evolutionary models (e.g.,~\citealp*{Nomoto2013};
\citealp{Sukhbold2016, Limongi2018}) typically predict
$y_\text{O}^\text{CC} = 0.005 - 0.015$ (see discussion in, e.g.,
\citealp{Weinberg2017} and~\citealp{Johnson2020}).
This value is~$\sim$1.75 times the solar O abundance of~$\sim$0.57\%
\citep{Asplund2009}, and if we had chosen a different alpha element (e.g., Mg),
then we would need to adjust accordingly to account for the intrinsically lower
abundance (e.g., $\yacc = 1.75 Z_{\text{Mg},\odot} \approx
\scinote{1.2}{-4}$).\footnote{
	The lighter alpha elements like O and Mg evolve similarly in GCE models due
	to metallicity-independent yields dominated by massive stars, so it is
	mathematically convenient to treat them as a single nuclear species
	under the assertion that [O/Mg]~$\approx 0$ (this assumption is indeed
	supported by empirical measurements in APOGEE; see, e.g., Fig. 8 of
	\citealt{Weinberg2019}).
	In practice, however, we use the~$\yacc = 0.01$ value for O and a solar
	abundance of~$Z_{\text{O},\odot} = 0.00572$~\citep{Asplund2009}.
}
The primary motivation behind this choice is to select a round number that
allows our best-fit values affected by this degeneracy to be scaled up or down
under different assumptions regarding the scale of effective yields.
We reserve further discussion of this topic for Appendix~\ref{sec:degeneracy}
where we also quantify the considerably strength of the yield-outflow
degeneracy in more detail.

\section{The Fitting Method}
\label{sec:fitting}

Our fitting method uses the abundances of an ensemble of stars, incorporating
age measurements as additional data where available, and without any binning,
accurately constructs the~\textit{likelihood function}
$L(\script{D} | \{\theta\})$ describing the probabiliy of observing the
data~$\script{D}$ given a set of model parameters $\{\theta\}$.
$L(\script{D} | \{\theta\})$ is related to the~\textit{posterior probability}
$\L(\{\theta\} | \script{D})$ according to Bayes' Theorem:
\begin{equation}
L(\{\theta\} | \script{D}) = \frac{
	L(\script{D} | \{\theta\}) L(\{\theta\})
}{
	L(\script{D})
},
\label{eq:bayes}
\end{equation}
where~$L(\{\theta\})$ is the likelihood of the parameters themselves (known as
the~\textit{prior}) and~$L(\script{D})$ is the likelihood of the data (known as
the~\textit{evidence}).
Although it is more desirable to measure the posterior probability, in practice
only the likelihood function can be robustly determined because the prior is
not directly quantifiable.
The prior requires quantitative information independent of the data on the
accuracy of a chosen set of parameters~$\{\theta\}$.
With no additional information on what the parameters should be, the best
practice is to assume a ``flat'' or ``uniform'' prior in which~$L(\{\theta\})$
is a constant, and therefore~$L(\{\theta\} | \script{D}) \approx
L(\script{D} | \{\theta\})$; we retain this convention here unless otherwise
stated.
\par
As mentioned in~\S~\ref{sec:intro}, the sampling of stars from an underlying
evolutionary track in abundance space proceeds according to an IPPP
\citep[e.g.,][]{Press2007}.
Due to its detailed nature, we reserve a full derivation of our likelihood
function for Appendix~\ref{sec:likelihood} and provide qualitative discussion
of its form here.
Though our use case in the present paper is in the context of one-zone GCE
models, our derivation assumes only that the chief prediction of the model is
a track of some arbitrary form in the observed space.
It is therefore highly generic and should be easily extensible to other
astrophysical models that predict tracks of some form (e.g., stellar streams
in kinematic space and stellar isochrones on CMDs).
\par
In practice, the evolutionary track predicted by a one-zone GCE model is
generally not known in some analytic functional form (unless specific
approximations are made as in, e.g.,~\citealp{Weinberg2017}).
Instead, it is most often quantified as a piece-wise linear form predicted by
some numerical code (in our case,~\vice).
For a sample~$\script{D} = \{\script{D}_1, \script{D}_2, \script{D}_3, ...,
\script{D}_N\}$ containing~$N$ abundance and age (where available) measurements
of individual stars and a track~$\script{M} = \{\script{M}_1, \script{M}_2,
\script{M}_3, ..., \script{M}_K\}$ sampled at~$K$ points in abundance space,
the likelihood function is given by
\begin{equation}
\ln L(\script{D} | \{\theta\}) = \sum_i^N \ln \left(
\sum_j^K w_j \exp\left(
\frac{-1}{2}\Delta_{ij}C_i^{-1}\Delta_{ij}^T
\right)
\right),
\label{eq:likelihood}
\end{equation}
where~$\Delta_{ij} = \script{D}_i - \script{M}_j$ is the vector difference
between the~$i$th datum and the~$j$th point on the predicted track,~$C_i^{-1}$
is the inverse covariance matrix of the~$i$th datum, and~$w_j$ is a weight to
be attached to~$\script{M}_j$ (we clarify our notation that~$ij$ refers to a
data-model pair and not a matrix element; the covariance matrix need not be
diagonal for this approach).
This functional form is appropriate for GCE models in which the normalization
of the SFH is inconsequential to the evolution of the abundances; in the
opposing case where the normalization does impact the predicted abundances,
one additional term subtracting the sum of the weights is required (see
discussion below).
\par
Equation~\refp{eq:likelihood} arises from marginalizing the likelihood of
observing each datum over the entire evolutionary track and has the more
general form of
\begin{subequations}\begin{align}
\ln L(\script{D} | \{\theta\}) &= \sum_i^N \left(
\int_\script{M} L(\script{D}_i | \script{M}) d\script{M}
\right)
\label{eq:likelihood_general_int}
\\
&\approx \sum_i^N \ln \left(
\sum_j^K L(\script{D}_i | \script{M}_j)\right).
\label{eq:likelihood_general}
\end{align}\end{subequations}
Equation~\refp{eq:likelihood_general} follows from equation
\refp{eq:likelihood_general_int} when the track is densely sampled by the
numerical integrator (see discussion below), and equation~\refp{eq:likelihood}
follows thereafter when the likelihood $L(\script{D}_i | \script{M}_j)$ of
observing the~$i$'th datum given the~$j$th point on the evolutionary track is
given by a weighted~$e^{-\chi^2/2}$ expression.
Mathematically, the requirement for this marginalization arises naturally from
the application of statistical likelihood and an IPPP to an evolutionary track
(see Appendix~\ref{sec:likelihood}).
Qualitatively, this requirement is due to observational uncertainties -- there
is no way of knowing which point on the evolutionary track the
datum~$\script{D}_i$ is truly associated with, and the only way to properly
account for its unknown position is to consider all pair-wise combinations
of~\script{D}~and~\script{M}.
\par
The mathematical requirement for a weighted as opposed to unweighted
$e^{-\chi^2/2}$ likelihood expression also arises naturally in our derivation.
Qualitatively, the weights arise because the likelihood of observing the datum
$\script{D}_i$ is proportionally higher for points on the evolutionary track
when the SFR is high or if the survey selection function is deeper.
For a selection function~\script{S}~and SFR~$\dot{M}_\star$, the weights should
scale as their product:
\begin{equation}
w_j \propto \script{S}(\script{M}_j | \{\theta\})
\dot{M}_\star(\script{M}_j | \{\theta\}).
\label{eq:weights}
\end{equation}
Whether or not the weights require an overall normalization is related to the
parametrization of the GCE model -- in particular, if the normalization of the
SFH impacts the abundances or not (see discussion below).
The selection function may be difficult to quantify, but one simple way to
characterize its form in chemical space would be to assess what fraction -- by
number -- of the stellar populations in the model would be incorporated into
the sample as a result of cuts in, e.g., color, surface gravity, effective
temperature, etc.
\par
The marginalization over the track and the weighted likelihood are of the
utmost importance to ensure accurate best-fit parameters.
In our tests against mock samples (see~\S~\ref{sec:mocks} below), we are unable
to recover the known evolutionary parameters of input models with discrepancies
at the many-$\sigma$ level if either are neglected.
While these details always remain a part of the likelihood function, equation
\refp{eq:likelihood} can change in form slightly if any one of a handful of
conditions are not met.
We discuss these conditions and the necessary modifications below, referring to
Appendix~\ref{sec:likelihood} for mathematical justification.
\par
\textit{The model track is infinitely thin.}
In the absence of measurement uncertainties, all of the data would fall
perfectly on a line in the observed space.
As discussed in the beginning of~\S~\ref{sec:onezone}, the fundamental
assumption of one-zone GCE models is instantaneous mixing of the various
nuclear species throughout the star forming reservoir.
Consequently, the ISM is chemically homogeneous and the model predicts a single
exact abundance for each element or isotope at any given time.
If the model in question instead predicts a track of some finite width, then
the likelihood function will have a different form requiring at least one
additional integral.
\par
\textit{Each observation is independent.}
When this condition is met, the total likelihood of observing the
data~\script{D}~can be expressed as the product of the likelihood of observing
each individual datum:
\begin{subequations}\begin{align}
L(\script{D} | \{\theta\}) &= \prod_i^N L(\script{D}_i | \script{M})
\\
\implies \ln L(\script{D} | \{\theta\}) &= \sum_i^N \ln
L(\script{D}_i | \script{M}).
\end{align}\end{subequations}
This condition plays an integral role in giving rise to the functional form of
equation~\refp{eq:likelihood}, and if violated, the likelihood function will
also have a fundamentally different form.
\par
\textit{The observational uncertainties are described by a multivariate
Gaussian.}
If this condition fails, the weighted~$\chi^2 = \Delta_{ij}C_i^{-1}\Delta_{ij}^T$
expression is no longer an accurate parametrization of~$L(\script{D}_i |
\script{M}_j)$ and it should be replaced with the more general form of
equation~\refp{eq:likelihood_general}.
In these cases, a common alternative would be to replace~$e^{-\chi^2 / 2}$ with
some kernel density estimate of the uncertainty at the point~$\script{M}_j$
while retaining the weight~$w_j$, but this substitution is only necessary for
the subset of~\script{D}~whose uncertainties are not adequately described by a
multivariate Gaussian.
\par
\textit{The track is densely sampled.}
That is, the spacing between the points on the track~\script{M}~is small
compared to the observational uncertainties in the data.
This assumption can be relaxed at the expense of including an additional
correction factor~$\beta_{ij}$ given by equation~\refp{eq:corrective_beta}
that integrates the likelihood between each pair of adjacent
points~$\script{M}_j$ and~$\script{M}_{j + 1}$ along the track (see discussion
in Appendix~\ref{sec:likelihood}).
If computing the evolutionary track is sufficiently expensive, relaxing the
number of points and including this correction factor may be the more
computationally efficient option.
\par
\textit{The normalization of the SFH does not impact the predicted abundances.}
Only the time-dependence of the SFH impacts the abundance evolution predicted
by the GCE model.
As mentioned above, the model-predicted SFH and the selection function of the
survey determine the weights to attach to each point~$\script{M}_j$ along the
track, and if the normalization of the SFH does not impact the abundance
evolution, then it must not impact the inferred likelihood either.
In our detailed derivation of equation~\refp{eq:likelihood}, we find that the
proper manner in which to assign the weights is to normalize then such that
they add up to 1 (see Appendix~\ref{sec:likelihood}).
Some GCE models, however, are parametrized such that the normalization of the
SFH~\textit{does} impact the abundance evolution.
One such example would be if the SFE timescale~$\tau_\star$ (see equation
\ref{eq:taustar} and discussion in~\S~\ref{sec:onezone}) depends on the gas
supply~$M_\text{g}$ in order to implement some version of a non-linear
Kennicutt-Schmidt relation\footnote{
	$\dot{\Sigma}_\star \propto \Sigma_\text{g}^N \implies \tau_\star \propto
	\Sigma_\text{g}^{1 - N}$ where~$N \neq 1$.
	\citet{Kennicutt1998} measured~$N = 1.4 \pm 0.15$ from the global gas
	densities and SFRs in star-forming spiral galaxies, although recent
	advancements suggest more sophisticated forms (e.g.,~\citealp{Krumholz2018};
	see discussion in~\S~2.6 of~\citealt{Johnson2021}).
} where the normalization of the SFH and size of the galaxy are taken into
account.
In these cases, the likelihood function is given by
equation~\refp{eq:corrective_beta} where the weights remain un-normalized and
their sum must be subtracted from equation~\refp{eq:likelihood}.
This requirement can be qualitatively understood as a penalty for models that
predict data in regions of the observed space where there are none -- a term
which encourages parsimony, rewarding parameter choices which explain the data
in as few predicted instances as possible.
This penalty is still included in models which normalize the weights, with the
tracks that extend too far in abundance space instead having a higher
\textit{fractional} weight from data at large~$\chi^2$, lowering the total
likelihood (see discussion near the end of Appendix~\ref{sec:likelihood}).

\begin{figure*}
\centering
\includegraphics[scale = 0.5]{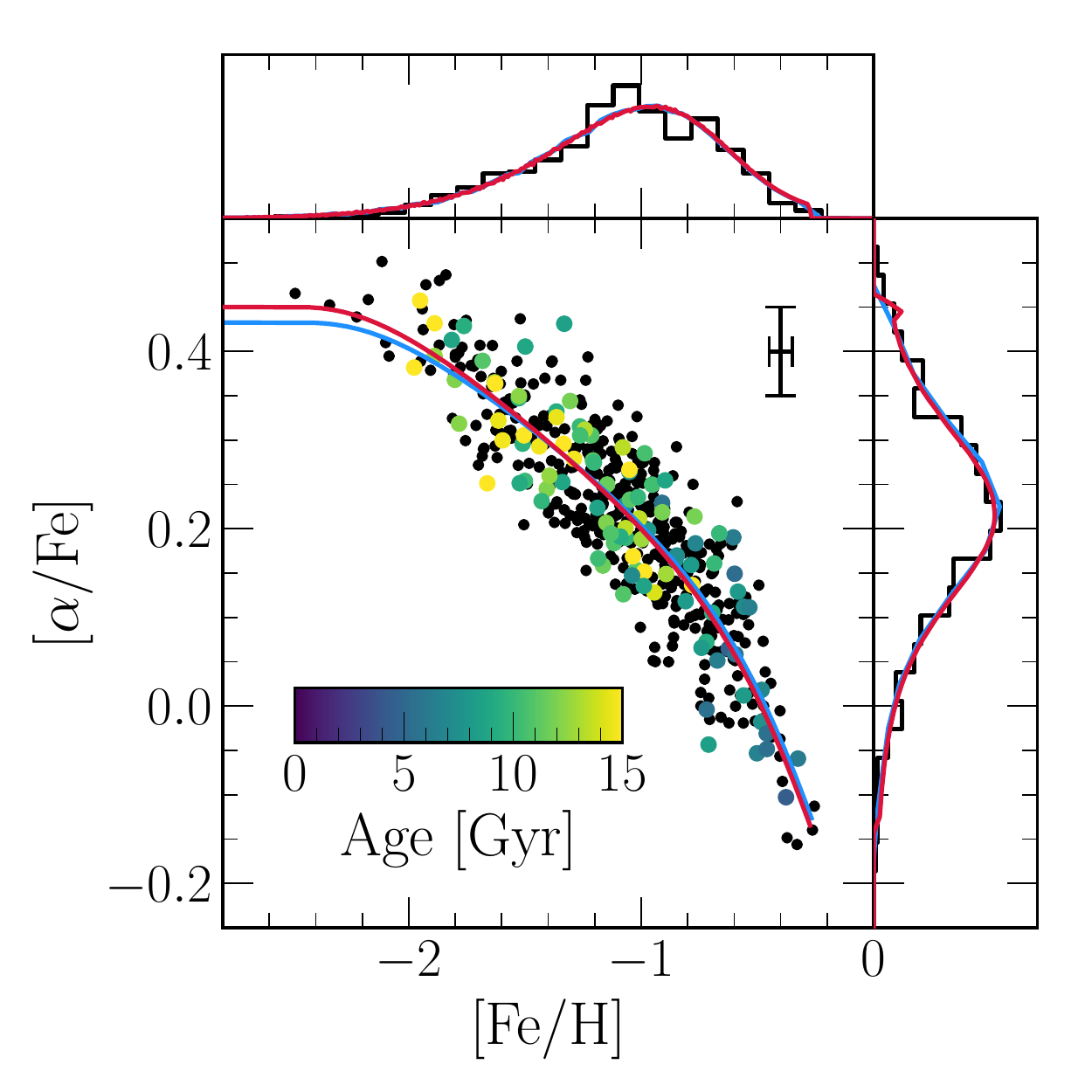}
\includegraphics[scale = 0.42]{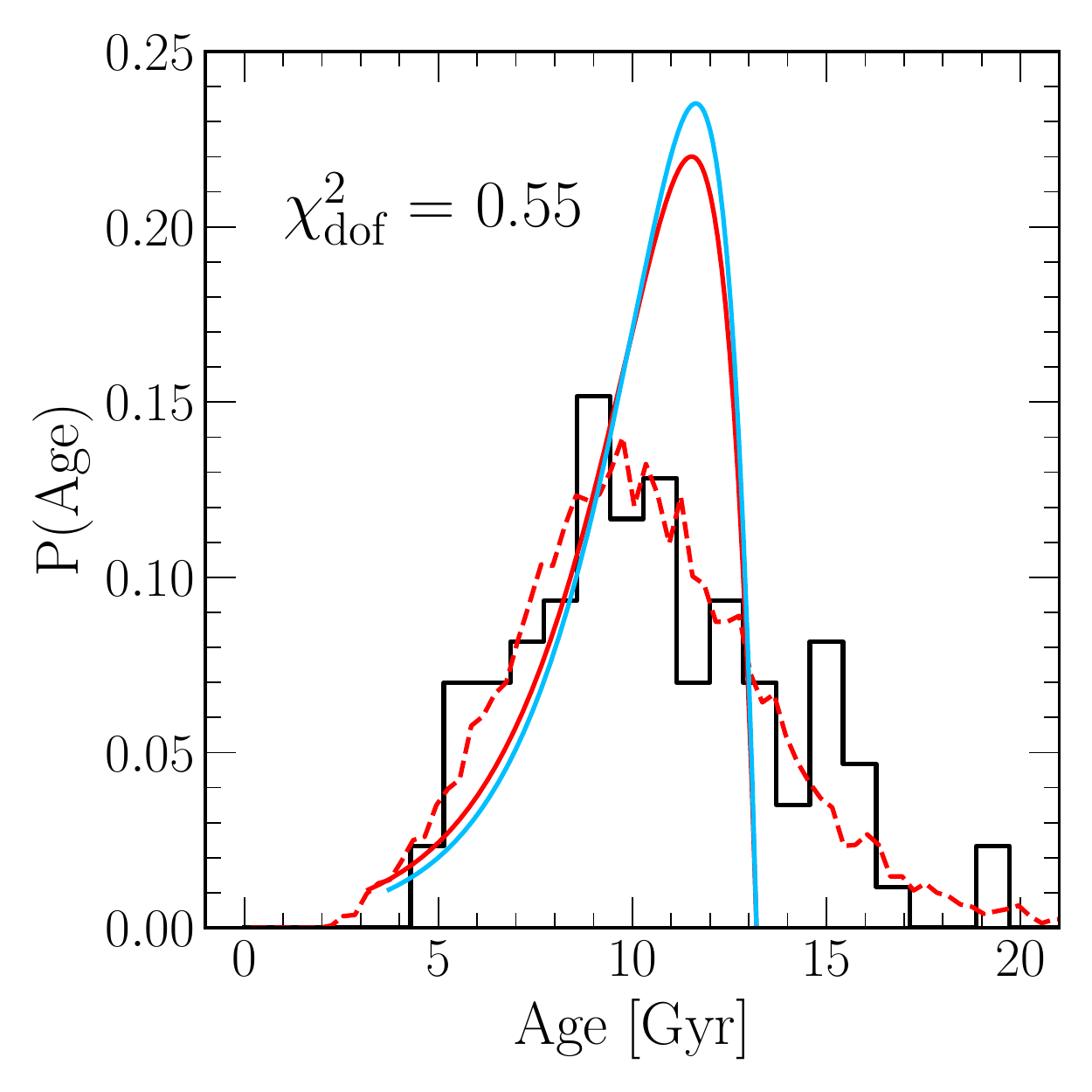}
\includegraphics[scale = 0.41]{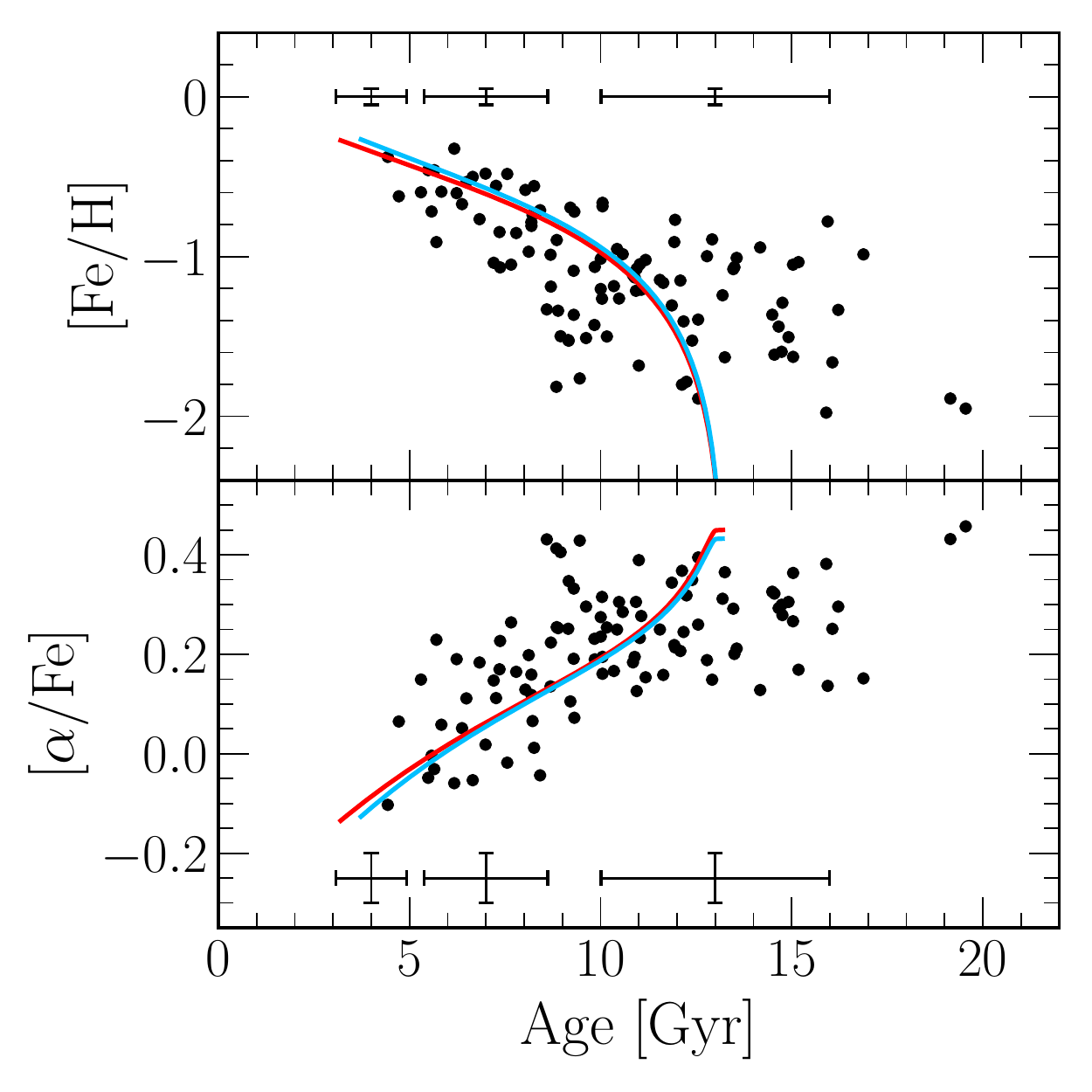}
\caption{
Our fiducial mock sample.
Red lines in all panels denote the input model while blue lines denote the
recovered best-fit model.
The mock sample has~$N = 500$ stars with abundance uncertainties of
$\sigma_\feh = \sigma_\afe = 0.05$ (marked by the errorbar in the left panel).
$N = 100$ of the stars have age information as indicated by the colourbar in
the left panel with an artificial uncertainty
of~$\sigma_{\log_{10}(\text{age})} = 0.1$.
\textbf{Left}: The mock sample in chemical space, with the marginalized
distributions in~\feh~and~\afe~shown on the top and right, respectively.
\textbf{Middle}: The age distribution of the mock sample (black, binned).
The dashed red line indicates the age distribution obtained by sampling
$N = 10^4$ rather than~$N = 500$ stars from the input model and assuming the
same age uncertainty.
\textbf{Right}: The age-\feh~(top) and age-\afe~(bottom) relations for the mock
sample.
Uncertainties at various ages are marked by the error bars at the top and
bottom of each panel.
}
\label{fig:fiducial_mock}
\end{figure*}

We demonstrate the accuracy of our likelihood function in~\S~\ref{sec:mocks}
below by means of tests against mock data samples.
Although our likelihood function does not include a direct fit to the
stellar distributions in age and abundances, weighting the inferred likelihood
by the SFR in the model indeed incorporates this information on how many stars
should form at which ages and abundances.
Our method therefore provides~\textit{implicit} fits to the age and abundance
distributions, even though this information is not directly included in the
likelihood calculation.
\par
There are a variety of ways to construct the likelihood distribution in
parameter space.
In the present paper, we employ the MCMC method, making use of
the~\mc~\python~package~\citep{ForemanMackey2013} to construct our Markov
chains.
Despite being more computationally expensive than other methods (e.g.,
maximum a posteriori estimation), MCMC offers a more generic solution by
sampling tails and multiple modes of the likelihood distribution which could
otherwise be missed or inaccurately characterized by the assumption of
Gaussianity.
Our method should nonetheless be extensible to additional data sets described
by GCE models with different parametrizations as well as different methods of
optimizing the likelihood distribution, such as maximum a posteriori estimates.

\section{Mock Samples}
\label{sec:mocks}

Using our parametrization of one-zone GCE models described
in~\S~\ref{sec:onezone}, here we define a set of parameter choices from which
mock samples of stars can be drawn.
We then demonstrate the validity of our likelihood function (Eq.
\ref{eq:likelihood}) in~\S~\ref{sec:mocks:recovered} by applying it to a
fiducial mock sample and comparing the best-fit values to the known parameters
of the input model.
In~\S~\ref{sec:mocks:variations}, we then explore variations in sample size,
measurement precision, and the availability of age information.

\subsection{A Fiducial Mock Sample}
\label{sec:mocks:fiducial}

We take an exponential infall history~$\dot{M}_\text{in} \propto e^{-t /
\tau_\text{in}}$ with an e-folding timescale of~$\tau_\text{in} = 2$ Gyr and an
initial ISM mass of~$M_\text{g} = 0$.
We select an SFE timescale of~$\tau_\star = 15$ Gyr, motivated by the
observational result that dwarf galaxies have generally inefficient star
formation (e.g.,~\citealp{Hudson2015}; though not necessarily halo dwarfs that
formed in denser environments -- see discussion in~\citealt{Naidu2022}).
We additionally select a mass-loading factor of~$\eta = 10$ because the
strength of outflows should, in principle, contain information on the depth of
the gravity well of a given galaxy, with lower mass systems being more
efficient at ejecting material from the ISM.
If the SFH in this model were constant, the analytic formulae of
\citet{Weinberg2017} suggest that the equilibrium alpha element abundance
should be~$\sim16$\% of the solar oxygen abundance, in qualitative agreement
with the empirical mass-metallicity relation for galaxies
(\citealp{Tremonti2004, Gallazzi2005};~\citealp*{Zahid2011};
\citealp{Andrews2013, Kirby2013, Zahid2014}).
\par
With these choices regarding~$\tau_\star$ and~$\eta$, our parameters are in
the regime where the normalization of the infall history, and consequently the
SFH, is inconsequential to the predicted evolution of the abundances.
The appropriate likelihood function is therefore equation~\refp{eq:likelihood}
with normalized weights, whereas equation~\refp{eq:lnL_minus_weights} with
un-normalized weights would be the proper form if we had selected a
parametrization in which the absolute scale of the SFH impacts the enrichment
history.
Inspection of the average SFHs predicted by the~\textsc{UniverseMachine}
semi-analytic model for galaxy formation~\citep{Behroozi2019} suggests that the
onset of star formation tends to occur a little over~$\sim$13 Gyr ago across
many orders of magnitude in stellar mass extending as low as
$\text{M}_\star \approx 10^{7.2}~\msun$.
We therefore assume that the onset of star formation occurred~$\sim$13.2 Gyr
ago, allowing~$\sim$500 Myr between the Big Bang and the first stars.
We evolve this model for 10 Gyr exactly (i.e., the youngest stars in the mock
sample have an exact age of 3.2 Gyr), stopping short of 13.2 Gyr because
surviving dwarf galaxies and stellar streams often have their star formation
quenched (e.g.,~\citealp{Monelli2010a, Monelli2010b, Sohn2013, Weisz2014a,
Weisz2014b, Weisz2015}).
These choices are not intended to resemble any one galaxy, but instead to
qualitatively resemble some disrupted dwarf galaxy whose evolutionary
parameters can be re-derived using our likelihood function as a check that it
produces accurate best-fit parameters.
\par
As discussed in~\S~\ref{sec:onezone}, thoughout this paper we assume that the
IMF-averaged alpha element yield is exactly~$\yacc = 0.01$ and
$y_\alpha^\text{Ia} = 0$.
While loosely motivated by nucleosynthesis models in massive stars
\citep[e.g.,][]{Nomoto2013, Sukhbold2016, Limongi2018}, this choice is intended
to set some normalization of the effective yields which can be scaled up or
down to accommodate alternative choices.
If no scale is assumed, then extremely strong degeneracies arise in the
inferred yields, the strength of outflows~$\eta$, and the SFE timescale
$\tau_\star$ due to the yield-outflow degeneracy (see discussion in
Appendix~\ref{sec:degeneracy}).
We do not distinguish between alpha elements in this validation of our
likelihood function because, from a modelling standpoint, they can all be
treated the same with a metallicity-independent yield from CCSNe and negligible
yields from all other sources (at least for the lighter alpha elements such as
O and Mg;~\citealp{Johnson2019}).
In practice, however, we take O as the canonical alpha element when integrating
these models with~\vice, adopting~$Z_{\text{O},\odot} = 0.00572$ as the
abundance of O in the sun according to~\citet{Asplund2009} and consistent with
the recent revisions of~\citet*{Asplund2021}, though similar~\afe~ratios would
arise anyway if we instead took, e.g., Mg and asserted that [O/Mg]~$\approx 0$.
\par
\citet{Weinberg2017} adopt~$\yacc = 0.015$,~$\yfecc = 0.0012$ and
$\yfeia = 0.0017$ (see discussion in their~\S~2.2).
This massive star yield of Fe is appropriate for nucleosynthesis models in
which most~$M > 8~\msun$ stars explode as a CCSN~\citep[e.g.,][]{Woosley1995,
Chieffi2004, Chieffi2013, Nomoto2013} assuming a~\citet{Kroupa2001} IMF.
This SN Ia yield of Fe is based on the W70 explosion model of
\citet{Iwamoto1999} which produces~$\sim$0.77~\msun~of Fe per SN Ia event and
assuming that~$\scinote{2.2}{-3}~\msun^{-1}$ SNe Ia arise per solar mass of
star formation based on~\citet{Maoz2012a}.
Following these arguments, we scale these yields down by factors of~$\sim$2/3
such that $\yacc = 0.01$, adopting~$\yfecc = \scinote{8}{-4}$ and
$\yfeia = \scinote{1.1}{-3}$ in our mock samples.
We retain the assumption that~$\yacc = 0.01$ in our fits to our mock samples
but otherwise let the Fe yields~\yfecc~and~\yfeia~be free parameters to be
recovered by our likelihood function.
We use this procedure in our application to the H3 survey in~\S~\ref{sec:h3}
below as well.
We then sample~$N = 500$ stars from the underlying SFH each of which have -- in
the interest of mimicking the typical precision achieved by a spectroscopic
survey of a local group dwarf galaxy --~$\sigma_\afe = \sigma_\feh = 0.05$.
100 of these stars have age measurements with an uncertainty of
$\sigma_{\log_{10}(\text{age})} = 0.1$ (i.e.,~$\sim$23\% precision).

\subsection{Recovered Parameters of the Fiducial Mock}
\label{sec:mocks:recovered}

\begin{figure*}
\centering
\includegraphics[scale = 0.42]{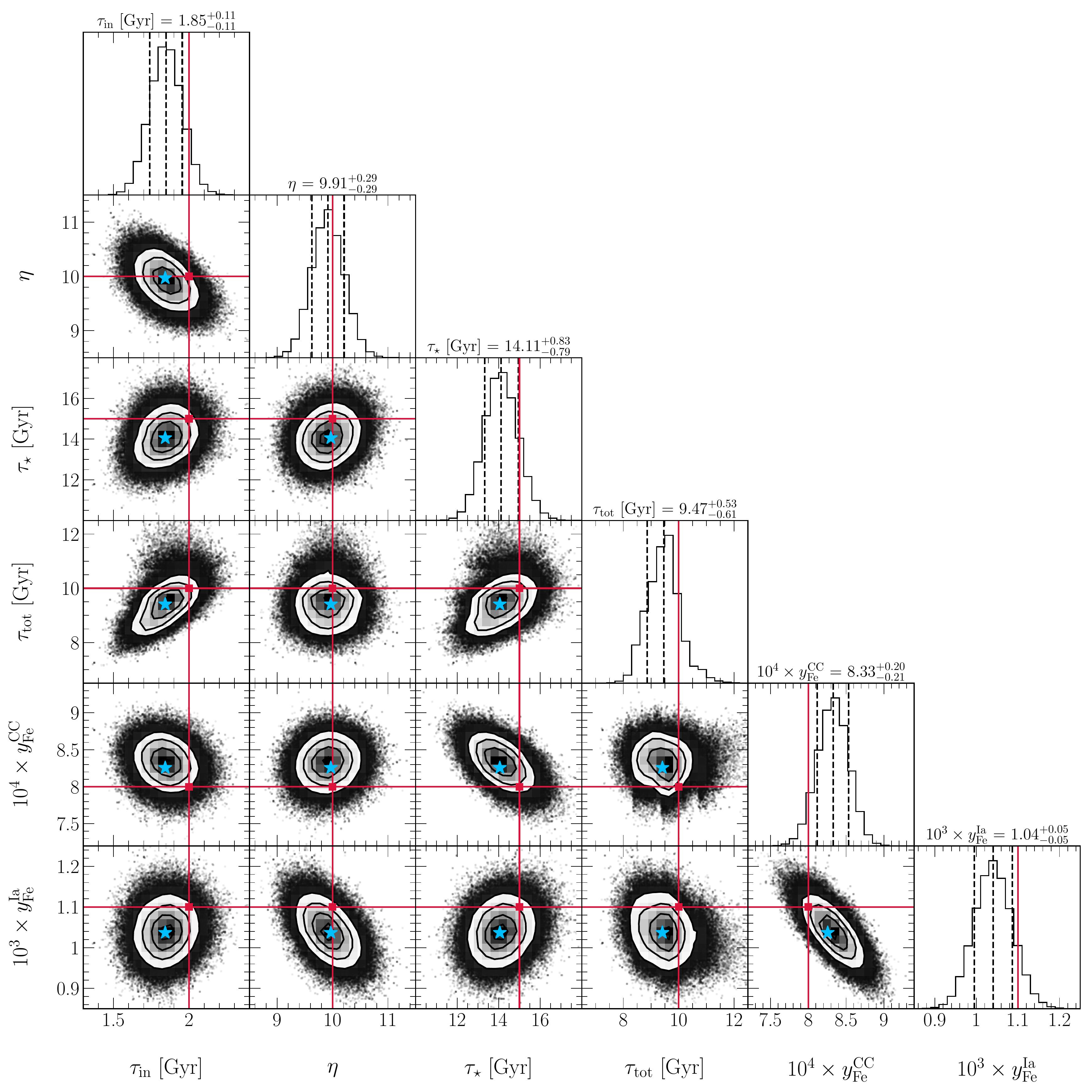}
\caption{
Posterior distributions obtained from applying our fitting method to our
fiducial mock sample (see Fig.~\ref{fig:fiducial_mock} and discussion
in~\S\S~\ref{sec:fitting} and~\ref{sec:mocks:fiducial}).
Panels below the diagonal show 2-dimensional cross-sections of the likelihood
function while panels along the diagonal show the marginalized distributions
along with the best-fit values and confidence intervals.
Blue stars mark the element of the Markov chain with the maximum likelihood.
Red ``cross-hairs'' denote the true, known values of the parameters from the
input model (see the top row of Table~\ref{tab:recovered_values}).
}
\label{fig:fiducial_mock_corner}
\end{figure*}

Fig.~\ref{fig:fiducial_mock} shows our fiducial mock in the observed space.
As intended by our parameter choices (see discussion
in~\S~\ref{sec:mocks:fiducial}), this sample qualitatively resembles a typical
disrupted dwarf galaxy -- dominated by old stars with metal-poor
($\feh \approx -1$) and alpha-enhanced ($\afe \approx +0.2$) modes in the MDF.
We now apply the method outline in~\S~\ref{sec:fitting} to recover the known
parameters of the input model.
Fig.~\ref{fig:fiducial_mock_corner} shows the resulting posterior distributions,
demonstrating that our likelihood function accurately recovers each parameter.
We include the predictions of the best-fit model in Fig.~\ref{fig:fiducial_mock},
finding excellent agreement with the input model.
To quantify the quality of the fit, for each datum~$\script{D}_i$ we find the
point along the track~$\script{M}_j$ with the maximum likelihood of observation
(i.e.,~$\{\script{D}_i,\script{M}_j~|~\ln L(\script{D}_i | \script{M}_j) =
\max(\ln L(\script{D}_i | \script{M}))\}$).
We then compute the chi-squared per degree of freedom diagnostic according to
\begin{equation}
\chi_\text{dof}^2 = \frac{1}{N_\text{obs} - N_\theta}
\sum_{i,j} \Delta_{ij} C_i^{-1} \Delta_{ij}^T,
\label{eq:chisquared_dof}
\end{equation}
where~$N_\text{obs}$ is the number of quantities in the observed sample,
$N_\theta$ is the number of model parameteres, and the summation is taken over
the pair-wise combinations of the data and model with the maximum likelihood of
observation.
Although marginalizing over the track~\script{M}~is necessary to derive
accurate best-fit parameters (see discussion below and in~\S~\ref{sec:fitting}),
it should be safe to estimate the quality of a fit by simply pairing each datum
with the most appropriate point on the track.
As noted in the middle panel of Fig.~\ref{fig:fiducial_mock}, our method
achieves~$\chi_\text{dof}^2 = 0.55$, indicating that we have perhaps
over-parametrized the data.
This result is unsurprising, however, because we have fit the mock data with
the exact, known parametrization of the evolutionary history and
nucleosynthetic yields of the input model in the interest of demonstrating
proof of concept that equation~\refp{eq:likelihood} provides accurate best-fit
values.
\par
Although it may appear that there are a worrying number of~$\gtrsim 1\sigma$
discrepancies in Fig.~\ref{fig:fiducial_mock_corner}, we demonstrate
in~\S~\ref{sec:mocks:variations} below that the differences between the known
and best-fit values here are consistent with randomly sampling from a Gaussian
distribution due to measurement uncertainty.
Although most cross sections of the posterior distribution are sufficiently
described by a multivariate Gaussian, there is some subtructure in the
likelihood distribution of~$\tau_\text{in}$, most noticeable in the
$\yfecc - \tau_\text{in}$ plane.
The MCMC algorithm naturally catches this structure, but it would be missed
under the assumption of Gaussianity as in, e.g., maximum a posteriori estimates.
There are a handful of degeneracies in the likelihood distribution of the
recovered parameters, which arise as a consequence of having an impact on the
same observable.
We discuss them individually below.
\par
\textit{The height of the ``plateau'' and position of the ``knee'' in the
evolutionary track.}
The plateau in the~\afe-\feh~plane occurs in our input model
at~$\afe_\text{CC} \approx +0.45$ and arises due to the IMF-averaged massive
star yields of alpha and iron-peak elements.
The knee occurs thereafter with the onset of SN Ia enrichment, a
nucleosynthetic source of Fe but negligible amounts of alpha elements like O
and Mg~\citep{Johnson2019}.
With fixed~\yacc, variations in~\yfecc~adjust the vertical height of the
plateau.
\citet{Weinberg2017} demonstrate that, to first order, the SFE timescale
$\tau_\star$ determines the metallicity~\feh~at which the knee occurs with
low~$\tau_\star$ models predicting a knee at high~\feh.
If a lowered plateau (i.e., higher~\yfecc) is accompanied by faster star
formation (i.e., lower~$\tau_\star$), the portion of the evolutionary track
in which~\afe~is decreasing occurs in a similar region of chemical space.
\yfecc~and~$\tau_\star$ are therefore inversely related when an overall scale
of nucleosynthetic yields is chosen.
When the overall scale is allowed to vary, we find a degeneracy of the opposite
sign (see discussion in Appendix~\ref{sec:degeneracy}).
\par
\textit{The endpoint of the model track and centroid of the MDF.}
These are the regions of chemical space where most of the
data are generally found, so for a given choice of~$\eta$, the~\textit{total}
Fe yield is well constrained observationally.
With only the total precisely determined,~\yfecc~and~\yfeia~are inversely
related.
On its own, adjusting~\yfeia~shifts the track vertically in the~\afe-\feh~plane
(there is horizontal movement as well, though the vertical movement is
stronger).
A downward shift in the predicted track (i.e., and increase in~$\yfeia$) can be
accompanied by a rightward shift (i.e., a decrease in~$\eta$) such that the
endpoint lies in the same location as the data.
$\yfeia$ and~$\eta$ are therefore inversely related, whereas the yield-outflow
degeneracy produces a direct relationship between these parameters
(see Appendix~\ref{sec:degeneracy}).
\par
\textit{The shape of the MDF.}
The~\afe~and~\feh~distributions are affected in a handful of ways by the
parameters of this input model.
The duration of star formation has the simplest effect of cutting off the MDF
at some abundance.
Inefficient star formation (i.e., high~$\tau_\star$) increases the frequency of
low metallicity stars because it takes significantly longer for the ISM to
reach the equilibrium abundance.
Sharp infall histories (i.e., low~$\tau_\text{in}$) predict wide MDFs because
the ISM mass declines with time through losses to star formation and the lack
of replenishment by accretion.
Metals are then deposited into a ``gas-starved'' reservoir, which then reaches
higher abundances due to a deficit of hydrogen and helium.
This effect is particularly strong for Fe because of the delayed nature of SN
Ia enrichment~\citep{Weinberg2017}.
These models achieve higher metallicities in the ISM, but their declining SFHs
produce a larger fraction of their stars early in their evolutionary history
when the abundances are lower than the late-time equilibrium abundance.
Consequently, the MDF that arises is wider for sharp infall histories but has
a peak in a similar position regardless of~$\tau_\text{in}$.
Folding these effects together, degeneracies arise in the inferred parameters
as a consequence of their effects on the MDF.
Between~$\tau_\text{in}$ and~$\tau_\text{tot}$, a sharp infall history can
broaden the MDF, but cutting off star formation earlier can allow the
distribution to remain peaked if the data suggest it.
Similarly, efficient star formation (i.e., low~$\tau_\star$) allows the ISM to
spend more time near its equilibrium abundance, enhancing the peak of the MDF,
but this change in shape can be reversed by cutting off star formation.
Between~$\tau_\text{in}$ and~$\eta$, a sharp infall history gives rise to a
high metallicity tail of the MDF, but increasing the strength of outflows
can lower the overall metallicity if this tail is too metal-rich compared to
the data.
\par
We emphasize that our fits achieve this level of precision by selecting an
overall scale for nucleosynthetic yields and outflows ($\yacc = 0.01$; see
discussion in~\S~\ref{sec:onezone} and Appendix~\ref{sec:degeneracy}).
Any GCE parameter that influences the centroid of the MDF or the position or
shape of the evolutionary track in abundance space is subject to the
yield-outflow degeneracy.
Given an overall scale of yields, set here by choosing~\yacc, a sample like
our fiducial mock gives quite precise constraints on all model parameters.
If we modify our choice of~\yacc, we would find similar predictions by
adjusting our Fe yields,~$\tau_\star$ and~$\eta$.
If~\yacc~is instead allowed to vary as a free parameter, then the degeneracies
are strong, but~$\tau_\text{in}$ and~$\tau_\text{tot}$ remain well constrained
due to their impact on the MDF shape.
\par
In conducting these tests against mock samples, we find that the two central
features of this method are essential to ensuring the accuracy of the best-fit
parameters.
When either the weighted likelihood or the marginalization over the track
(see discussion in~\S~\ref{sec:fitting}) are omitted, the fit fails to recover
the parameters of the input model with discrepancies at the many-$\sigma$ level
between the best-fit and known values.
For this reason, we caution against the reliability of GCE parameters inferred
from simplified likelihood estimates, such as matching each datum with the
nearest point on the track.

{\renewcommand{\arraystretch}{1.8}
\begin{table*}
\caption{
Known (top row) and recovered best-fit values of the evolutionary parameters
of the input GCE model to out mock samples.
From left to right: the variation of our fiducial mock sample, the e-folding
timescale of the infall history~$\tau_\text{in}$, the outflow mass-loading
factor~$\eta$, the SFE timescale~$\tau_\star$, the duration of star formation
$\tau_\text{tot}$, the IMF-averaged Fe yield from CCSNe~\yfecc~and the
DTD-integrated Fe yield from SNe Ia~\yfeia.
Each variation has the same evolutionary parameters as the input model, but
has either a different sample size (top block), measurement uncertainty
in~\feh~and~\afe~abundances (top-middle block), measurement uncertainty
in~$\log_{10}(\text{age})$ (bottom-middle block), or fraction of the
sample with available age measurements (bottom block).
The values taken in the fiducial mock sample are marked in bold.
We provide illustrations of the accuracy and precision of these fits in
Figs.~\ref{fig:accuracy} and~\ref{fig:precision}, respectively.
}
\begin{tabularx}{\textwidth}{l @{\extracolsep{\fill}} c c c c c c}
\hline
Mock Sample & $\tau_\text{in}$ & $\eta$ & $\tau_\star$ & $\tau_\text{tot}$ &
\yfecc & \yfeia
\\
\hline
\hline
\null &
2 Gyr &
10  &
15 Gyr &
10 Gyr &
\scinote{8.00}{-4} &
\scinote{1.10}{-3}
\\
\hline
\hline
$N = 20$ &
$2.55^{+0.75}_{-0.45}$ Gyr &
$8.39^{+1.11}_{-1.30}$  &
$14.35^{+5.56}_{-3.32}$ Gyr &
$10.60^{+1.65}_{-1.09}$ Gyr &
$\scinote{7.90^{+1.20}_{-1.90}}{-4}$ &
$\scinote{1.36^{+0.33}_{-0.23}}{-3}$
\\
$N = 50$ &
$2.13^{+0.42}_{-0.36}$ Gyr &
$10.39^{+0.80}_{-0.76}$  &
$13.75^{+2.79}_{-2.38}$ Gyr &
$11.25^{+1.37}_{-1.76}$ Gyr &
$\scinote{(8.30 \pm 0.60)}{-4}$ &
$\scinote{(0.95 \pm 0.14)}{-3}$
\\
$N = 100$ &
$2.06^{+0.27}_{-0.26}$ Gyr &
$9.88^{+0.64}_{-0.62}$  &
$15.06^{+2.00}_{-1.79}$ Gyr &
$11.52^{+1.06}_{-1.30}$ Gyr &
$\scinote{(8.10 \pm 0.40)}{-4}$ &
$\scinote{(1.08 \pm 0.09)}{-3}$
\\
$N = 200$ &
$2.10^{+0.18}_{-0.17}$ Gyr &
$10.11^{+0.45}_{-0.43}$  &
$14.61^{+1.34}_{-1.18}$ Gyr &
$10.60^{+1.07}_{-0.86}$ Gyr &
$\scinote{(7.70 \pm 0.30)}{-4}$ &
$\scinote{(1.14 \pm 0.07)}{-3}$
\\
$\bm{N = 500}$ &
$\bm{1.85 \pm 0.11}$ Gyr &
$\bm{9.91 \pm 0.29}$  &
$\bm{14.11^{+0.83}_{-0.79}}$ \textbf{Gyr} &
$\bm{9.47^{+0.53}_{-0.61}}$ \textbf{Gyr} &
$\bm{\scinote{8.30^{+0.20}_{-0.21}}{-4}}$ &
$\bm{\scinote{(1.04 \pm 0.05)}{-3}}$
\\
$N = 1000$ &
$2.05^{+0.09}_{-0.08}$ Gyr &
$9.72 \pm 0.20$  &
$14.62^{+0.57}_{-0.56}$ Gyr &
$9.83^{+0.38}_{-0.39}$ Gyr &
$\scinote{(8.10 \pm 0.10)}{-4}$ &
$\scinote{(1.14 \pm 0.03)}{-3}$
\\
$N = 2000$ &
$2.00 \pm 0.05$ Gyr &
$10.26 \pm 0.15$  &
$15.82^{+0.44}_{-0.42}$ Gyr &
$10.30^{+0.25}_{-0.32}$ Gyr &
$\scinote{(8.00 \pm 0.10)}{-4}$ &
$\scinote{(1.09 \pm 0.02)}{-3}$
\\
\hline
\hline
$\sigma_\text{[X/Y]} = 0.01$ &
$1.89 \pm 0.10$ Gyr &
$10.25 \pm 0.28$  &
$15.06^{+0.52}_{-0.47}$ Gyr &
$9.70^{+0.51}_{-0.59}$ Gyr &
$\scinote{(8.00 \pm 0.10)}{-4}$ &
$\scinote{(1.09 \pm 0.02)}{-3}$
\\
$\sigma_\text{[X/Y]} = 0.02$ &
$1.92^{+0.10}_{-0.09}$ Gyr &
$10.10 \pm 0.25$  &
$14.71^{+0.56}_{-0.55}$ Gyr &
$9.79^{+0.45}_{-0.40}$ Gyr &
$\scinote{(8.10 \pm 0.10)}{-4}$ &
$\scinote{1.08^{+0.02}_{-0.03}}{-3}$
\\
$\bm{\sigma_\textbf{[X/Y]} = 0.05}$ &
$\bm{1.85 \pm 0.11}$ Gyr &
$\bm{9.91 \pm 0.29}$  &
$\bm{14.11^{+0.83}_{-0.79}}$ \textbf{Gyr} &
$\bm{9.47^{+0.53}_{-0.61}}$ \textbf{Gyr} &
$\bm{\scinote{8.30^{+0.20}_{-0.21}}{-4}}$ &
$\bm{\scinote{(1.04 \pm 0.05)}{-3}}$
\\
$\sigma_\text{[X/Y]} = 0.1$ &
$2.00^{+0.13}_{-0.12}$ Gyr &
$9.88^{+0.31}_{-0.33}$  &
$13.39 \pm 1.02$ Gyr &
$11.10^{+1.00}_{-0.84}$ Gyr &
$\scinote{8.50^{+0.40}_{-0.30}}{-4}$ &
$\scinote{(1.01 \pm 0.07)}{-3}$
\\
$\sigma_\text{[X/Y]} = 0.2$ &
$2.22 \pm 0.21$ Gyr &
$9.83^{+0.58}_{-0.67}$  &
$18.21^{+2.19}_{-2.02}$ Gyr &
$10.32^{+1.05}_{-0.67}$ Gyr &
$\scinote{(8.70 \pm 0.70)}{-4}$ &
$\scinote{(1.05 \pm 0.14)}{-3}$
\\
$\sigma_\text{[X/Y]} = 0.5$ &
$2.73^{+0.82}_{-0.60}$ Gyr &
$10.05^{+1.22}_{-1.26}$  &
$12.52^{+3.75}_{-3.35}$ Gyr &
$9.00^{+1.26}_{-0.95}$ Gyr &
$\scinote{7.50^{+1.80}_{-1.60}}{-4}$ &
$\scinote{(1.12 \pm 0.31)}{-3}$
\\
\hline
\hline
$\sigma_{\log_{10}(\text{age})} = 0.02$ &
$2.08^{+0.09}_{-0.08}$ Gyr &
$9.84^{+0.24}_{-0.26}$  &
$14.69^{+0.50}_{-0.46}$ Gyr &
$10.41^{+0.47}_{-0.41}$ Gyr &
$\scinote{(8.10 \pm 0.20)}{-4}$ &
$\scinote{1.11^{+0.05}_{-0.04}}{-3}$
\\
$\sigma_{\log_{10}(\text{age})} = 0.05$ &
$1.96 \pm 0.11$ Gyr &
$9.88^{+0.32}_{-0.30}$  &
$15.70^{+0.71}_{-0.68}$ Gyr &
$9.95^{+0.63}_{-0.53}$ Gyr &
$\scinote{(8.00 \pm 0.20)}{-4}$ &
$\scinote{1.11^{+0.05}_{-0.04}}{-3}$
\\
$\bm{\sigma_{\log_{10}(\textbf{age})} = 0.1}$ &
$\bm{1.85 \pm 0.11}$ Gyr &
$\bm{9.91 \pm 0.29}$  &
$\bm{14.11^{+0.83}_{-0.79}}$ \textbf{Gyr} &
$\bm{9.47^{+0.53}_{-0.61}}$ \textbf{Gyr} &
$\bm{\scinote{8.30^{+0.20}_{-0.21}}{-4}}$ &
$\bm{\scinote{(1.04 \pm 0.05)}{-3}}$
\\
$\sigma_{\log_{10}(\text{age})} = 0.2$ &
$2.20^{+0.18}_{-0.17}$ Gyr &
$9.83^{+0.28}_{-0.27}$  &
$15.19 \pm 1.11$ Gyr &
$10.76^{+0.85}_{-0.93}$ Gyr &
$\scinote{(8.00 \pm 0.20)}{-4}$ &
$\scinote{1.11^{+0.05}_{-0.04}}{-3}$
\\
$\sigma_{\log_{10}(\text{age})} = 0.5$ &
$2.25^{+0.20}_{-0.25}$ Gyr &
$9.86^{+0.28}_{-0.30}$  &
$16.24^{+1.44}_{-1.62}$ Gyr &
$11.38^{+1.00}_{-1.34}$ Gyr &
$\scinote{(8.00 \pm 0.20)}{-4}$ &
$\scinote{(1.10 \pm 0.05)}{-3}$
\\
$\sigma_{\log_{10}(\text{age})} = 1$ &
$1.69^{+0.35}_{-0.32}$ Gyr &
$9.53 \pm 0.29$  &
$12.38^{+2.27}_{-2.08}$ Gyr &
$8.66^{+1.86}_{-1.74}$ Gyr &
$\scinote{(8.30 \pm 0.30)}{-4}$ &
$\scinote{(1.15 \pm 0.06)}{-3}$
\\
\hline
\hline
$f_\text{age} = 0$ &
$1.65^{+0.55}_{-0.37}$ Gyr &
$9.39^{+0.30}_{-0.29}$  &
$11.80^{+3.36}_{-2.44}$ Gyr &
$7.35^{+2.62}_{-1.74}$ Gyr &
$\scinote{(8.30 \pm 0.40)}{-4}$ &
$\scinote{1.19^{+0.08}_{-0.07}}{-3}$
\\
$f_\text{age} = 0.1$ &
$1.75^{+0.16}_{-0.17}$ Gyr &
$10.06^{+0.29}_{-0.28}$  &
$13.65^{+1.22}_{-1.12}$ Gyr &
$8.84 \pm 0.87$ Gyr &
$\scinote{(8.40 \pm 0.20)}{-4}$ &
$\scinote{(1.06 \pm 0.05)}{-3}$
\\
$\bm{f_\textbf{age} = 0.2}$ &
$\bm{1.85 \pm 0.11}$ Gyr &
$\bm{9.91 \pm 0.29}$  &
$\bm{14.11^{+0.83}_{-0.79}}$ \textbf{Gyr} &
$\bm{9.47^{+0.53}_{-0.61}}$ \textbf{Gyr} &
$\bm{\scinote{8.30^{+0.20}_{-0.21}}{-4}}$ &
$\bm{\scinote{(1.04 \pm 0.05)}{-3}}$
\\
$f_\text{age} = 0.3$ &
$1.94^{+0.11}_{-0.10}$ Gyr &
$9.80^{+0.27}_{-0.28}$  &
$14.26^{+0.74}_{-0.67}$ Gyr &
$9.89^{+0.54}_{-0.48}$ Gyr &
$\scinote{(8.00 \pm 0.20)}{-4}$ &
$\scinote{(1.10 \pm 0.04)}{-3}$
\\
$f_\text{age} = 0.4$ &
$1.91^{+0.09}_{-0.10}$ Gyr &
$10.07^{+0.32}_{-0.30}$  &
$16.79^{+0.81}_{-0.83}$ Gyr &
$10.34^{+0.61}_{-0.50}$ Gyr &
$\scinote{(7.80 \pm 0.20)}{-4}$ &
$\scinote{(1.12 \pm 0.05)}{-3}$
\\
$f_\text{age} = 0.5$ &
$2.00 \pm 0.10$ Gyr &
$10.16^{+0.30}_{-0.29}$  &
$15.46^{+0.70}_{-0.69}$ Gyr &
$9.83^{+0.48}_{-0.40}$ Gyr &
$\scinote{(7.80 \pm 0.20)}{-4}$ &
$\scinote{1.12^{+0.05}_{-0.04}}{-3}$
\\
$f_\text{age} = 0.6$ &
$2.18 \pm 0.09$ Gyr &
$9.65^{+0.27}_{-0.25}$  &
$14.25^{+0.67}_{-0.64}$ Gyr &
$10.49^{+0.44}_{-0.37}$ Gyr &
$\scinote{(7.80 \pm 0.20)}{-4}$ &
$\scinote{(1.15 \pm 0.04)}{-3}$
\\
$f_\text{age} = 0.7$ &
$1.99 \pm 0.08$ Gyr &
$9.81^{+0.28}_{-0.27}$  &
$14.92^{+0.68}_{-0.62}$ Gyr &
$10.25^{+0.46}_{-0.37}$ Gyr &
$\scinote{(8.10 \pm 0.20)}{-4}$ &
$\scinote{(1.08 \pm 0.04)}{-3}$
\\
$f_\text{age} = 0.8$ &
$2.06 \pm 0.09$ Gyr &
$9.53^{+0.29}_{-0.26}$  &
$15.18^{+0.63}_{-0.59}$ Gyr &
$9.76^{+0.36}_{-0.33}$ Gyr &
$\scinote{(7.90 \pm 0.20)}{-4}$ &
$\scinote{(1.15 \pm 0.05)}{-3}$
\\
$f_\text{age} = 0.9$ &
$1.93 \pm 0.08$ Gyr &
$10.41 \pm 0.31$  &
$16.23^{+0.73}_{-0.70}$ Gyr &
$10.03^{+0.39}_{-0.33}$ Gyr &
$\scinote{(7.70 \pm 0.20)}{-4}$ &
$\scinote{(1.14 \pm 0.04)}{-3}$
\\
$f_\text{age} = 1$ &
$2.13 \pm 0.09$ Gyr &
$9.44^{+0.28}_{-0.27}$  &
$15.67^{+0.64}_{-0.60}$ Gyr &
$10.21^{+0.35}_{-0.31}$ Gyr &
$\scinote{(8.00 \pm 0.20)}{-4}$ &
$\scinote{(1.15 \pm 0.05)}{-3}$
\\
\hline
\hline
\end{tabularx}
\label{tab:recovered_values}
\end{table*}
}

\begin{figure*}
\centering
\includegraphics[scale = 0.42]{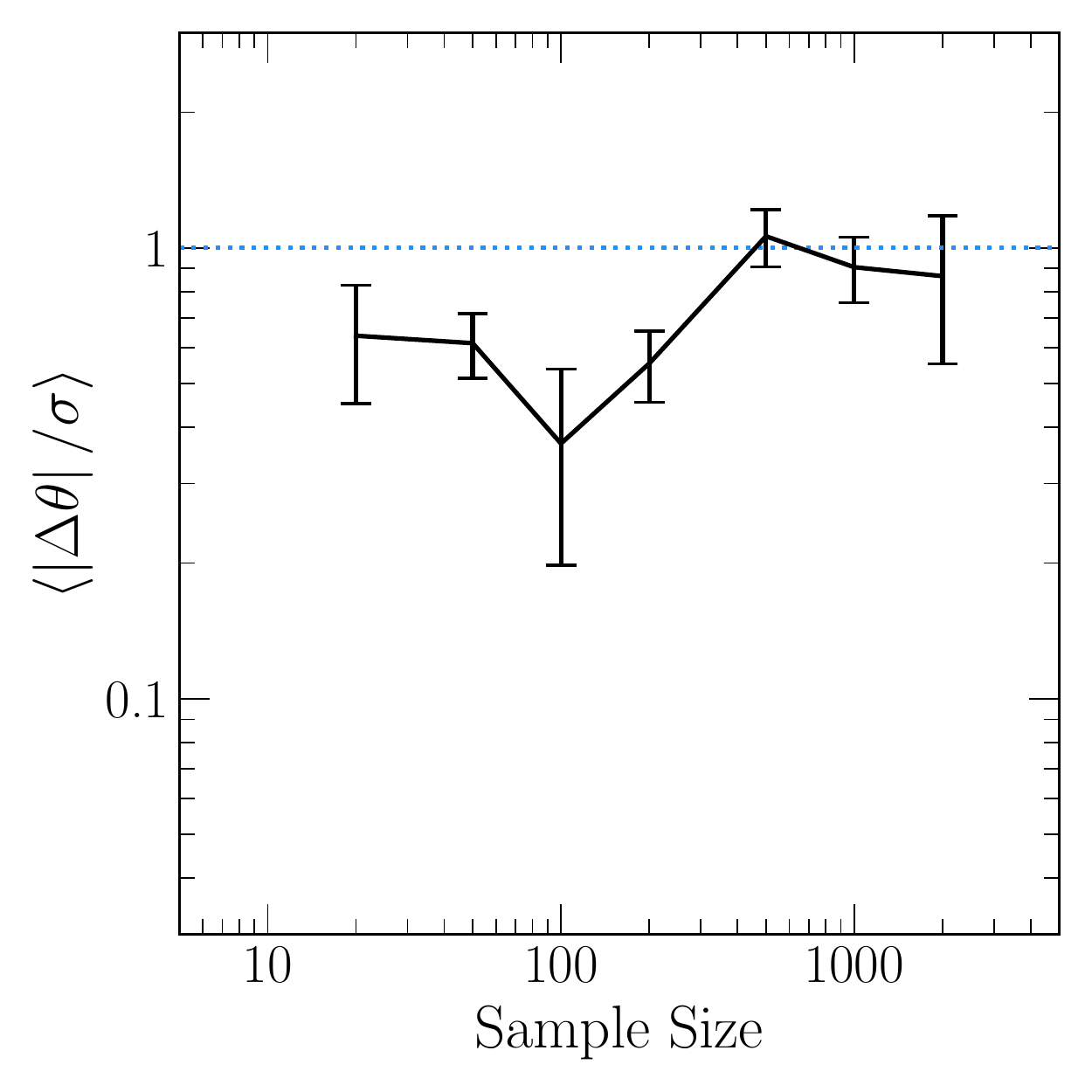}
\includegraphics[scale = 0.42]{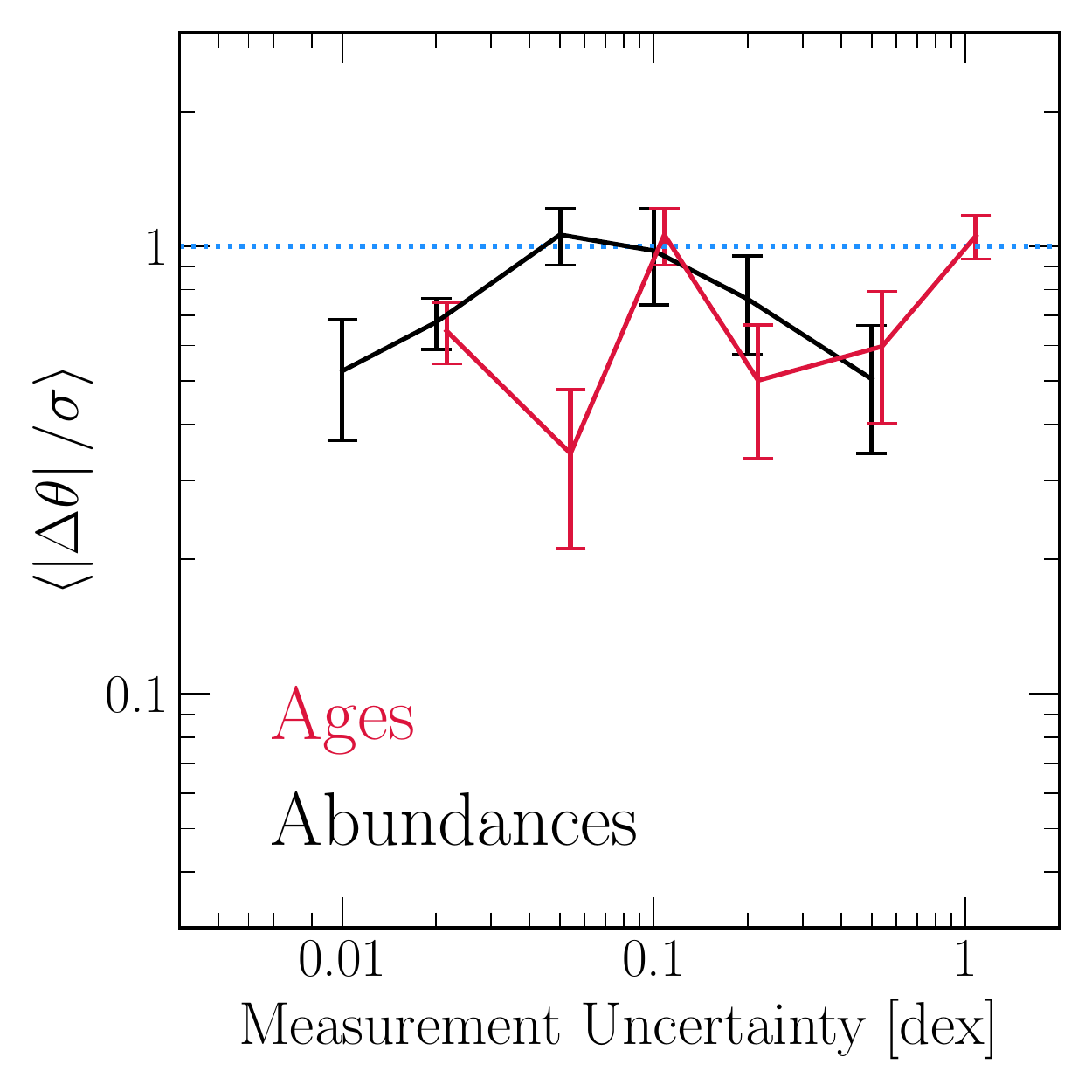}
\includegraphics[scale = 0.42]{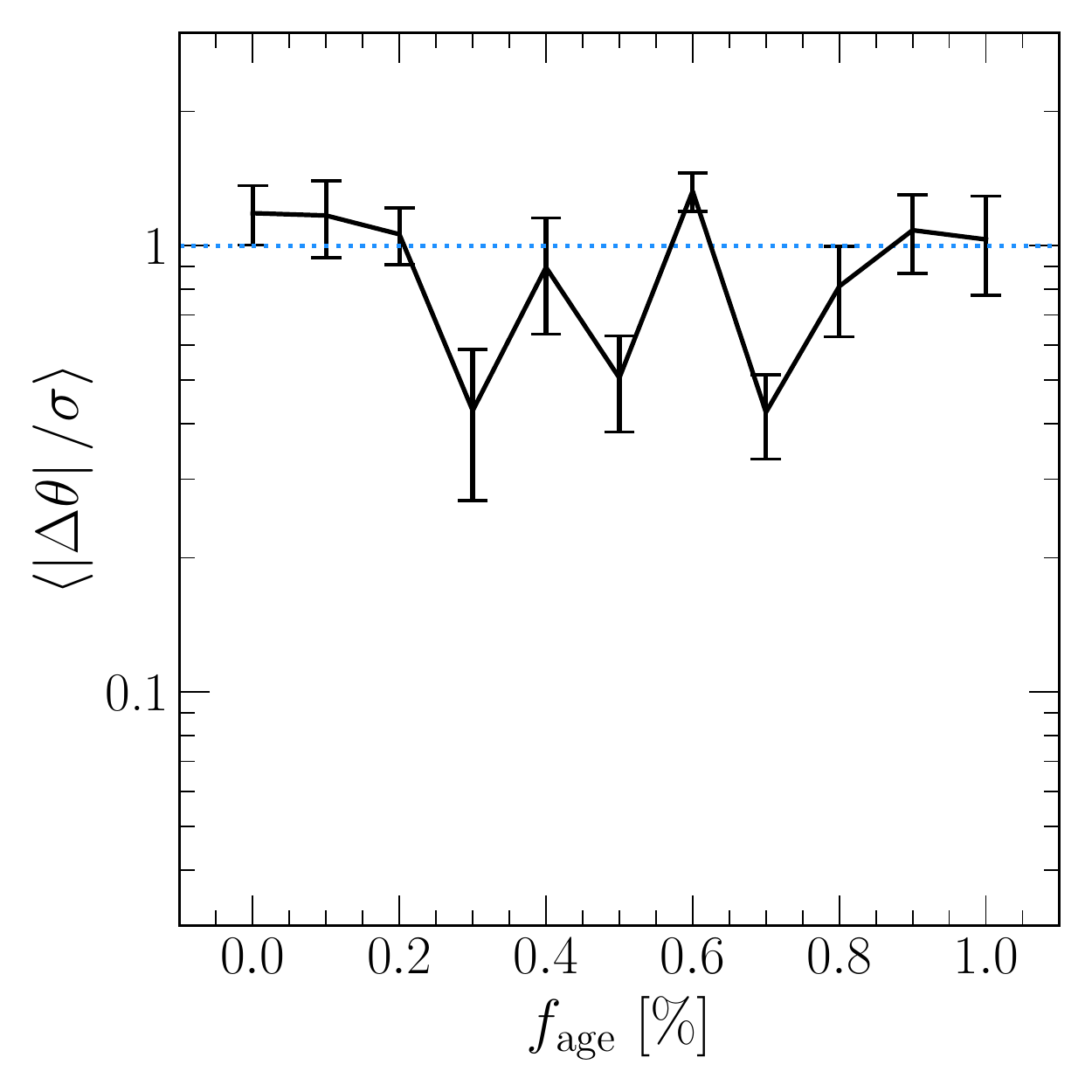}
\caption{
Differences between input model parameters and recovered best-fit values.
Each point is the mean deviation~$\left|\Delta\theta\right|$ for each of the
six free parameters in Table~\ref{tab:recovered_values} (i.e.,
$\{\theta\} = \{\tau_\text{in}, \eta, \tau_\star, \tau_\text{tot}, \yfecc,
\yfeia\}$) in units of the best-fit uncertainty~$\sigma$.
Our mock samples vary in terms of their sample size (left), measurement
precision in~\feh~and~\afe~abundances (middle, black), measurement precision in
$\log_{10}(\text{age})$ (middle, red), and the fraction of the sample with
available age measurements (right).
Error bars denote the error in the mean deviation of the six free parameters.
Blue dotted lines mark~$\langle \Delta \theta / \sigma \rangle = 1$, the
expected mean offset due to randomly sampling from a Gaussian distribution.
}
\label{fig:accuracy}
\end{figure*}

\subsection{Variations in Sample Size, Measurement Precision and the
Availability of Age Information}
\label{sec:mocks:variations}

We now explore variations of our fiducial mock sample.
We retain the same evolutionary parameters of the input model (see discussion
in~\S~\ref{sec:mocks:fiducial}), but each variant differs in one of the
following:
\begin{itemize}

	\item Sample size.

	\item Measurement precision in~\feh~and~\afe.

	\item Measurement precision in~$\log_{10}(\text{age})$.

	\item The fraction of the sample that has age measurements.

\end{itemize}
The left-hand column of Table~\ref{tab:recovered_values} provides a summary of
the values we take as exploratory cases with the fiducial mock marked in bold.
In the remaining columns, we provide the associated values derived for each
GCE parameter~$\theta$ along with their~$1\sigma$ confidence intervals.
The sample sizes we consider are intended to reflect the range that is
typically achieved in disrupted dwarf galaxies where the proximity might
allow individual age estimates for main sequence turnoff stars.
Because of their distance and low stellar mass, dwarf galaxies are considerably
less conducive to the large sample sizes achieved by Milky Way surveys like
APOGEE~\citep{Majewski2017} and GALAH~\citep{DeSilva2015, Martell2017}.
Our choices in measurement precision are intended to reflect typical values
achieved by modern spectroscopic surveys.
Although deriving elemental abundances through spectroscopy is a nontrivial
problem known to be affected by systematics~\citep[e.g.,][]{Anguino2018},
stellar age measurements
are generally the more difficult of the two~\citep{Soderblom2010, Chaplin2013}.
The age measurements may therefore be available for only a small portion of the
sample and are often less precise than the abundances ($f_\text{age} = 20$\%
and~$\sigma_\feh = \sigma_\afe = 0.05$ versus
$\sigma_{\log_{10}(\text{age})} = 0.1$ in our fiducial mock).
In practice, however, uncertainties vary with stellar mass; for example, hot
main sequence turnoff stars have precise ages but poorly constrained
abundances due to the lack of lines in their spectra.
\par
\begin{figure*}
\centering
\includegraphics[scale = 0.55]{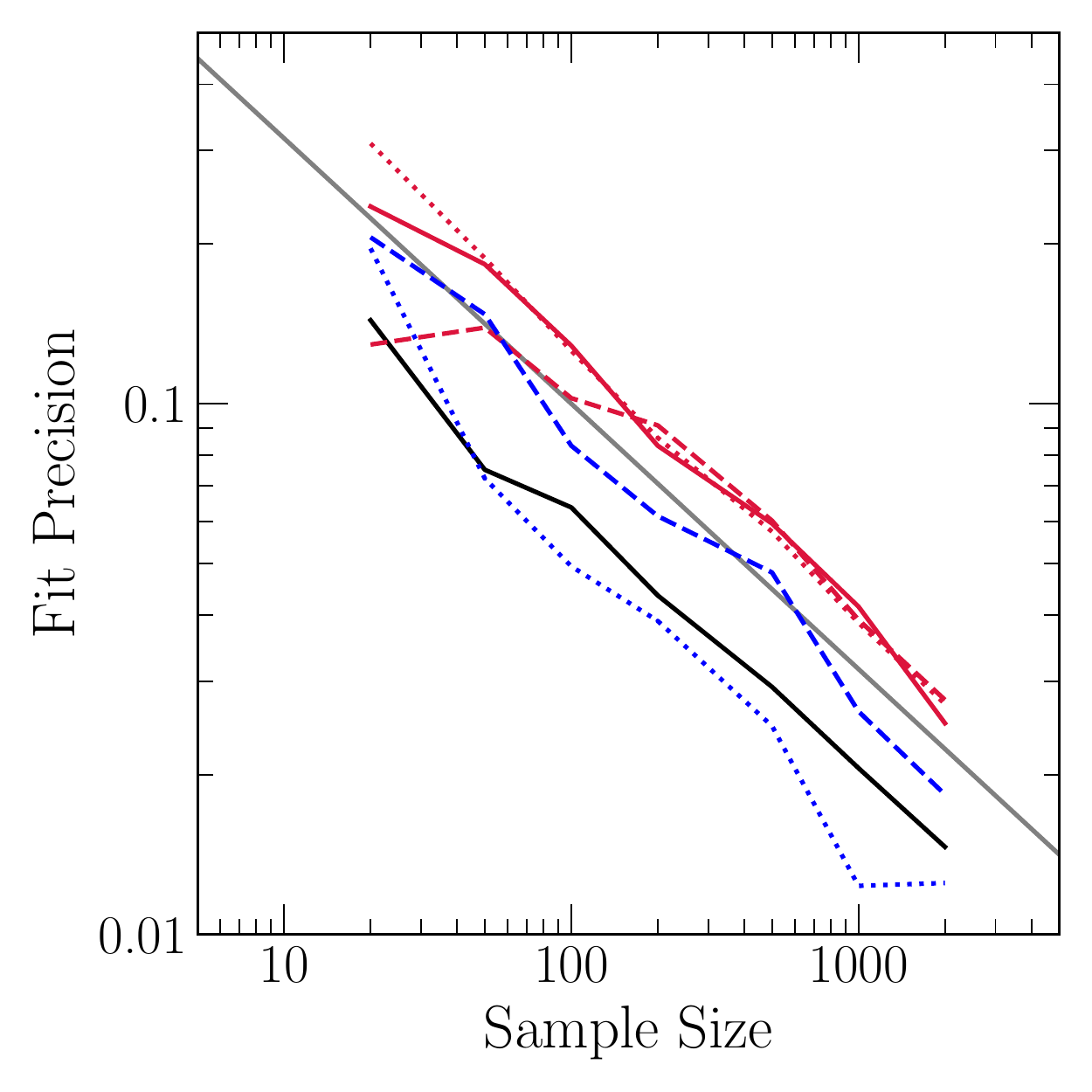}
\includegraphics[scale = 0.55]{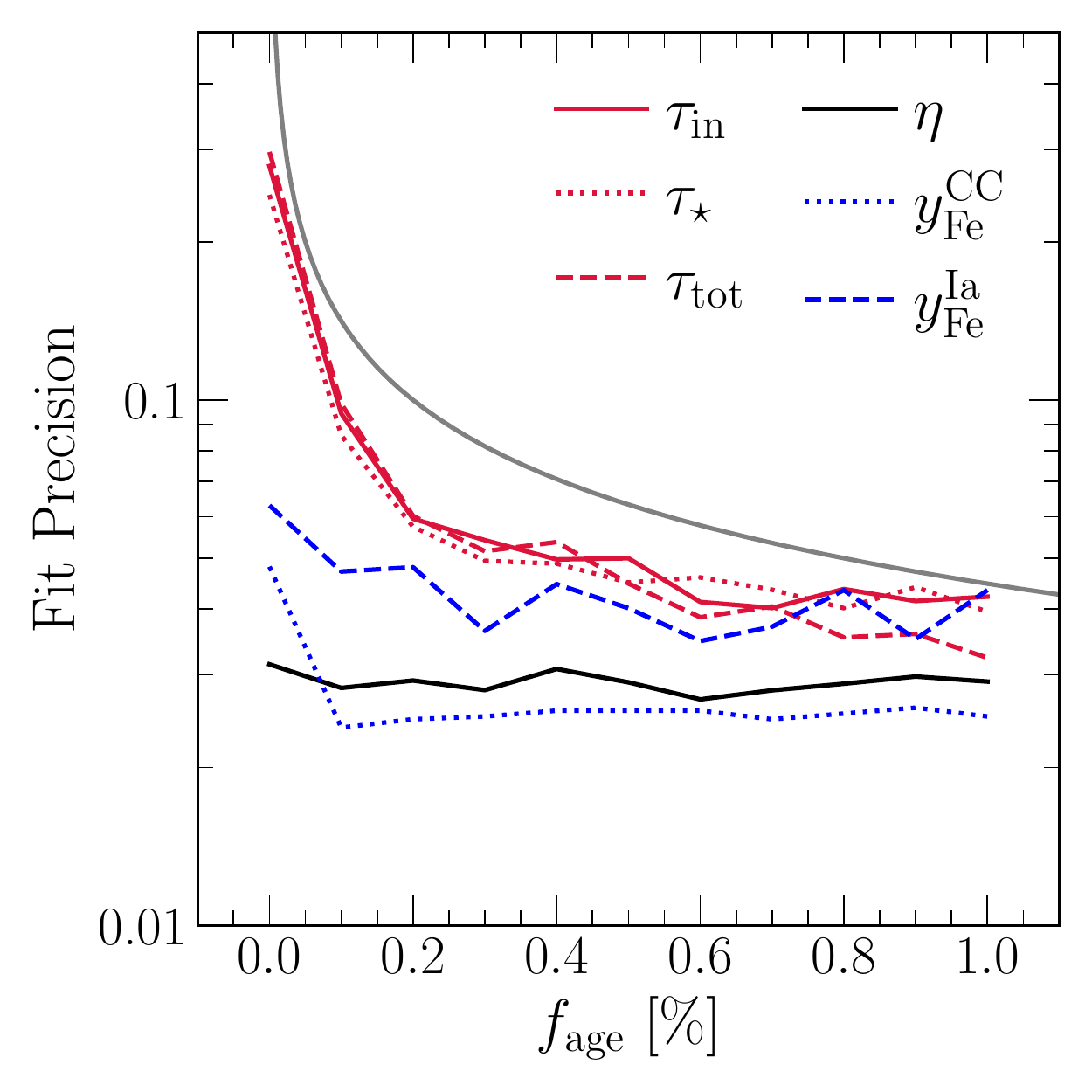}
\includegraphics[scale = 0.55]{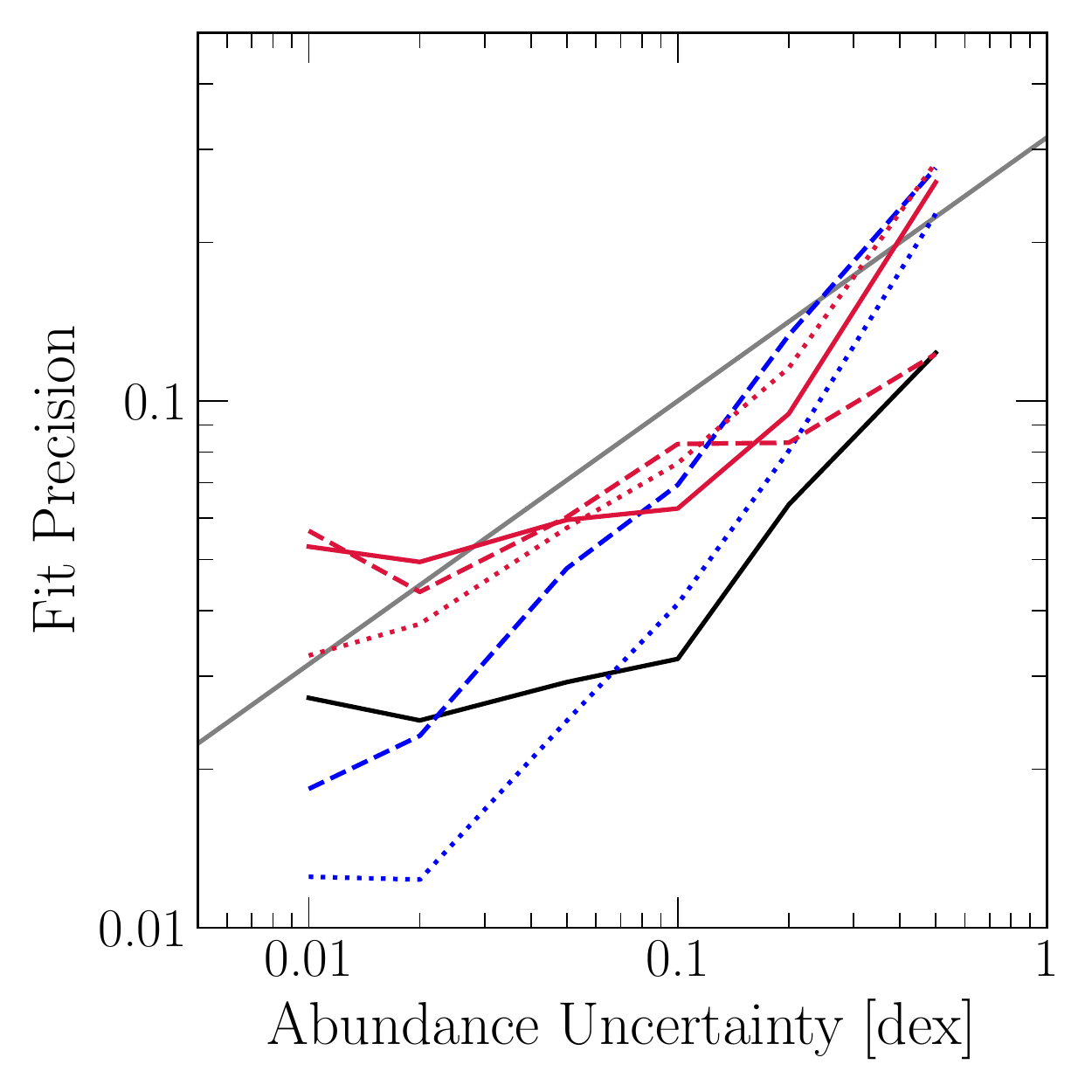}
\includegraphics[scale = 0.55]{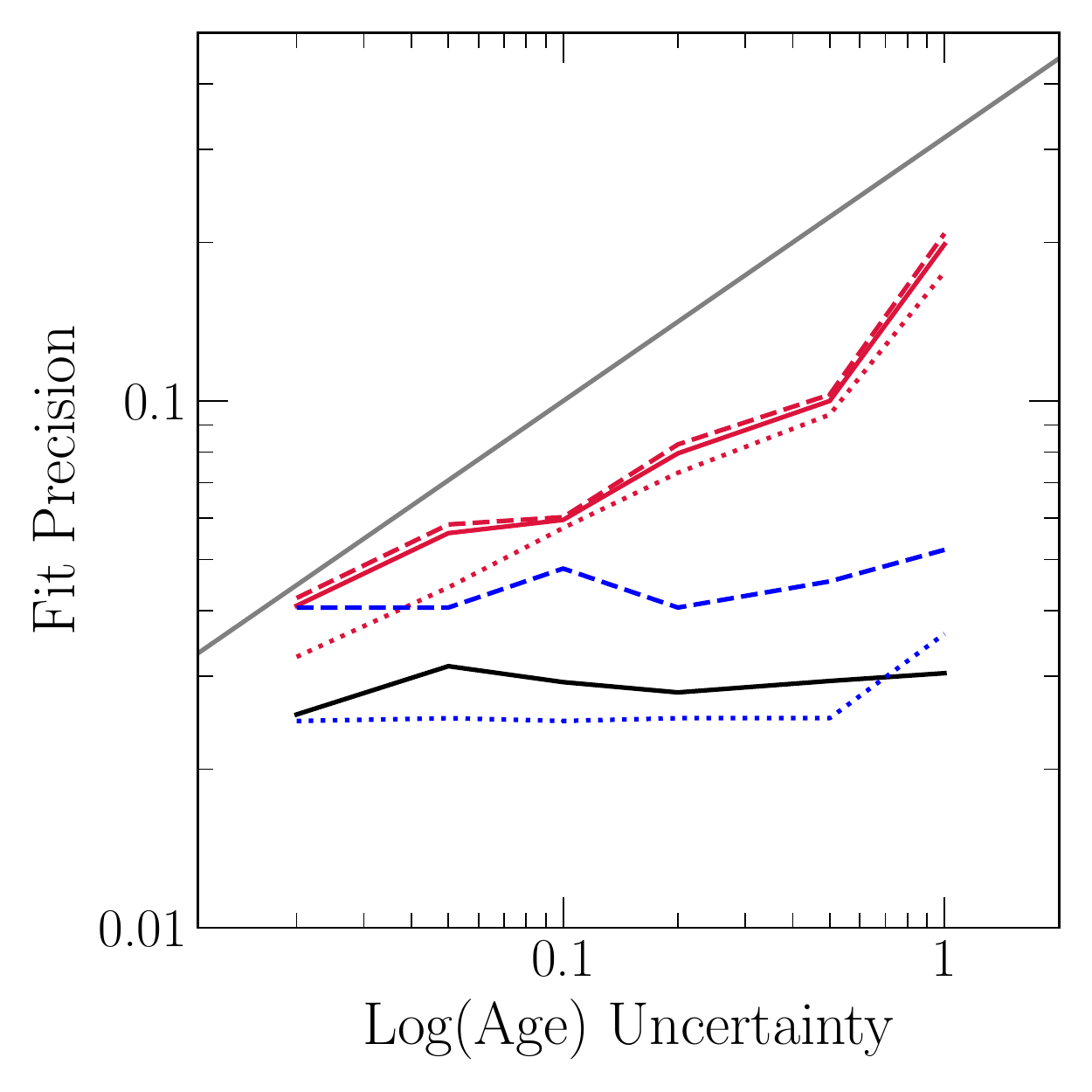}
\caption{
Precision of our fitting method.
For a fit uncertainty~$\sigma$ and deviation from the known value
$\Delta\theta$, we compute precision according to
$\left|\Delta\theta\right| / \sigma$ for each of the six free parameters
in Table~\ref{tab:recovered_values} and plot them as a function
of sample size (top left), the fraction of the sample with age information (top
right), abundance uncertainties (bottom left), and age uncertainties (bottom
right).
Grey lines in each panel denote~$x^{\pm0.5}$ scaling where~$x$ is the quantity
on the~$y$-axis.
We plot timescales in red, Fe yields in blue, and the mass-loading
factor~$\eta$ in black in all panels according to the legend.
}
\label{fig:precision}
\end{figure*}

Fig.~\ref{fig:accuracy} demonstrates the accuracy of our fitting method with
respect to variations in these details surrounding the data.
We compute the deviation between each re-derived parameter~$\theta$
(i.e.,~$\tau_\text{in}$,~$\eta$,~$\tau_\star$, etc.) and its known value from
the input model, then divide by the fit uncertainty~$\sigma_\theta$ and plot
the mean on the y-axis.
Under all variants that we explore, our likelihood function accurately recovers
the input parameters to~$\sim1\sigma$ or slightly better.
This deviation is exactly as expected when the uncertainties are described by a
Gaussian random process, wherein the most likely deviation from the true value
is exactly~$1\sigma$.
This expectation holds even with infinite data, though in that limit
the~$1\sigma$ uncertainty interval becomes arbitrarily small.
This demonstrates that equation~\refp{eq:likelihood} provides accurate best-fit
parameters even when the sample size is as low as~$N \approx 20$, when the
measurement uncertainties are as imprecise as~$\sigma_\text{[X/Y]} \approx 0.5$
and~$\sigma_{\log_{10}(\text{age})} \approx 1$, or even when there is no age
information available at all.
The precision of the fit will indeed suffer in such cases (see Fig.
\ref{fig:precision} and associated discussion below), but the inferred
parameters will remain accurate nonetheless.

We have explored alternate parametrizations of our mock sample's evolutionary
history and indeed found that our method accurately recovers the parameters
in all cases.
For example, one is a case in which we build in a significant starburst,
finding that we accurately recover both the timing and the strength of the
burst.
We have also explored an infall rate that varies sinusoidally about some mean
value, mimicking natural fluctuations in the accretion history or a series of
minor starbursts.
Although idealized and potentially unrealistic, our likelihood function
accurately recovers the amplitude, phase and frequency in this case as well.
Of course, the parametrization itself must allow for such possibilities, but
we stick to smooth SFHs for the remainder of these tests.
\par
Fig.~\ref{fig:precision} demonstrates how the uncertainty of each best-fit
parameter is affected by these details of the sample.
With differences in the normalization, the precision of each inferred parameter
scales with sample size approximately as~$N^{-0.5}$.
In general, the mass-loading factor~$\eta$ and the Fe yields are constrained
more precisely than the timescales.
The primary exception to this rule is when the abundance uncertainties are
large compared to the age uncertainties, in which case the Fe yields are
constrained to a similar precision as~$\tau_\text{in}$ and~$\tau_\star$
but~$\tau_\text{tot}$ is determined more precisely.
The Fe yields are, unsurprisingly, the most sensitive parameters to the
abundance uncertainties, while~$\eta$ can be determined with~$\sim$10\% precision
even with highly imprecise measurements ($\sigma_\text{[X/Y])} \approx 0.5$).
Even with imprecise abundances, the centroid of the MDF can still be robustly
determined with a sufficiently large sample, which allows a precise inference
of the strength of winds due to its impact on the equilibrium metallicity (for
an assumed scale of nucleosynthetic yields such as~$\yacc = 0.01$ in this
paper).
\par
Only the inferred timescales are impacted by the availability of age
information and the uncertainties thereof.
Even with order of magnitude uncertainties in stellar ages, however, the
evolutionary timescales of our mock samples are recovered to~$\sim$20\%
precision.
Interestingly the introduction of age information to the sample impacts the
fit uncertainty only for~$f_\text{age} \lesssim 30$\%.
Above this value, there is only marginal gain in the precision of best-fit
timescales.
These results suggest that authors seeking to determine best-fit evolutionary
parameters for one-zone models applied to any sample should focus their efforts
on sample size and precise abundance measurements with age information being
a secondary consideration.
Thankfully, abundances are generally easier than ages to measure on a
star-by-star basis~\citep{Soderblom2010, Chaplin2013}.

\section{Application to Observations}
\label{sec:h3}

We now apply our likelihood function (Eq.~\ref{eq:likelihood}) to two
disrupted dwarf galaxies in the Milky Way stellar halo.
The first is a relatively well-studied system: GSE~\citep{Belokurov2018,
Helmi2018, Haywood2018, Myeong2018, Mackereth2019}, believed to be responsible
for a major merger event early in the Milky Way's history~\citep{Gallart2019,
Bonaca2020, Chaplin2020, Montalban2021, Xiang2022} which
contributed~$\sim$$10^9~\msun$ of total stellar mass to the Galaxy
\citep{Deason2019, Fattahi2019, Mackereth2019, Vincenzo2019, Kruijssen2020,
Han2022}, including eight globular clusters in the stellar
halo~\citep{Myeong2018, Massari2019, Kruijssen2019, Forbes2020}.
GSE is a good test case for this method both because it is the dominant
structure in the Milky Way's inner halo~\citep{Helmi2018} and because we can
compare to independent constraints thanks to the amount of attention it has
received in the literature.

{\renewcommand{\arraystretch}{1.8}
\begin{table*}
\caption{
Inferred best-fit parameters for the fits to our GSE and Wukong/LMS-1 samples.
The parametrization is the same as the input GCE model to our mock samples (see
discussion in~\S~\ref{sec:mocks}).
The quality of each fit~$\chi_\text{dof}^2$ computed according to
equation~\refp{eq:chisquared_dof} is noted at the bottom.
}
\begin{tabularx}{\textwidth}{c @{\extracolsep{\fill}} c c c c}
\hline
Parameter & GSE (with ages) & GSE (without ages) & Wukong/LMS-1 (yields are
fixed) & Wukong/LMS-1 (yields are free parameters)
\\
\hline
\hline
$\tau_\text{in}$ &
$1.01 \pm 0.13$ Gyr &
$2.18^{+0.43}_{-0.56}$ Gyr &
$3.08^{+3.19}_{-1.16}$ Gyr &
$14.80^{+22.19}_{-11.10}$ Gyr
\\
$\eta$ &
$8.84^{+0.83}_{-0.89}$ &
$9.56^{+0.72}_{-0.77}$ &
$47.99^{+4.76}_{-4.98}$ &
$18.26^{+15.63}_{-12.59}$
\\
$\tau_\star$ &
$16.08^{+1.33}_{-1.26}$ Gyr &
$26.60^{+4.83}_{-6.11}$ Gyr &
$44.97^{+7.85}_{-6.77}$ Gyr &
$43.98^{+24.85}_{-12.48}$ Gyr
\\
$\tau_\text{tot}$ &
$5.40^{+0.32}_{-0.31}$ Gyr &
$10.73^{+1.76}_{-2.69}$ Gyr &
$3.36^{+0.55}_{-0.47}$ Gyr &
$2.33^{+1.92}_{-0.78}$ Gyr
\\
\yfecc &
$\scinote{7.78^{+0.37}_{-0.38}}{-4}$ &
$\scinote{7.25^{+0.55}_{-0.57}}{-4}$ &
N/A &
$\scinote{6.17^{+0.55}_{-0.70}}{-4}$
\\
\yfeia &
$\scinote{1.23^{+0.11}_{-0.10}}{-3}$ &
$\scinote{1.06^{+0.10}_{-0.09}}{-3}$ &
N/A &
$\scinote{2.42^{+0.88}_{-0.65}}{-3}$
\\
\hline
\hline
$\chi_\text{dof}^2$ & $1.34$ & $2.18$ & $0.98$ & $0.84$
\\
\hline
\hline
\end{tabularx}
\label{tab:results}
\end{table*}
}

The second is a less well-studied system: Wukong/LMS-1, a structure chemically
distinct from GSE which sits between it and the Helmi stream~\citep{Helmi1999}
in energy-angular momentum space~\citep{Naidu2020, Yuan2020} that
formed from an M$_\star \approx \scinote{1.3}{7}~\msun$ disrupted
galaxy~\citep{Naidu2022}.
Wukong/LMS-1 is an interesting system to investigate with our method
because it displays a ``classic'' enrichment history with an obvious ``knee''
in the evolutionary track near~$\feh \approx -2.8$ (see Fig.~\ref{fig:wukong}
below).
It has been associated~\citep{Malhan2022} with the most metal-poor streams in
the halo~\citep[e.g.,][]{Roederer2019, Wan2020, Martin2022} and a high fraction
of carbon-enhanced metal-poor stars given its low stellar mass
\citep{Shank2022, Zepeda2022}, marking it as a disrupted dwarf with a
potentially remarkable chemical history.
We make use of data from the H3 survey (see discussion
in~\S~\ref{sec:h3:survey} below) and discuss our GCE model fits to GSE and
Wukong/LMS-1 in~\S\S~\ref{sec:h3:gse} and~\ref{sec:h3:wukong}, comparing our
results for the two galaxies in~\S~\ref{sec:h3:comparison}.

\subsection{The H3 Survey}
\label{sec:h3:survey}

The H3 survey~\citep{Conroy2019} is collecting medium-resolution spectra
of~$\sim$300,000 stars in high-latitude fields ($\left|b\right| > 20^\circ$).
Spectra are collected from the Hectochelle instrument on the MMT
\citep{Szentgyorgyi2011}, which delivers~$R \approx$~32,000 spectra over the
wavelength range of~$5150 - 5300$~\AA.
Spectral lines in this wavelength range are dominated by iron-peak elements and
the MgI triplet (see Fig. 6 of~\citealt{Conroy2019}).
Throughout this section, the alpha element abundances we refer to are therefore
Mg abundances specifically, whereas in previous sections an alpha element
refers to any species where the only statistically significant enrichment
source is a metallicity-dependent yield from massive stars.
\par
The survey selection function is deliberately simple: the primary sample
consists of stars with~$r$ band magnitudes of~$15 < r < 18$
and~\gaia~\citep{Gaia2016} parallaxes~$<$ 0.3 mas (this threshold has evolved
over the course of the survey as the~\gaia~astrometry has become more precise).
Stellar parameters are estimated by the~\textsc{MINESweeper} program
\citep{Cargile2020}, which fits grids of isochrones, synthetic spectra and
photometry to the Hectochelle spectrum and broadband photometry from~\gaia,
Pan-STARRS~\citep{Chambers2016}, SDSS~\citep{York2000}, 2MASS
\citep{Skrutskie2006} and WISE~\citep{Wright2010} with the~\gaia~parallax
used as a prior.
The fitted parameters include radial velocity, spectrophotometric distance,
reddening,~\feh,~\afe~and age.
The default analysis includes a complicated prior on age and distance
(see~\citealt{Cargile2020} for details).
We have also re-fit high signal-to-noise data with a flat age prior for cases
where ages play an important role.
In this paper we use the catalog which uses this flat age prior.

\subsection{\gaia-Sausage Enceladus}
\label{sec:h3:gse}

\begin{figure*}
\centering
\includegraphics[scale = 0.65]{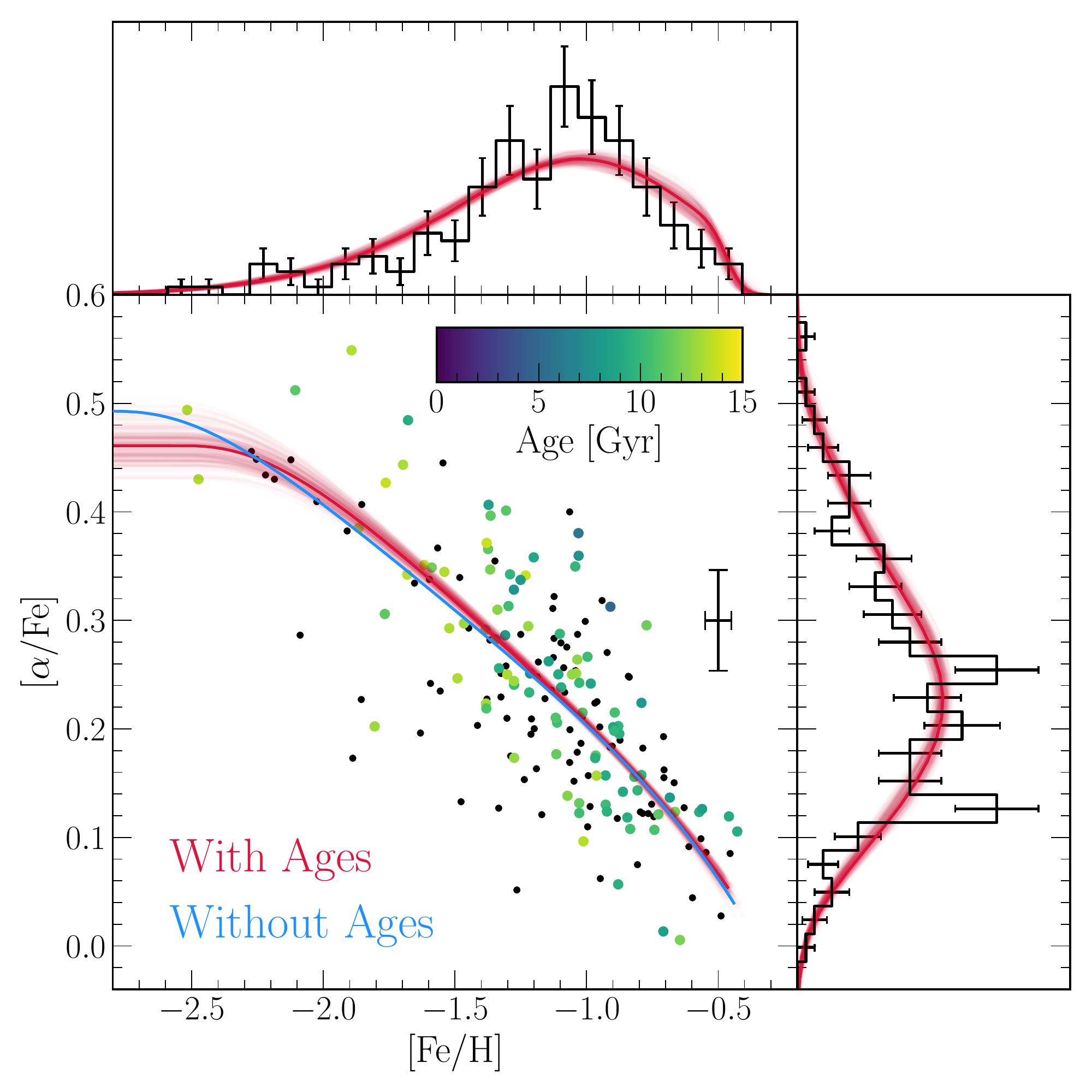}
\includegraphics[scale = 0.54]{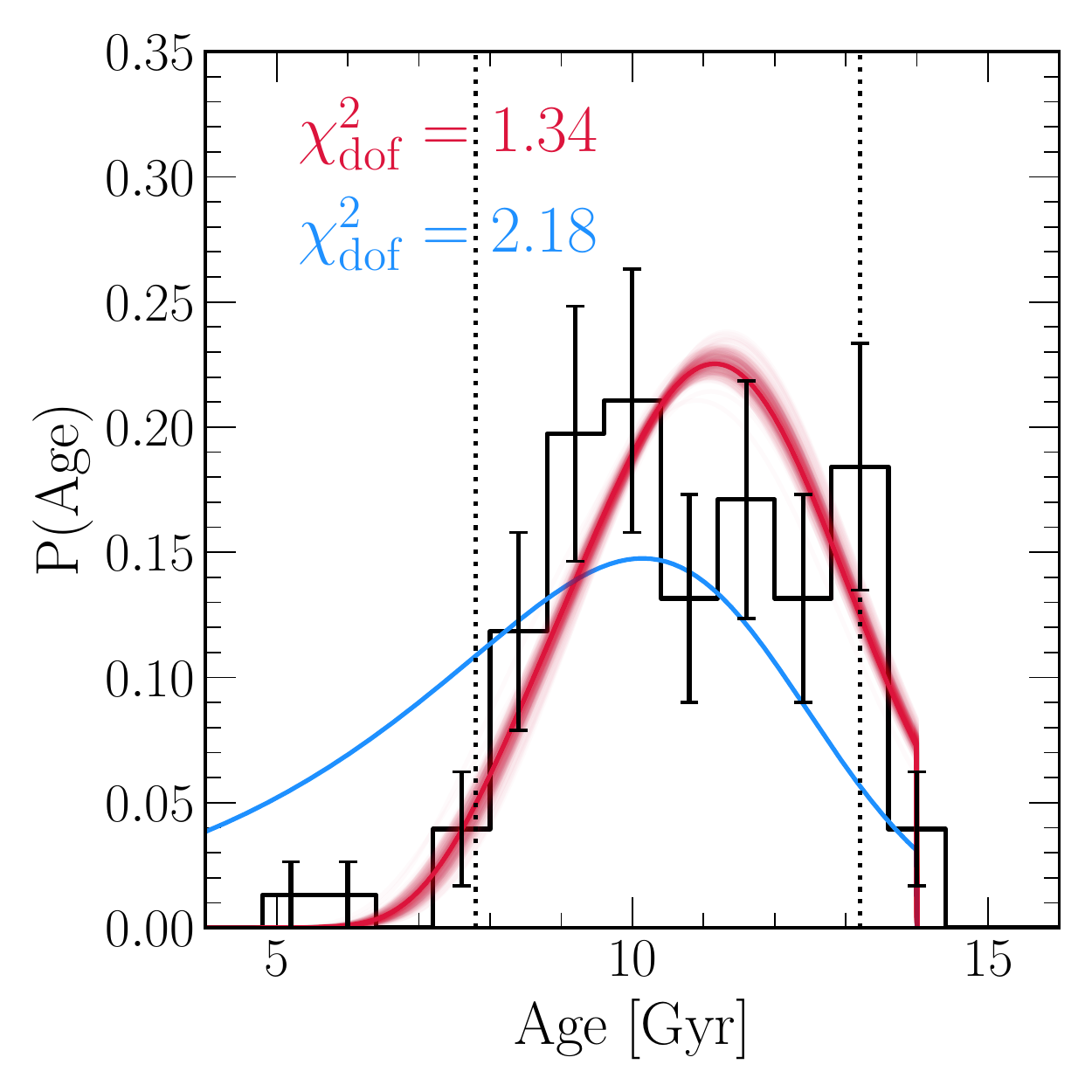}
\includegraphics[scale = 0.53]{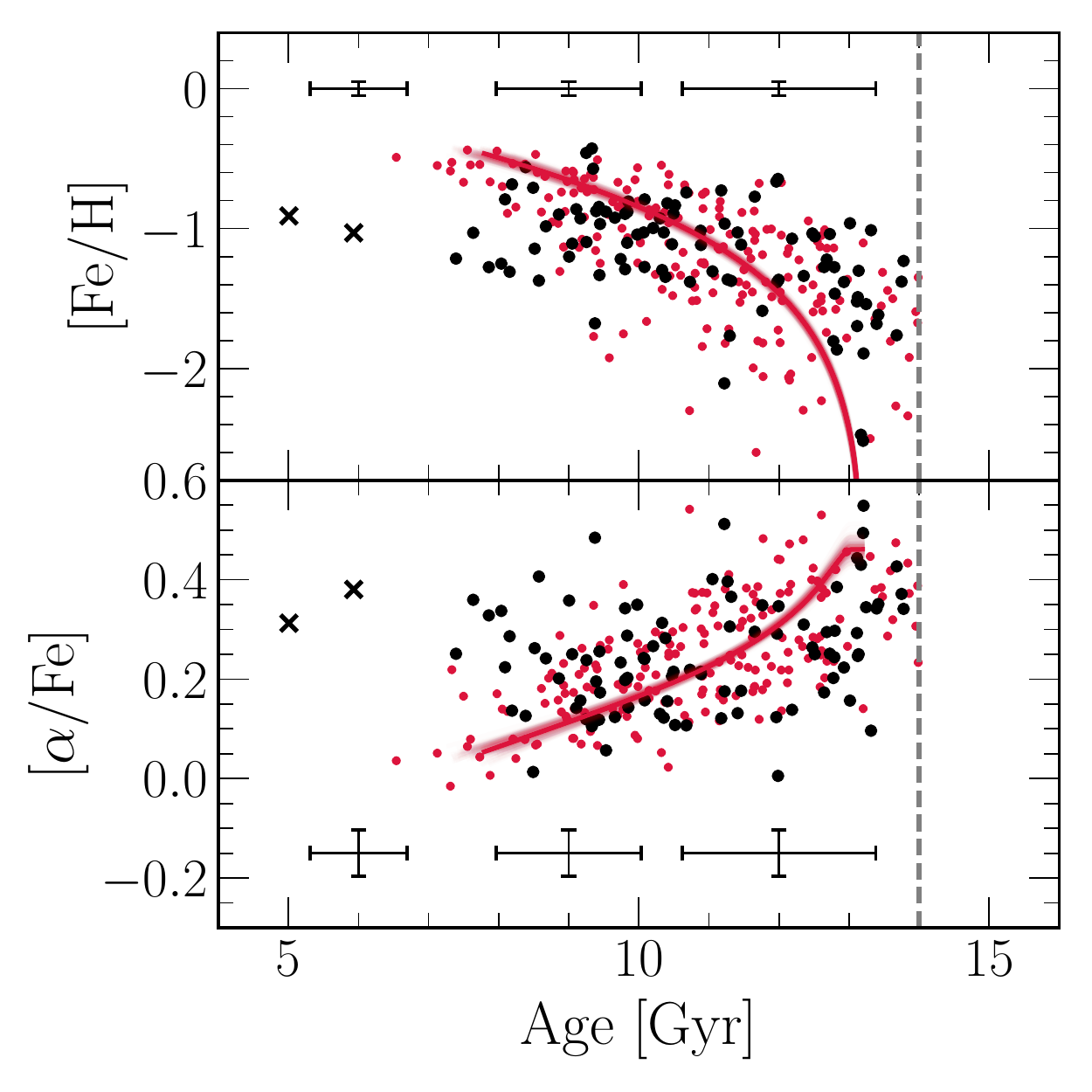}
\caption{
Our GSE sample.
Red lines in all panels denote the best-fit one-zone model, while the
blue lines in the top and bottom left panels denote the best-fit model obtained
when excluding age measurements from the fit.
Distributions in~\feh,~\afe~and age are convolved with the median uncertainty
of the sample (see discussion in~\S~\ref{sec:h3:gse}).
We additionally subsample 200 sets of parameter choices from our Markov chain
and plot their predictions as highly transparent lines to offer a sense of the
fit uncertainty.
Error bars in each distribution indicate a~$\sqrt{N}$ uncertainty associated
with random sampling.
\textbf{Top}: Our sample in chemical space and the associated marginalized
distributions.
Stars with age measurements are colour coded accordingly and are otherwise
plotted in black.
The median~\feh~and~\afe~uncertainty in the sample is shown by the error bar
to the right of the data.
\textbf{Bottom left}: The age distribution of our GSE sample (black,
binned).
\textbf{Bottom right}: Age-\feh~(top) and age-\afe~(bottom) relations
The median~\feh,~\afe~and age uncertainties are shown by the error bars at the
top and bottom of each panel.
We plot the two stars that we exclude from our fit as black X's (likely
blue stragglers; see discussion in~\S~\ref{sec:h3:gse}).
Red points denote~$N = 95$ stars (the same size as the stars with ages in our
GSE sample) drawn from out best-fit model and perturbed by the median age
uncertainty of the sample.
}
\label{fig:gse}
\end{figure*}

\begin{figure*}
\centering
\includegraphics[scale = 0.45]{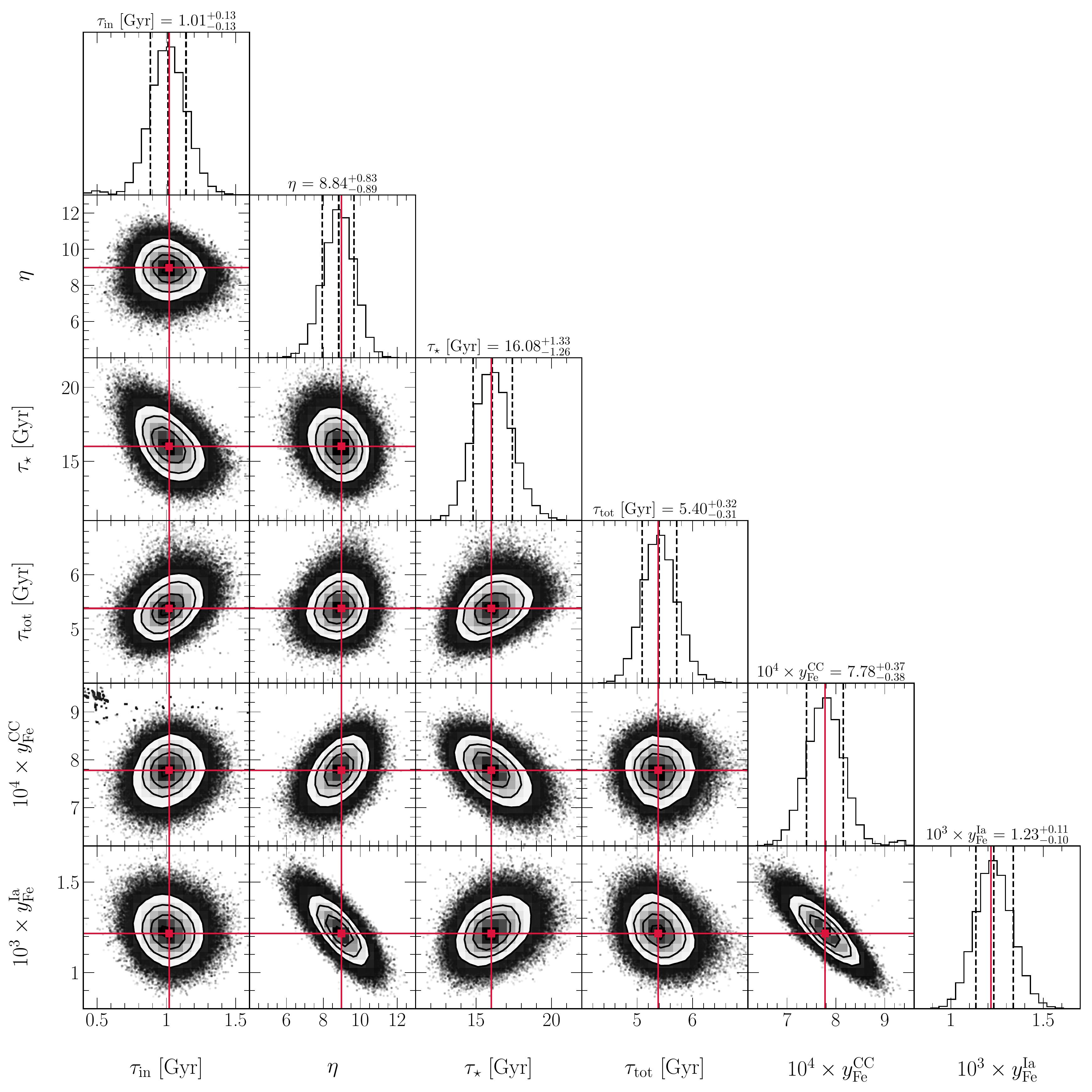}
\caption{
Posterior distributions for an exponential infall history applied to our GSE
sample.
The parametrization is the same as the input model to our mock samples (see
discussion in~\S~\ref{sec:mocks:fiducial}).
Panels below the diagonal show 2-dimensional cross-sections of the likelihood
function while panels along the diagonal show the marginalized distributions
along with the best-fit values and confidence intervals.
Red ``cross-hairs'' mark the element of the Markov chain with the maximum
statistical likelihood.
The points in the upper left corner of the~$\yfecc - \tau_\text{in}$ plane
are a part of an extended tail of the likelihood distribution which does
not appear in other panels when zoomed in on the peak.
}
\label{fig:gse_corner}
\end{figure*}

\begin{figure}
\centering
\includegraphics[scale = 0.65]{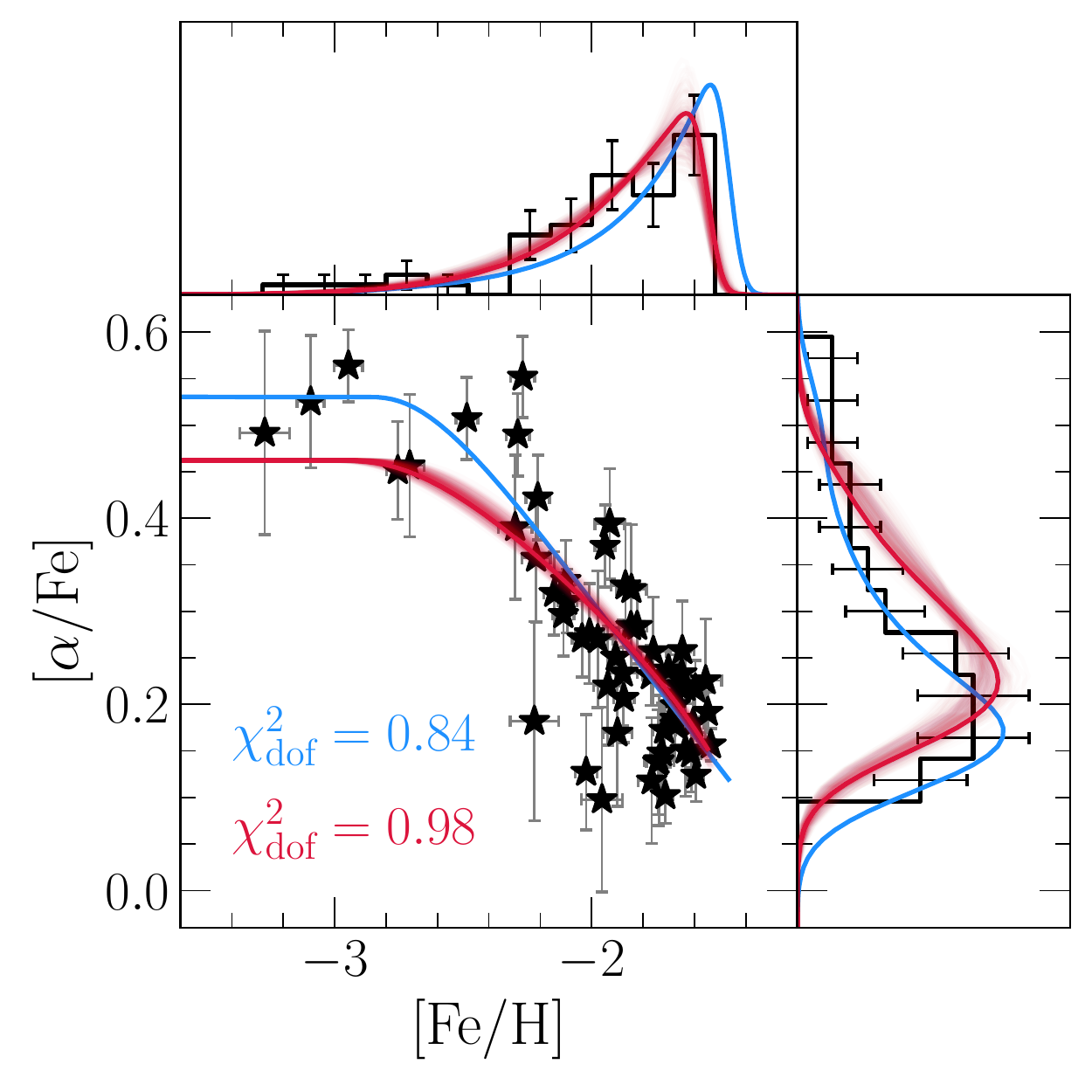}
\caption{
Our Wukong/LMS-1 sample in the~\afe-\feh~plane and the associated marginalized
distributions.
Error bars indicate uncertainties on individual abundances in the central panel
and a~$\sigma = \sqrt{N}$ uncertainty from sampling noise in the top and right
panels.
Red lines denote our best-fit chemical evolution model (see discussion
in~\S~\ref{sec:h3:wukong}), with 200 additional sets of parameter choices
subsampled from our Markov chain to give a sense of the fit precision.
Blue lines denote an alternate fit in which we allow the Fe yields to vary as
free parameters.
}
\label{fig:wukong}
\end{figure}

\begin{figure*}
\centering
\includegraphics[scale = 0.6]{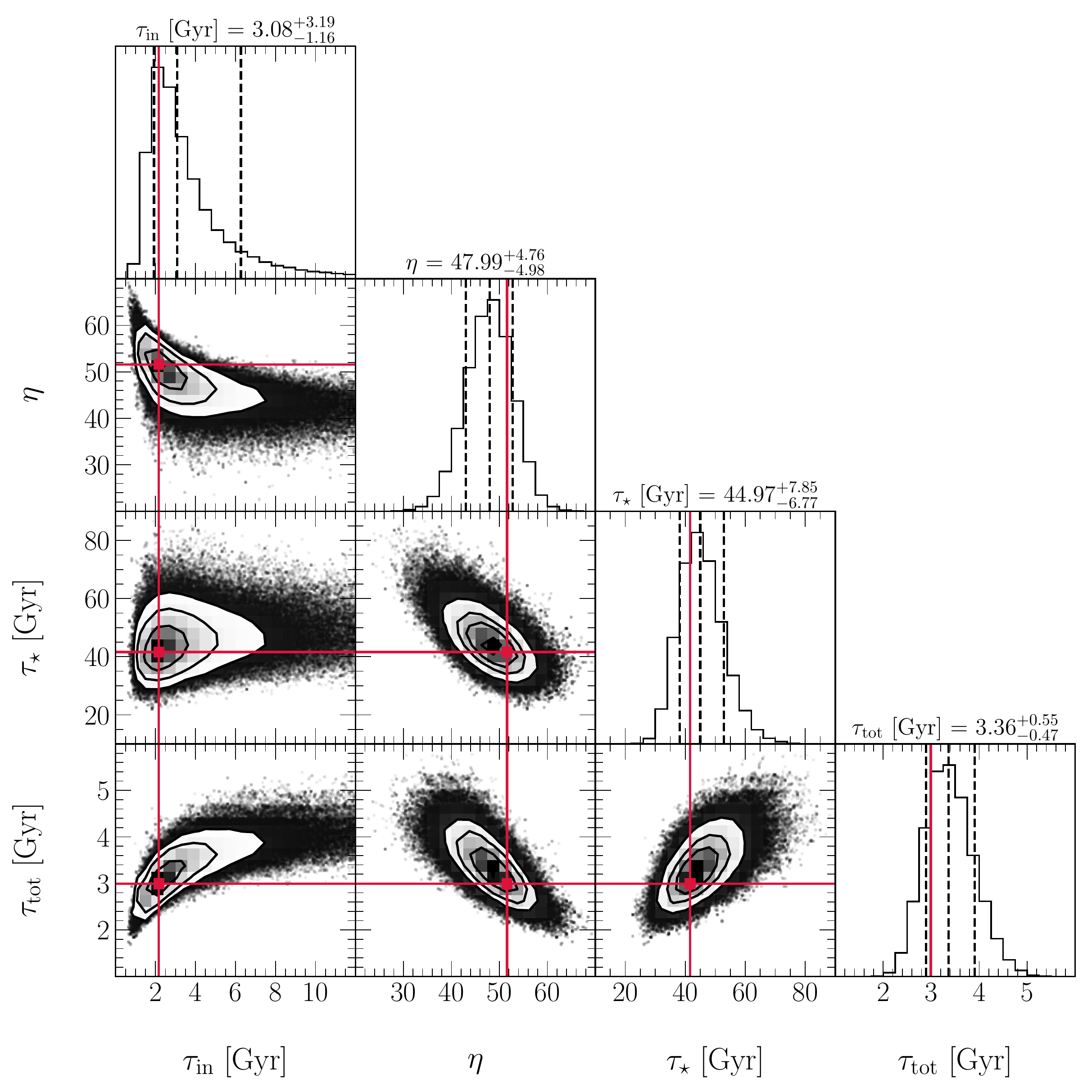}
\caption{
Posterior distributions for an exponential infall history applied to our
Wukong/LMS-1 sample.
The parametrization is the same as the input model to our mock samples (see
discussion in~\S~\ref{sec:mocks:fiducial}) but with the Fe yields held fixed
at the values determined by the fit to our GSE sample
($\yfecc = \scinote{7.78}{-4}$ and~$\yfeia = \scinote{1.23}{-3}$).
Panels below the diagonal show 2-dimensional cross-sections of the likelihood
function while panels along the diagonal show the marginalized distributions
along with the best-fit values and confidence intervals.
Red ``cross-hairs'' mark the element of the Markov chain with the maximum
statistical likelihood.
}
\label{fig:wukong_corner}
\end{figure*}

We select our GSE sample based on the criteria in~\citet{Conroy2022}, which
yields a sample of 189 stars with spectroscopic signal-to-noise
SNR~$> 15$ and~\gaia~RUWE~$< 1.5$.
95 of them are main sequence turnoff and subgiant stars with surface gravities
of~$3.8 < \log g < 4.2$ with reliable age measurements.
Abundance uncertainties range from~$\sim$0.02 to 0.12 dex in
both~\feh~and~\afe~with median values near~$\sim$0.05.
Every age measurement has a statistical uncertainty
$\sigma_{\log_{10}(\text{age})} \leq 0.05$, corresponding to a measurement
precision of~$\lesssim$12\%.
However, due to the difficulty associated with measuring stellar ages both
accurately and precisely~\citep[e.g.,][]{Soderblom2010, Chaplin2013, Angus2019},
we adopt~$0.05$ as the age uncertainty for the entire sample to account for any
systematic errors that may be present.
\par
We illustrate our sample in Fig.~\ref{fig:gse} along with our best-fit GCE
models (see discussion below).
We note the presence of two outliers at ages of~$\sim$5 and~$\sim$6 Gyr, marked
by X's in the right panel of Fig.~\ref{fig:gse}.
With abundances typical of the rest of the GSE population but anomalously young
ages, these stars are likely blue stragglers, which are thought to be made
hotter and more luminous by accretion from a binary companion and biasing their
age measurements to low values~\citep[e.g.,][]{Bond1971, Stryker1993}.
It is also possible that these stars are high-eccentricity contaminants kicked
out of the disk by Sagittarius~\citep[e.g.,][]{Donlon2020}.
The smooth decline of~\afe~with~\feh~and the unimodal nature of the
distributions in~\feh,~\afe~and age indicate that the GSE did not experience
any significant starburst events.
If it had, we would expect to see a multi-peaked age distribution
as well as an increase in~\afe~at a distinct~\feh~due to the perturbed ratio of
CCSN to SN Ia rates~\citep{Johnson2020}.
We therefore fit the GSE with an exponential infall history (the same as our
mock samples explored in~\S~\ref{sec:mocks}), omitting the two~$\sim$5
and~$\sim$6 Gyr old stars from the procedure and retaining the assumption that
star formation commenced 13.2 Gyr ago.
Because H3 selects targets based only on a magnitude range and a maximum
parallax, the selection function in chemical space should be nearly uniform
(i.e.,~$\script{S}(\script{M}_j | \{\theta\}) \approx 1$ for all points
$\script{M}_j$ along the evolutionary track.
We therefore take weights that are proportional to the SFR alone (see
equations~\ref{eq:likelihood} and~\ref{eq:weights} and discussion
in~\S~\ref{sec:fitting}).
\par
We report our best-fit evolutionary parameters in Table~\ref{tab:results}
with Fig.~\ref{fig:gse_corner} illustrating the posterior distributions.
These values suggest strong outflows ($\eta \approx 9$) and inefficient star
formation ($\tau_\star \approx 16$ Gyr).
Invoking the equilibrium arguments of~\citet{Weinberg2017}, strong outflows and
slow star formation are consistent with the metal-poor mode of the MDF and the
``knee'' in the evolutionary track occurring at low~\feh, respectively.
These results are expected for a dwarf galaxy where the gravity well is
intrinsically shallow and the stellar-to-halo mass ratios are known empirically
to be smaller than their higher mass counterparts~\citep{Hudson2015}.
The alpha-enhanced mode of the MDF reflects the short duration of star
formation, stopping before SN Ia enrichment could produce enough Fe to reach
solar~\afe.
The associated truncation of the age distribution (shown in the bottom left
panel of Fig.~\ref{fig:gse}) likely reflects the quenching of star
formation in the GSE progenitor as a consequence of ram pressure stripping by
the hot halo of the Milky Way after its first infal~$\sim$10 Gyr ago
\citep{Bonaca2020}.
The inferred Fe yields suggest that massive stars account for
$\yfecc / (\yfecc + \yfeia) \approx 40$\% of the Fe in the universe.
These values may however be influenced by the H3 pipeline~\textsc{MINESweeper}
\citep{Cargile2020}, which includes a prior enforcing~$\afe \leq +0.6$ -- if
the~\afe~plateau occurs near this value in nature, this prior could bias the
most alpha-rich stars in our sample to slightly lower~\afe~ratios.
\par
Red lines in Fig.~\ref{fig:gse} illustrate our best-fit model compared to the
data
Visually, this model is a reasonable description of the data, though in detail
it predicts a slightly broader~\feh~distribution and a slightly more peaked age
distribution.
We assess the quality of the fit with equation~\refp{eq:chisquared_dof} and
find~$\chi_\text{dof}^2 = 1.34$, suggesting that this fit is indeed accurate
but that there may be some marginal room for improvement.
The substantial scatter in the age-metallicity relation (lower right panel)
arises due to the age uncertainties -- to clarify this point, we subsample 95
stars (the same number in our sample with age measurements) from our best-fit
SFH and perturb their implied ages and abundances by the median observational
uncertainties.
These random draws (red points) occupy a very similar region of the age-\feh~and
age-\afe~planes.
We do however note an additional~$\sim$6 or 7 potential blue stragglers with
ages of~$\sim$$8 - 9$ Gyr,~$\feh \approx -1.2$ and~$\afe \approx +0.4$.
These stars are less obviously blue stragglers than the~$\sim$5 and~$\sim$6 Gyr
old ones and would not have stood out without this comparison.
These stars likely play a role in increasing the~$\chi_\text{dof}^2$ of our
fit, and removing them from our sample would also bring the observed age
distribution into better agreement with our best-fit model.
We however do not explore more detailed investigations of individual stars for
fits to carefully tailored populations here, and the fit we obtain is
statistically reasonble anyway.
\par
In~\S~\ref{sec:mocks:variations}, we found that our model accurately recovered
the evolutionary timescales of the input model even in the absence of age
information due to their impact on the shape of the MDF.
To assess the feasibility of deducing these parameters from abundances alone,
we conduct an additional fit to our GSE sample omitting the age measurements.
We report the best-fit parameters in Table~\ref{tab:results}.
This procedure results in accurate fits to the~\feh~and~\afe~distributions, and
the SN yields and mass-loading factor~$\eta$ are generally consistent with
and without ages.
The inferred timescales are biased toward higher values and are discrepant
by~$\sim$$2\sigma$, with the duration of star formation showing the largest
difference.
These results indicate that such an approach is theoretically possible, but in
practice age information in some form is essential to pinning down these
timescales.
In~\S~\ref{sec:mocks}, we fit our mock samples with the exact underlying GCE
model and same numerical code which integrated the input model, placing the
same systematic effects in the data as the model.
It is also never guaranteed that the evolutionary history built into the model
is an accurate description of the galaxy.

\subsection{Wukong/LMS-1}
\label{sec:h3:wukong}

\begin{figure}
\centering
\includegraphics[scale = 0.63]{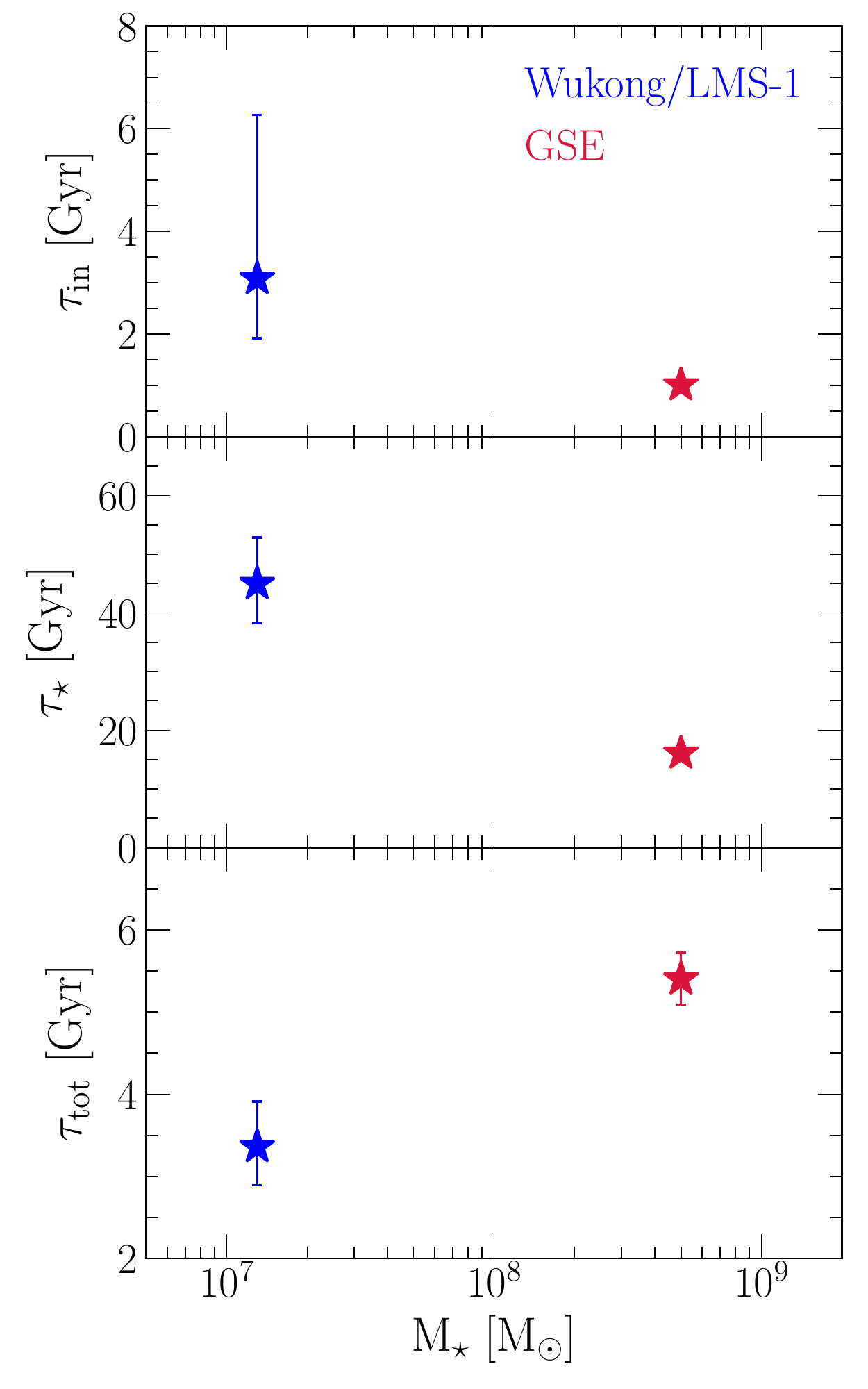}
\caption{
Our best-fit evolutionary timescales for Wukong/LMS-1 (blue) and GSE (red) as a
function of their stellar mass (taken from~\citealt{Naidu2022}; values are
tabulated in Table~\ref{tab:results}).
The uncertainties in the infall timescale~$\tau_\text{in}$ and the SFE
timescales~$\tau_\star$ for GSE are smaller than the point.
}
\label{fig:gse_wukong_timescales}
\end{figure}

\begin{figure}
\centering
\includegraphics[scale = 0.6]{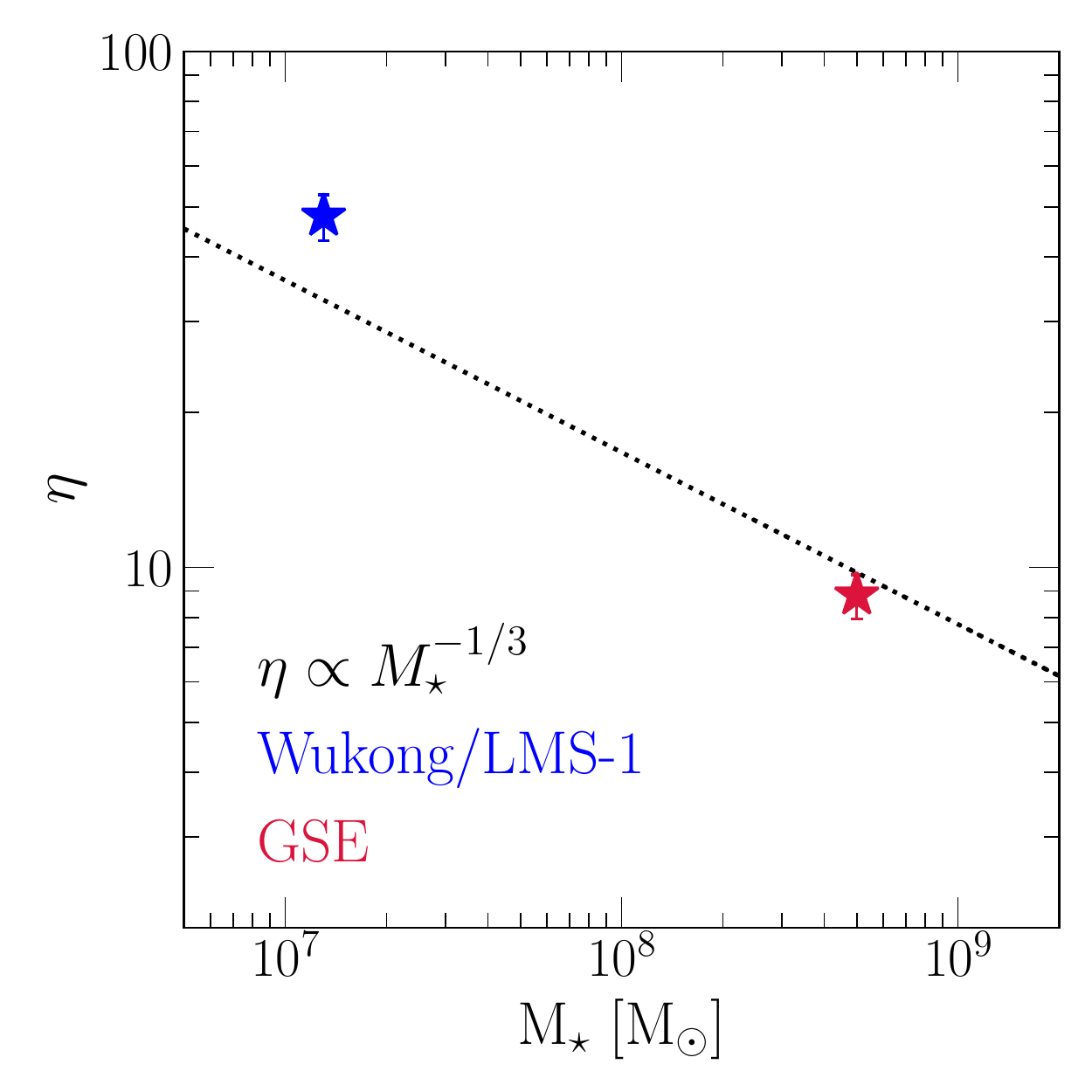}
\caption{
Our best-fit mass-loading factors~$\eta$ for Wukong/LMS-1 (blue) and GSE (red) as
a function of their stellar mass (taken from~\citealt{Naidu2022}; values are
tabulated in Table~\ref{tab:results}).
The black dashed line denotes~$\eta \propto M_\star^{-1/3}$ as suggested by
\citet{Finlator2008} and~\citet{Peeples2011} with the normalization of
$\eta = 3.6$ at~$M_\star = 10^{10}~M_\odot$ taken from~\citet{Muratov2015}.
}
\label{fig:gse_wukong_eta}
\end{figure}

We select Wukong/LMS-1 stars following the criteria in~\citet{Naidu2020}, with
the following additional cuts for high purity (inspired by the orbits of the
accomparnying globular clusters, NGC 5024 and NGC 5053, and~\citealt{Yuan2020}
and~\citealt{Malhan2021} who made selections based on the orbital plane):
\begin{itemize}

	\item[\textbf{1.}] $(J_z - J_r) / J_\text{tot}$ > 0.7, where~$J$ is action.

	\item[\textbf{2.}] $90^\circ < \theta < 120^\circ$, where~$\theta$ and
	$\phi$ are angles defining the angular momentum unit vector.

\end{itemize}
The~\citet{Naidu2020} selection features a hard cut at~$\feh < -1.45$ to avoid
GSE contamination, but visual inspection of the Wukong/LMS-1 sequence in
the~\afe-\feh~plane indicates that it drops off around~$\feh \approx -1.5$,
(see Fig.~\ref{fig:wukong}) and high [$\alpha$/Fe] GSE stars appear at higher
metallicities.
Our sample consists of 57 stars with spectroscopic SNR~$> 10$
and~\gaia~RUWE~$< 1.5$, none of which have age information as they are all
distant halo stars.
Within this sample, 23 stars are at SNR~$> 20$ and the remaining 34 are at
$10 <$ SNR~$< 20$.
Abundance uncertainties range from~$\sim$0.02 to~$\sim$0.10 dex in
both~\afe~and~\feh~with median values near~$\sim$0.045.
\par
Fig.~\ref{fig:wukong} illustrates this sample in chemical space along with our
best-fit GCE model (see discussion below).
Similar to the GSE, the lack of discontinuities in the age and abundance trends
indicates a smooth SFH devoid of any starburst events.
We therefore fit this sample with the same exponential infall history as the
input model to our mock samples, which we also applied to our GSE data.
We retain the assumption that star formation began 13.2 Gyr ago and that the H3
selection function is uniform in chemical space (see discussion
in~\S~\ref{sec:h3:gse}).
However, due to the smaller sample size and the lack of age information, we
initially hold our Fe yields fixed at~$\yfecc = \scinote{7.78}{-3}$ and
$\yfeia = \scinote{1.23}{-3}$ as suggested by the fit to our GSE sample.
It is reasonable to expect SN yields to be the same from galaxy-to-galaxy since
they are set by stellar as opposed to galactic physics, though we explore the
impact of relaxing this assumption below.
\par
Table~\ref{tab:results} reports the inferred best-fit parameters and
Fig.~\ref{fig:wukong_corner} illustrates the posterior distributions.
The degeneracies between parameters are noticeably more asymmetric than in our
GSE sample, a result of the lack of age information (we found similar effects
in our tests against mock data in~\S~\ref{sec:mocks}, though we did not discuss
it there).
The e-folding timescale of the accretion rate in particular has a highly skewed
likelihood distribution ($\tau_\text{in} = 3.08^{+3.19}_{-1.16}$ Gyr).
We have also had reasonable success describing Wukong/LMS-1 with a constant
star formation history.
Consequently, the likelihood function has a tail that extends
to~$\tau_\text{in} \rightarrow \infty$.
The exponential infall history is indeed a statistically better fit, so
throughout this section we include a prior that
enforces~$\tau_\text{in} \leq 50$ Gyr to focus on this portion of parameter
space.
This tail is significantly more extended if the Fe yields are allowed to vary
as a free parameter (see Table~\ref{tab:results} and discussion below).
\par
An exponential infall history yields a statistically good fit
($\chi_\text{dof}^2 = 0.98$; equation~\ref{eq:chisquared_dof}) for Wukong/LMS-1,
though visually it appears that the SN yields implied by our GSE data
underestimate the height of the [$\alpha$/Fe] plateau, which we indirectly
held fixed via the Fe yields.
Although we asserted above that it is reasonable expect SN yields to be the
same between Wukong/LMS-1 and GSE, variations in the plateau height could
indicate either metallicity-dependent yields or variations in the IMF.
To investigate this hypothesis, we conduct an additional fit where we allow
the Fe yields to vary as free parameters, reporting the results in
Table~\ref{tab:results} and illustrating the deduced model for comparison in
Fig.~\ref{fig:wukong}.
A higher plateau indeed provides an even better fit
($\chi_\text{dof}^2 = 0.84$), but with~$\chi_\text{dof}^2$ less than 1, this
could be an overparametrization of the data.
This possibility is not necessarily to a worrisome extent though; we cannot
rule out either model.
The best-fit SFE timescales between the two fits are in excellent agreement,
indicating that~$\tau_\star$ does not significantly impact the height of the
plateau (to first-order, it determines the position of the knee in the
track;~\citealp{Weinberg2017}).

\subsection{Comparison}
\label{sec:h3:comparison}

Fig.~\ref{fig:gse_wukong_timescales} compares the best-fit evolutionary
timescales between GSE and Wukong/LMS-1 as a function of their stellar mass (we
adopt the stellar masses inferred by~\citealt{Naidu2021, Naidu2022}; our GCE
models as we have parametrized them do not offer any constraints on this
quantity).
Due to the yield-outflow degeneracy (see Appendix~\ref{sec:degeneracy}), only
relative values of~$\tau_\star$ carry meaning, while the absolute values of
$\tau_\text{in}$ and~$\tau_\text{tot}$ do.
Qualitatively consistent with semi-analytic models of galaxy formation
\citep[e.g.,][]{Baugh2006, Somerville2015a, Behroozi2019} and results from
hydrodynamical simulations~\citep[e.g.,][]{GarrisonKimmel2019}, the less
massive of the two galaxies experienced the more extended accretion history.
Star formation in Wukong/LMS-1, however, was less efficient and did not last as
long as in GSE -- sensible results given the empirical correlation between
stellar-to-halo mass ratioes and stellar mass~\citep{Hudson2015}.
To the extent that our one-zone model framework is accurate, we have
constrained the duration of star formation in Wukong/LMS-1 and GSE to 15.2\%
and 5.8\%, respectively.
However, our Wukong/LMS-1 sample has no age measurements, and we have not
derived an SFH from its CMD here.
The failure of our fit to GSE omitting all ages (see Table~\ref{tab:results})
suggests that these best-fit parameters may be biased to high values.
\par
As expected given Wukong/LMS-1's shallower gravity well, it experienced
stronger mass-loading than GSE.
Fig.~\ref{fig:gse_wukong_eta} shows the inferred mass-loading factors in
comparison to the scaling of~$\eta \propto M_\star^{-1/3}$ as suggested by
\citet{Finlator2008} and~\citet{Peeples2011} modelling the impact of galactic
winds on the mass-metallicity realtion for galaxies.
We take the normalization of~$\eta = 3.6$ at~$M_\star = 10^{10}~M_\odot$ from
\citet{Muratov2015} who find a similar scaling in the FIRE simulations
($\eta \propto M_\star^{-0.35}$;~\citealp{Hopkins2014}).
There is excellent agreement between this predicted scaling and our one-zone
model fits -- rather remarkably so given that we have made no deliberate
choices for either the normalization or the slope to agree.
\par
\begin{figure*}
\centering
\includegraphics[scale = 0.45]{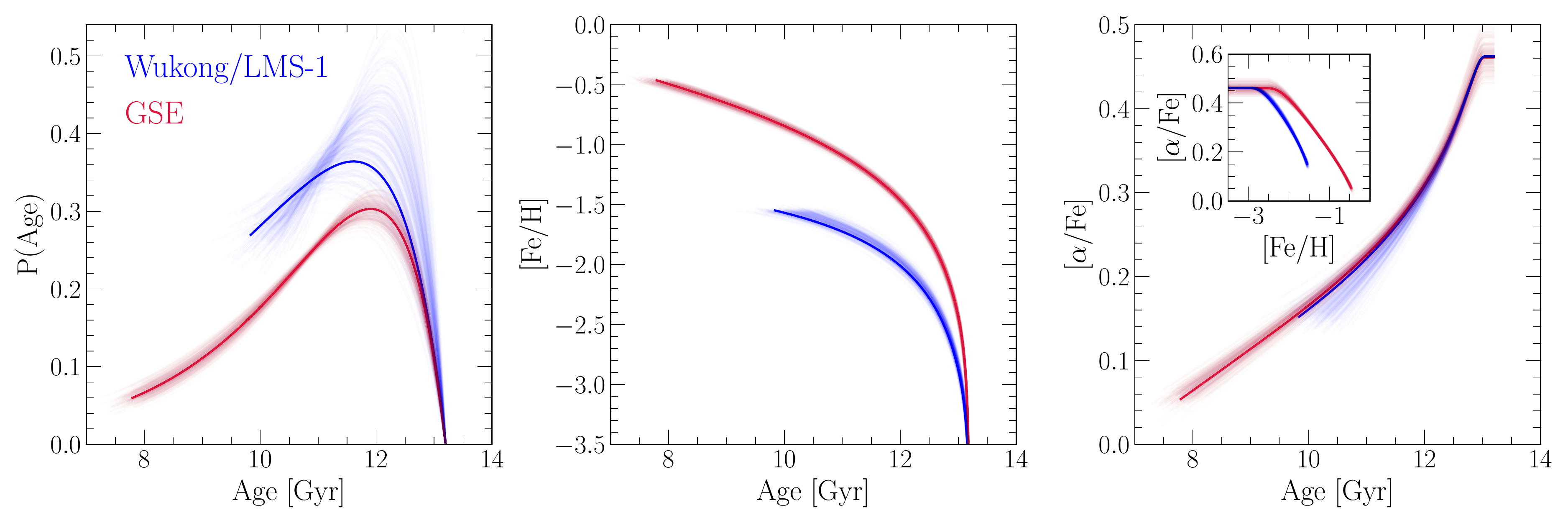}
\caption{
A comparison of our best-fit models for GSE (red) and Wukong/LMS-1 (blue): the
age distributions (left), the age-\feh~relations (middle) and age-\afe~relations
(right).
The inset in the right hand panel shows the tracks in the~\afe-\feh~plane.
In all panels, we subsample 200 additional parameter choices from our Markov
chains and plot the predictions as high transparency lines to provide a sense
of the fit uncertainty.
Due to the lack of age information for Wukong/LMS-1, the centroid of the age
distribution is determined by our assumption that star formation began 13.2
Gyr ago (see discussion in~\S~\ref{sec:mocks:fiducial}).
}
\label{fig:comparison}
\end{figure*}

In Fig.~\ref{fig:comparison}, we compare our best-fit models for GSE and
Wukong/LMS-1.
The intrinsic age distribution of GSE is predicted with considerably higher
precision than for Wukong/LMS-1, a consequence of the lack of age information
in our Wukong/LMS-1 sample.
The uncertainties in the Wukong/LMS-1 age distribution are noticeably
asymmetric due to the skewed posterior distribution of the infall timescale
($\tau_\text{in} = 3.08^{+3.19}_{-1.16}$ Gyr).
If our assumption that star formation began~$T \approx 13.2$ Gyr ago (see
discussion in~\S~\ref{sec:mocks:fiducial}) is accurate for Wukong/LMS-1, then
it experienced quenching~$\sim$2 Gyr earlier than the GSE ($\sim$9.8 versus
$\sim$7.8 Gyr ago).
However, because we do not have age information for Wukong/LMS-1, this
distribution could shift uniformly to lower values with affecting the quality
of the fit.
Constraints on the centroid of the distribution could be derived by
analysing the CMD as in, e.g.,~\citet{Dolphin2002} and~\citet{Weisz2014b}, but
we do not pursue this method in the present paper as it involves an entirely
separate mathematical framework.
\par
Also as a consequence of the lack of age information, our fits constrain the
intrinsic age-\feh~and age-\afe~relations to somewhat higher precision for GSE
than Wukong/LMS-1.
While the age-\feh~relations are significantly offset from one another, the
predicted age-\afe~relations are remarkably consistent with one another.
A portion of this agreement can likely be traced back to our fixing the Fe
yields in our fit to Wukong/LMS-1 to the values inferred in our fit to GSE.
Nonetheless, it is reasonable to assume that the SN yields are the same between
the two galaxies because this should be set by stellar physics, sufficiently
decoupled from the galactic environment.
The evolution of~\afe~with time is in principle impacted by the various
evolutionary timescales at play, so their consistency with one another is still
noteworthy.

\section{Discussion and Conclusions}
\label{sec:conclusions}

We use statistically robust methods to derive best-fit parameters of
one-zone GCE models for two disrupted dwarf galaxies in the Mily Way stellar
halo: GSE~\citep{Belokurov2018, Helmi2018}, and Wukong/LMS-1~\citep{Naidu2020,
Naidu2022, Yuan2020}.
We fit both galaxies with an exponential accretion history
(see~\S~\ref{sec:mocks}), deriving e-folding timescales and durations of star
formation of~$(\tau_\text{in}, \tau_\text{tot}) \approx (1~\text{Gyr},
5.4~\text{Gyr})$ for GSE and~$(\tau_\text{in}, \tau_\text{tot}) \approx
(3.1~\text{Gyr}, 3.4~\text{Gyr})$ for Wukong/LMS-1 (we refer to table
\ref{tab:results} for exact values).
These differences in evolutionary parameters are qualitatively consistent with
predictions from hydrodynamical simulations~\citep[e.g.,][]{GarrisonKimmel2019}
and semi-analytic models of galaxy formation~\citep[e.g.,][]{Baugh2006,
Somerville2015a, Behroozi2019}.
\par
Quantitatively, we arrive at a longer duration of star formation than
\citet{Gallart2019}, who derived an age distribution for GSE by analysing its
CMD according to the method described in~\citet{Dolphin2002} and found a
median age of 12.37 Gyr.
Consistent with their results,~\citet{Vincenzo2019} infer a sharply declining
infall history with a timescale of~$\tau_\text{in} = 0.24$ Gyr.
However, the star-by-star age measurements provided by H3~\citep{Conroy2019}
suggest that GSE's SFH was more extended (see Fig.~\ref{fig:gse}).
The peak of the age distribution is near~$\sim$11 Gyr (Fig.~\ref{fig:gse}),
consistent with Feuillet et al.'s~\citeyearpar{Feuillet2021} results
from~\gaia~\citep{Gaia2016} and APOGEE~\citep{Majewski2017}.
Consequently, we deduce a higher value of~$\tau_\text{in}$ of~$1.01 \pm 0.13$
Gyr.
If its first infall into the Milky Way halo was~$\sim$10 Gyr ago
\citep[e.g.,][]{Helmi2018, Bonaca2020}, then depending on exactly how long ago
it started forming stars, the duration of star formation we derive
($\tau_\text{tot} = 5.4$ Gyr) implies that GSE formed stars for~$\sim$$1.5 - 2$
Gyr after its first infall.
\par
To our knowledge, this is the first detailed modelling of multi-element stellar
abundances in Wukong/LMS-1.
Wukong/LMS-1 experienced a more extended accretion history
($\tau_\text{in} = 3.08^{+3.19}_{-1.16}$ Gyr), but the duration of star
formation was~$\sim$2 Gyr shorter than in GSe.
If they started forming stars around the same time, then Wukong/LMS-1 was
quenched at approximately the time of GSE's first infall.
However, our sample includes no age information for Wukong/LMS-1, so the
centroid of the age distribution is a prediction of our model as opposed to an
empirical constraint.
We find no statistically significant evidence of IMF variability or
metallicity-dependent Fe yields comparing GSE and Wukong/LMS-1.
A pathway to investigate this hypothesis further and potentially pin down the
yield-outflow degeneracy as well (see discussion in
Appendix~\ref{sec:degeneracy}) is to perform a hierarchical analysis of a
sample of galaxies where the yields are free parameters but are required to be
the same for all systems.
\par
Although these models are statistically good descriptions of our GSE and
Wukong/LMS-1 data, they are simplified in nature.
In particular, we have assumed a linear relation between the gas supply and the
SFR while empirical results would suggest a non-linear relation
\citep[e.g.,][]{Kennicutt1998, Kennicutt2012, delosReyes2019, Kennicutt2021}.
We have also taken a constant outflow mass-loading factor~$\eta$, when in
principle this parameter could vary with time as the potential well of the
galaxy deepens as in, e.g.,~\citet{Conroy2022}.
The primary motivation of these choices, however, is to provide proof of
concept for our fitting method with an example application to observations.
We reserve more detailed modelling of galaxies with both simple and complex
evolutionary histories for future work.
\par
Our method is built around a likelihood function which requires no binning of
the data (Eq.~\ref{eq:likelihood}) and has two central features.
First, the likelihood of observing some datum~$\script{D}_i$ must be
marginalized over the entire evolutionary track~\script{M}.
This requirement arises due to measurement uncertainties: for any given datum,
it is impossible to know where on the track the observation truly arose from,
and mathematically accounting for this requires considering all pair-wise
combinations between~\script{M} and~\script{D}.
Second, the likelihood of observing a datum~$\script{D}_i$ given a point on
the evolutionary track~$\script{M}_j$ must be weighted by the SFR at that time
in the model, simultaneously folding in any selection effects introduced by the
survey.
This requirement arises because an observed star is proportionally more likely
to have been sampled from an epoch of a galaxy's history in which the SFR was
large and/or if the survey designed is biased toward certain epochs.
\par
We establish the accuracy of our method by means of tests against mock data,
demonstrating that the known evolutionary parameters of subsampled input models
are accurately re-derived across a broad range of sample sizes
($N = 20 - 2000$), abundance uncertainties ($\sigma_\text{[X/Y]} = 0.01 - 0.5$),
age uncertainties ($\sigma_{\log_{10}(\text{age})} = 0.02 - 1$) and the
fraction of the sample with age information ($f_\text{age} = 0 - 1$; see
discussion in~\S~\ref{sec:mocks}).
The fit precision of the inferred parameters generally scales with sample size
as~$\sim$$N^{-0.5}$.
We demonstrate that evolutionary timescales can theoretically be derived with
abundances alone, but in practice age information helps reduce the effect of
systematic differences between the data and model, improving both the
accuracy and the precision.
Our likelihood function requires no binning of the data, and we derive it
in Appendix~\ref{sec:likelihood} assuming only that the model predicts an
evolutionary track of some unknown shape in the observed space.
It should therefore be applicable to one-zone models of any parametrization as
well as easily extensible to other astrophysical models in which the chief
prediction is a track of some form (e.g., stellar streams and isochrones).
\par
Having provided proof of concept for our method, a promising direction for
future work is to apply it to a much broader sample of disrupted dwarf galaxies
in the Milky Way stellar halo to take a ``chemical census'' of the accreted
systems.
This approach is also of interest to authors seeking to derive quenching
times (i.e., the lookback time to when star formation stopped) for intact and 
disrupted dwarf galaxies.
At present, the most reliable method to empirically determine a dwarf galaxy's
quenching time is via a direct reconstruction of its SFH through some method,
such as analysing its CMD~\citep[e.g.,][]{Dolphin2002, Weisz2015}.
Consequently, the most precise SFH measurements are for nearby systems with
resolved stars, a considerable limitation even with modern instrumentation.
To our knowledge, there are only four quenched galaxies outside of the Milky
Way subgroup with well-constrained SFHs: Andromeda II, Andromeda XIV
\citep{Weisz2014a}, Cetus~\citep{Monelli2010a} and Tucana~\citep{Monelli2010b}.
Some authors have connected quenching timescales to observed galaxy properties
in N-body simulations (e.g.,~\citealp*{Rocha2012};~\citealp{Slater2013,
Slater2014, Phillips2014, Phillips2015, Wheeler2014}), but unfortunately
simulation outcomes are strongly dependent on the details of the adopted
sub-grid models~\citep[e.g.,][]{Li2020} as well as how feedback and the grid
itself are implemented~\citep{Hu2022}.
Our results suggest that chemical abundances can provide valuable additional
information for these methods.
\par
However, with current instrumentation, spectroscopic measurements of
multi-element abundances in dwarf galaxies are limited to the local group
\citep[e.g.,][]{Kirby2011, Kirby2020}, and sample sizes are small even for
these relatively nearby systems.
Larger sample sizes could potentially be achieved with a high
angular resolution integral field unit such as the Multi Unit Spectroscopic
Explorer~\citep[MUSE;][]{Bacon2014}.
Alternatively, photometry is more conducive to larger sample sizes due to the
lower observational overhead, and the MDF can still be constrained using the
CMD~\citep[e.g.,][]{Lianou2011}.
One possibility is to forward-model the CMDs of dwarf galaxies using the SFHs
and MDFs predicted by one-zone GCE models, simultaneously constraining both
quantities photometrically.
The high angular resolution of the James Webb Space Telescope
\citep[JWST;][]{Gardner2006} should provide a considerable increase in the
number of resolved stars in nearby galaxies, making it a promising instrument
to pursue this potential pathway.
Farther in the future, the upcoming Nancy Grace Roman Space Telescope
(\citealp{Spergel2013, Spergel2015}; formerly WFIRST) will revolutionize
stellar populations in nearby galaxies.
In the era of next-generation telescopes, statistically robust methods such as
the one detailed in this paper will be essential to deduce the lessons the
community can learn about dwarf galaxy evolution.

\section{Acknowledgments}
\label{sec:acknowledgments}

We thank David H. Weinberg for comments on this manuscript.
JWJ is grateful for the hospitality of Harvard University and the 
Center for Astrophysics | Harvard \& Smithsonian.
JWJ acknowledges valuable discussion with Jennifer A. Johnson, Adam K. Leroy,
Todd A. Thompson, and other members of the Ohio State University Gas, Galaxies,
and Feedback Group.
JWJ thanks John F. Beacom and the support staff at the Ohio Supercomputer
Center for computational resources.
JWJ also acknowledges financial support from an Ohio State University
Presidential Fellowship.
CC and PAC acknowledge support from National Science Foundation (NSF) Grant
No. AST-2107253.
AHGP acknowledges support from the NSF Grant No. AST-1813628 and AST-2008110.
RPN ackowledges support for this work provided by NASA through the NASA Hubble
Fellowship grant HST-HF2-51515.001-A awarded by the Space Telescope Science
Institute, which is operated by the Association of Universities for Research in
Astronomy, Incorporated, under NASA contract NAS5-26555.
Y.-S.T. acknowledges financial support from the Australian Research Council
through DE- CRA Fellowship DE220101520.
\par\null\par\noindent
\textit{Software}:~\vice~\citep{Johnson2020},
\textsc{NumPy}~\citep{Harris2020},
\textsc{Matplotlib}~\citep{Hunter2007},
\mc~\citep{ForemanMackey2013},
\textsc{corner}~\citep{ForemanMackey2016}

\section{Data Availability}
\label{sec:data_availability}

The data in this paper will be made available upon reasonable request to the
corresponding author.

\bibliographystyle{mnras}
\bibliography{ms}

\begin{thebibliography}{}
\makeatletter
\relax
\def\mn@urlcharsother{\let\do\@makeother \do\$\do\&\do\#\do\^\do\_\do\%\do\~}
\def\mn@doi{\begingroup\mn@urlcharsother \@ifnextchar [ {\mn@doi@}
  {\mn@doi@[]}}
\def\mn@doi@[#1]#2{\def\@tempa{#1}\ifx\@tempa\@empty \href
  {http://dx.doi.org/#2} {doi:#2}\else \href {http://dx.doi.org/#2} {#1}\fi
  \endgroup}
\def\mn@eprint#1#2{\mn@eprint@#1:#2::\@nil}
\def\mn@eprint@arXiv#1{\href {http://arxiv.org/abs/#1} {{\tt arXiv:#1}}}
\def\mn@eprint@dblp#1{\href {http://dblp.uni-trier.de/rec/bibtex/#1.xml}
  {dblp:#1}}
\def\mn@eprint@#1:#2:#3:#4\@nil{\def\@tempa {#1}\def\@tempb {#2}\def\@tempc
  {#3}\ifx \@tempc \@empty \let \@tempc \@tempb \let \@tempb \@tempa \fi \ifx
  \@tempb \@empty \def\@tempb {arXiv}\fi \@ifundefined
  {mn@eprint@\@tempb}{\@tempb:\@tempc}{\expandafter \expandafter \csname
  mn@eprint@\@tempb\endcsname \expandafter{\@tempc}}}

\bibitem[\protect\citeauthoryear{{Adams}, {Kochanek}, {Gerke}, {Stanek}  \&
  {Dai}}{{Adams} et~al.}{2017}]{Adams2017}
{Adams} S.~M.,  {Kochanek} C.~S.,  {Gerke} J.~R.,  {Stanek} K.~Z.,   {Dai} X.,
  2017, \mn@doi [\mnras] {10.1093/mnras/stx816}, \href
  {https://ui.adsabs.harvard.edu/abs/2017MNRAS.468.4968A} {468, 4968}

\bibitem[\protect\citeauthoryear{{Andrews} \& {Martini}}{{Andrews} \&
  {Martini}}{2013}]{Andrews2013}
{Andrews} B.~H.,  {Martini} P.,  2013, \mn@doi [\apj]
  {10.1088/0004-637X/765/2/140}, \href
  {https://ui.adsabs.harvard.edu/abs/2013ApJ...765..140A} {765, 140}

\bibitem[\protect\citeauthoryear{{Andrews}, {Weinberg}, {Sch{\"o}nrich}  \&
  {Johnson}}{{Andrews} et~al.}{2017}]{Andrews2017}
{Andrews} B.~H.,  {Weinberg} D.~H.,  {Sch{\"o}nrich} R.,   {Johnson} J.~A.,
  2017, \mn@doi [\apj] {10.3847/1538-4357/835/2/224}, \href
  {https://ui.adsabs.harvard.edu/abs/2017ApJ...835..224A} {835, 224}

\bibitem[\protect\citeauthoryear{{Anguiano} et~al.,}{{Anguiano}
  et~al.}{2018}]{Anguino2018}
{Anguiano} B.,  et~al., 2018, \mn@doi [\aap] {10.1051/0004-6361/201833387},
  \href {https://ui.adsabs.harvard.edu/abs/2018A&A...620A..76A} {620, A76}

\bibitem[\protect\citeauthoryear{{Angus} et~al.,}{{Angus}
  et~al.}{2019}]{Angus2019}
{Angus} R.,  et~al., 2019, \mn@doi [\aj] {10.3847/1538-3881/ab3c53}, \href
  {https://ui.adsabs.harvard.edu/abs/2019AJ....158..173A} {158, 173}

\bibitem[\protect\citeauthoryear{{Asplund}, {Grevesse}, {Sauval}  \&
  {Scott}}{{Asplund} et~al.}{2009}]{Asplund2009}
{Asplund} M.,  {Grevesse} N.,  {Sauval} A.~J.,   {Scott} P.,  2009, \mn@doi
  [\araa] {10.1146/annurev.astro.46.060407.145222}, \href
  {https://ui.adsabs.harvard.edu/abs/2009ARA&A..47..481A} {47, 481}

\bibitem[\protect\citeauthoryear{{Asplund}, {Amarsi}  \& {Grevesse}}{{Asplund}
  et~al.}{2021}]{Asplund2021}
{Asplund} M.,  {Amarsi} A.~M.,   {Grevesse} N.,  2021, \mn@doi [\aap]
  {10.1051/0004-6361/202140445}, \href
  {https://ui.adsabs.harvard.edu/abs/2021A&A...653A.141A} {653, A141}

\bibitem[\protect\citeauthoryear{{Bacon} et~al.,}{{Bacon}
  et~al.}{2014}]{Bacon2014}
{Bacon} R.,  et~al., 2014, The Messenger, \href
  {https://ui.adsabs.harvard.edu/abs/2014Msngr.157...13B} {157, 13}

\bibitem[\protect\citeauthoryear{{Balser} \& {Bania}}{{Balser} \&
  {Bania}}{2018}]{Balser2018}
{Balser} D.~S.,  {Bania} T.~M.,  2018, \mn@doi [\aj]
  {10.3847/1538-3881/aaeb2b}, \href
  {https://ui.adsabs.harvard.edu/abs/2018AJ....156..280B} {156, 280}

\bibitem[\protect\citeauthoryear{{Basinger}, {Kochanek}, {Adams}, {Dai}  \&
  {Stanek}}{{Basinger} et~al.}{2021}]{Basinger2021}
{Basinger} C.~M.,  {Kochanek} C.~S.,  {Adams} S.~M.,  {Dai} X.,   {Stanek}
  K.~Z.,  2021, \mn@doi [\mnras] {10.1093/mnras/stab2620}, \href
  {https://ui.adsabs.harvard.edu/abs/2021MNRAS.508.1156B} {508, 1156}

\bibitem[\protect\citeauthoryear{{Baugh}}{{Baugh}}{2006}]{Baugh2006}
{Baugh} C.~M.,  2006, \mn@doi [Reports on Progress in Physics]
  {10.1088/0034-4885/69/12/R02}, \href
  {https://ui.adsabs.harvard.edu/abs/2006RPPh...69.3101B} {69, 3101}

\bibitem[\protect\citeauthoryear{{Behroozi}, {Wechsler}, {Hearin}  \&
  {Conroy}}{{Behroozi} et~al.}{2019}]{Behroozi2019}
{Behroozi} P.,  {Wechsler} R.~H.,  {Hearin} A.~P.,   {Conroy} C.,  2019,
  \mn@doi [\mnras] {10.1093/mnras/stz1182}, \href
  {https://ui.adsabs.harvard.edu/abs/2019MNRAS.488.3143B} {488, 3143}

\bibitem[\protect\citeauthoryear{{Belokurov}, {Erkal}, {Evans}, {Koposov}  \&
  {Deason}}{{Belokurov} et~al.}{2018}]{Belokurov2018}
{Belokurov} V.,  {Erkal} D.,  {Evans} N.~W.,  {Koposov} S.~E.,   {Deason}
  A.~J.,  2018, \mn@doi [\mnras] {10.1093/mnras/sty982}, \href
  {https://ui.adsabs.harvard.edu/abs/2018MNRAS.478..611B} {478, 611}

\bibitem[\protect\citeauthoryear{{Bertelli Motta} et~al.,}{{Bertelli Motta}
  et~al.}{2018}]{BertelliMotta2018}
{Bertelli Motta} C.,  et~al., 2018, \mn@doi [\mnras] {10.1093/mnras/sty1011},
  \href {https://ui.adsabs.harvard.edu/abs/2018MNRAS.478..425B} {478, 425}

\bibitem[\protect\citeauthoryear{{Bonaca} et~al.,}{{Bonaca}
  et~al.}{2020}]{Bonaca2020}
{Bonaca} A.,  et~al., 2020, \mn@doi [\apjl] {10.3847/2041-8213/ab9caa}, \href
  {https://ui.adsabs.harvard.edu/abs/2020ApJ...897L..18B} {897, L18}

\bibitem[\protect\citeauthoryear{{Bond} \& {MacConnell}}{{Bond} \&
  {MacConnell}}{1971}]{Bond1971}
{Bond} H.~E.,  {MacConnell} D.~J.,  1971, \mn@doi [\apj] {10.1086/150875},
  \href {https://ui.adsabs.harvard.edu/abs/1971ApJ...165...51B} {165, 51}

\bibitem[\protect\citeauthoryear{{Bovy}}{{Bovy}}{2016}]{Bovy2016}
{Bovy} J.,  2016, \mn@doi [\apj] {10.3847/0004-637X/817/1/49}, \href
  {https://ui.adsabs.harvard.edu/abs/2016ApJ...817...49B} {817, 49}

\bibitem[\protect\citeauthoryear{{Cameron} et~al.,}{{Cameron}
  et~al.}{2021}]{Cameron2021}
{Cameron} A.~J.,  et~al., 2021, \mn@doi [\apjl] {10.3847/2041-8213/ac18ca},
  \href {https://ui.adsabs.harvard.edu/abs/2021ApJ...918L..16C} {918, L16}

\bibitem[\protect\citeauthoryear{{Cargile}, {Conroy}, {Johnson}, {Ting},
  {Bonaca}, {Dotter}  \& {Speagle}}{{Cargile} et~al.}{2020}]{Cargile2020}
{Cargile} P.~A.,  {Conroy} C.,  {Johnson} B.~D.,  {Ting} Y.-S.,  {Bonaca} A.,
  {Dotter} A.,   {Speagle} J.~S.,  2020, \mn@doi [\apj]
  {10.3847/1538-4357/aba43b}, \href
  {https://ui.adsabs.harvard.edu/abs/2020ApJ...900...28C} {900, 28}

\bibitem[\protect\citeauthoryear{{Casamiquela}, {Tarricq}, {Soubiran},
  {Blanco-Cuaresma}, {Jofr{\'e}}, {Heiter}  \& {Tucci Maia}}{{Casamiquela}
  et~al.}{2020}]{Casamiquela2020}
{Casamiquela} L.,  {Tarricq} Y.,  {Soubiran} C.,  {Blanco-Cuaresma} S.,
  {Jofr{\'e}} P.,  {Heiter} U.,   {Tucci Maia} M.,  2020, \mn@doi [\aap]
  {10.1051/0004-6361/201936978}, \href
  {https://ui.adsabs.harvard.edu/abs/2020A&A...635A...8C} {635, A8}

\bibitem[\protect\citeauthoryear{{Chabrier}}{{Chabrier}}{2003}]{Chabrier2003}
{Chabrier} G.,  2003, \mn@doi [\pasp] {10.1086/376392}, \href
  {https://ui.adsabs.harvard.edu/abs/2003PASP..115..763C} {115, 763}

\bibitem[\protect\citeauthoryear{{Chambers} et~al.,}{{Chambers}
  et~al.}{2016}]{Chambers2016}
{Chambers} K.~C.,  et~al., 2016, arXiv e-prints, \href
  {https://ui.adsabs.harvard.edu/abs/2016arXiv161205560C} {p. arXiv:1612.05560}

\bibitem[\protect\citeauthoryear{{Chaplin} \& {Miglio}}{{Chaplin} \&
  {Miglio}}{2013}]{Chaplin2013}
{Chaplin} W.~J.,  {Miglio} A.,  2013, \mn@doi [\araa]
  {10.1146/annurev-astro-082812-140938}, \href
  {https://ui.adsabs.harvard.edu/abs/2013ARA&A..51..353C} {51, 353}

\bibitem[\protect\citeauthoryear{{Chaplin} et~al.,}{{Chaplin}
  et~al.}{2020}]{Chaplin2020}
{Chaplin} W.~J.,  et~al., 2020, \mn@doi [Nature Astronomy]
  {10.1038/s41550-019-0975-9}, \href
  {https://ui.adsabs.harvard.edu/abs/2020NatAs...4..382C} {4, 382}

\bibitem[\protect\citeauthoryear{{Chiappini}, {Matteucci}  \&
  {Gratton}}{{Chiappini} et~al.}{1997}]{Chiappini1997}
{Chiappini} C.,  {Matteucci} F.,   {Gratton} R.,  1997, \mn@doi [\apj]
  {10.1086/303726}, \href
  {https://ui.adsabs.harvard.edu/abs/1997ApJ...477..765C} {477, 765}

\bibitem[\protect\citeauthoryear{{Chieffi} \& {Limongi}}{{Chieffi} \&
  {Limongi}}{2004}]{Chieffi2004}
{Chieffi} A.,  {Limongi} M.,  2004, \mn@doi [\apj] {10.1086/392523}, \href
  {https://ui.adsabs.harvard.edu/abs/2004ApJ...608..405C} {608, 405}

\bibitem[\protect\citeauthoryear{{Chieffi} \& {Limongi}}{{Chieffi} \&
  {Limongi}}{2013}]{Chieffi2013}
{Chieffi} A.,  {Limongi} M.,  2013, \mn@doi [\apj]
  {10.1088/0004-637X/764/1/21}, \href
  {https://ui.adsabs.harvard.edu/abs/2013ApJ...764...21C} {764, 21}

\bibitem[\protect\citeauthoryear{{Chisholm}, {Tremonti}  \&
  {Leitherer}}{{Chisholm} et~al.}{2018}]{Chisholm2018}
{Chisholm} J.,  {Tremonti} C.,   {Leitherer} C.,  2018, \mn@doi [\mnras]
  {10.1093/mnras/sty2380}, \href
  {https://ui.adsabs.harvard.edu/abs/2018MNRAS.481.1690C} {481, 1690}

\bibitem[\protect\citeauthoryear{{Conroy} et~al.,}{{Conroy}
  et~al.}{2019}]{Conroy2019}
{Conroy} C.,  et~al., 2019, \mn@doi [\apj] {10.3847/1538-4357/ab38b8}, \href
  {https://ui.adsabs.harvard.edu/abs/2019ApJ...883..107C} {883, 107}

\bibitem[\protect\citeauthoryear{{Conroy} et~al.,}{{Conroy}
  et~al.}{2022}]{Conroy2022}
{Conroy} C.,  et~al., 2022, arXiv e-prints, \href
  {https://ui.adsabs.harvard.edu/abs/2022arXiv220402989C} {p. arXiv:2204.02989}

\bibitem[\protect\citeauthoryear{{Cooke}, {Noterdaeme}, {Johnson}, {Pettini},
  {Welsh}, {Peroux}, {Murphy}  \& {Weinberg}}{{Cooke} et~al.}{2022}]{Cooke2022}
{Cooke} R.~J.,  {Noterdaeme} P.,  {Johnson} J.~W.,  {Pettini} M.,  {Welsh} L.,
  {Peroux} C.,  {Murphy} M.~T.,   {Weinberg} D.~H.,  2022, \mn@doi [\apj]
  {10.3847/1538-4357/ac6503}, \href
  {https://ui.adsabs.harvard.edu/abs/2022ApJ...932...60C} {932, 60}

\bibitem[\protect\citeauthoryear{{C{\^o}t{\'e}}, {O'Shea}, {Ritter}, {Herwig}
  \& {Venn}}{{C{\^o}t{\'e}} et~al.}{2017}]{Cote2017}
{C{\^o}t{\'e}} B.,  {O'Shea} B.~W.,  {Ritter} C.,  {Herwig} F.,   {Venn} K.~A.,
   2017, \mn@doi [\apj] {10.3847/1538-4357/835/2/128}, \href
  {https://ui.adsabs.harvard.edu/abs/2017ApJ...835..128C} {835, 128}

\bibitem[\protect\citeauthoryear{{Dalcanton}}{{Dalcanton}}{2007}]{Dalcanton2007}
{Dalcanton} J.~J.,  2007, \mn@doi [\apj] {10.1086/508913}, \href
  {https://ui.adsabs.harvard.edu/abs/2007ApJ...658..941D} {658, 941}

\bibitem[\protect\citeauthoryear{{Davies} et~al.,}{{Davies}
  et~al.}{2016}]{Davies2016}
{Davies} L.~J.~M.,  et~al., 2016, \mn@doi [\mnras] {10.1093/mnras/stw1342},
  \href {https://ui.adsabs.harvard.edu/abs/2016MNRAS.461..458D} {461, 458}

\bibitem[\protect\citeauthoryear{{Deason}, {Belokurov}  \& {Sanders}}{{Deason}
  et~al.}{2019}]{Deason2019}
{Deason} A.~J.,  {Belokurov} V.,   {Sanders} J.~L.,  2019, \mn@doi [\mnras]
  {10.1093/mnras/stz2793}, \href
  {https://ui.adsabs.harvard.edu/abs/2019MNRAS.490.3426D} {490, 3426}

\bibitem[\protect\citeauthoryear{{\noopsort{Delosreyes}{de los Reyes}} \&
  {Kennicutt}}{{\noopsort{Delosreyes}{de los Reyes}} \&
  {Kennicutt}}{2019}]{delosReyes2019}
{\noopsort{Delosreyes}{de los Reyes}} M. A.~C.,  {Kennicutt} Robert~C. J.,
  2019, \mn@doi [\apj] {10.3847/1538-4357/aafa82}, \href
  {https://ui.adsabs.harvard.edu/abs/2019ApJ...872...16D} {872, 16}

\bibitem[\protect\citeauthoryear{{\noopsort{Delosreyes}{de los Reyes}},
  {Kirby}, {Ji}  \& {Nu{\~n}ez}}{{\noopsort{Delosreyes}{de los Reyes}}
  et~al.}{2022}]{delosReyes2022}
{\noopsort{Delosreyes}{de los Reyes}} M. A.~C.,  {Kirby} E.~N.,  {Ji} A.~P.,
  {Nu{\~n}ez} E.~H.,  2022, \mn@doi [\apj] {10.3847/1538-4357/ac332b}, \href
  {https://ui.adsabs.harvard.edu/abs/2022ApJ...925...66D} {925, 66}

\bibitem[\protect\citeauthoryear{{\noopsort{Desilva}{De Silva}}, {Sneden},
  {Paulson}, {Asplund}, {Bland-Hawthorn}, {Bessell}  \&
  {Freeman}}{{\noopsort{Desilva}{De Silva}} et~al.}{2006}]{DeSilva2006}
{\noopsort{Desilva}{De Silva}} G.~M.,  {Sneden} C.,  {Paulson} D.~B.,
  {Asplund} M.,  {Bland-Hawthorn} J.,  {Bessell} M.~S.,   {Freeman} K.~C.,
  2006, \mn@doi [\aj] {10.1086/497968}, \href
  {https://ui.adsabs.harvard.edu/abs/2006AJ....131..455D} {131, 455}

\bibitem[\protect\citeauthoryear{{\noopsort{Desilva}{De Silva}}
  et~al.,}{{\noopsort{Desilva}{De Silva}} et~al.}{2015}]{DeSilva2015}
{\noopsort{Desilva}{De Silva}} G.~M.,  et~al., 2015, \mn@doi [\mnras]
  {10.1093/mnras/stv327}, \href
  {https://ui.adsabs.harvard.edu/abs/2015MNRAS.449.2604D} {449, 2604}

\bibitem[\protect\citeauthoryear{{Dolphin}}{{Dolphin}}{2002}]{Dolphin2002}
{Dolphin} A.~E.,  2002, \mn@doi [\mnras] {10.1046/j.1365-8711.2002.05271.x},
  \href {https://ui.adsabs.harvard.edu/abs/2002MNRAS.332...91D} {332, 91}

\bibitem[\protect\citeauthoryear{{Donlon}, {Newberg}, {Sanderson}  \&
  {Widrow}}{{Donlon} et~al.}{2020}]{Donlon2020}
{Donlon} Thomas I.,  {Newberg} H.~J.,  {Sanderson} R.,   {Widrow} L.~M.,  2020,
  \mn@doi [\apj] {10.3847/1538-4357/abb5f6}, \href
  {https://ui.adsabs.harvard.edu/abs/2020ApJ...902..119D} {902, 119}

\bibitem[\protect\citeauthoryear{{Driver} et~al.,}{{Driver}
  et~al.}{2018}]{Driver2018}
{Driver} S.~P.,  et~al., 2018, \mn@doi [\mnras] {10.1093/mnras/stx2728}, \href
  {https://ui.adsabs.harvard.edu/abs/2018MNRAS.475.2891D} {475, 2891}

\bibitem[\protect\citeauthoryear{{Dutta}, {Begum}, {Bharadwaj}  \&
  {Chengalur}}{{Dutta} et~al.}{2009}]{Dutta2009}
{Dutta} P.,  {Begum} A.,  {Bharadwaj} S.,   {Chengalur} J.~N.,  2009, \mn@doi
  [\mnras] {10.1111/j.1365-2966.2009.15105.x}, \href
  {https://ui.adsabs.harvard.edu/abs/2009MNRAS.398..887D} {398, 887}

\bibitem[\protect\citeauthoryear{{Ertl}, {Janka}, {Woosley}, {Sukhbold}  \&
  {Ugliano}}{{Ertl} et~al.}{2016}]{Ertl2016}
{Ertl} T.,  {Janka} H.~T.,  {Woosley} S.~E.,  {Sukhbold} T.,   {Ugliano} M.,
  2016, \mn@doi [\apj] {10.3847/0004-637X/818/2/124}, \href
  {https://ui.adsabs.harvard.edu/abs/2016ApJ...818..124E} {818, 124}

\bibitem[\protect\citeauthoryear{{Fattahi} et~al.,}{{Fattahi}
  et~al.}{2019}]{Fattahi2019}
{Fattahi} A.,  et~al., 2019, \mn@doi [\mnras] {10.1093/mnras/stz159}, \href
  {https://ui.adsabs.harvard.edu/abs/2019MNRAS.484.4471F} {484, 4471}

\bibitem[\protect\citeauthoryear{{Feuillet}, {Sahlholdt}, {Feltzing}  \&
  {Casagrande}}{{Feuillet} et~al.}{2021}]{Feuillet2021}
{Feuillet} D.~K.,  {Sahlholdt} C.~L.,  {Feltzing} S.,   {Casagrande} L.,  2021,
  \mn@doi [\mnras] {10.1093/mnras/stab2614}, \href
  {https://ui.adsabs.harvard.edu/abs/2021MNRAS.508.1489F} {508, 1489}

\bibitem[\protect\citeauthoryear{{Finlator} \& {Dav{\'e}}}{{Finlator} \&
  {Dav{\'e}}}{2008}]{Finlator2008}
{Finlator} K.,  {Dav{\'e}} R.,  2008, \mn@doi [\mnras]
  {10.1111/j.1365-2966.2008.12991.x}, \href
  {https://ui.adsabs.harvard.edu/abs/2008MNRAS.385.2181F} {385, 2181}

\bibitem[\protect\citeauthoryear{{Forbes}}{{Forbes}}{2020}]{Forbes2020}
{Forbes} D.~A.,  2020, \mn@doi [\mnras] {10.1093/mnras/staa245}, \href
  {https://ui.adsabs.harvard.edu/abs/2020MNRAS.493..847F} {493, 847}

\bibitem[\protect\citeauthoryear{Foreman-Mackey}{Foreman-Mackey}{2016}]{ForemanMackey2016}
Foreman-Mackey D.,  2016, \mn@doi [The Journal of Open Source Software]
  {10.21105/joss.00024}, 1, 24

\bibitem[\protect\citeauthoryear{{Foreman-Mackey}, {Hogg}, {Lang}  \&
  {Goodman}}{{Foreman-Mackey} et~al.}{2013}]{ForemanMackey2013}
{Foreman-Mackey} D.,  {Hogg} D.~W.,  {Lang} D.,   {Goodman} J.,  2013, \mn@doi
  [\pasp] {10.1086/670067}, \href
  {https://ui.adsabs.harvard.edu/abs/2013PASP..125..306F} {125, 306}

\bibitem[\protect\citeauthoryear{{Freundlich} \& {Maoz}}{{Freundlich} \&
  {Maoz}}{2021}]{Freundlich2021}
{Freundlich} J.,  {Maoz} D.,  2021, \mn@doi [\mnras] {10.1093/mnras/stab493},
  \href {https://ui.adsabs.harvard.edu/abs/2021MNRAS.502.5882F} {502, 5882}

\bibitem[\protect\citeauthoryear{{Fu} et~al.,}{{Fu} et~al.}{2022}]{Fu2022}
{Fu} S.~W.,  et~al., 2022, \mn@doi [\apj] {10.3847/1538-4357/ac3665}, \href
  {https://ui.adsabs.harvard.edu/abs/2022ApJ...925....6F} {925, 6}

\bibitem[\protect\citeauthoryear{{Gaia Collaboration} et~al.,}{{Gaia
  Collaboration} et~al.}{2016}]{Gaia2016}
{Gaia Collaboration} et~al., 2016, \mn@doi [\aap]
  {10.1051/0004-6361/201629272}, \href
  {https://ui.adsabs.harvard.edu/abs/2016A&A...595A...1G} {595, A1}

\bibitem[\protect\citeauthoryear{{Gallart}, {Bernard}, {Brook}, {Ruiz-Lara},
  {Cassisi}, {Hill}  \& {Monelli}}{{Gallart} et~al.}{2019}]{Gallart2019}
{Gallart} C.,  {Bernard} E.~J.,  {Brook} C.~B.,  {Ruiz-Lara} T.,  {Cassisi} S.,
   {Hill} V.,   {Monelli} M.,  2019, \mn@doi [Nature Astronomy]
  {10.1038/s41550-019-0829-5}, \href
  {https://ui.adsabs.harvard.edu/abs/2019NatAs...3..932G} {3, 932}

\bibitem[\protect\citeauthoryear{{Gallazzi}, {Charlot}, {Brinchmann}, {White}
  \& {Tremonti}}{{Gallazzi} et~al.}{2005}]{Gallazzi2005}
{Gallazzi} A.,  {Charlot} S.,  {Brinchmann} J.,  {White} S. D.~M.,   {Tremonti}
  C.~A.,  2005, \mn@doi [\mnras] {10.1111/j.1365-2966.2005.09321.x}, \href
  {https://ui.adsabs.harvard.edu/abs/2005MNRAS.362...41G} {362, 41}

\bibitem[\protect\citeauthoryear{{Gardner} et~al.,}{{Gardner}
  et~al.}{2006}]{Gardner2006}
{Gardner} J.~P.,  et~al., 2006, \mn@doi [\ssr] {10.1007/s11214-006-8315-7},
  \href {https://ui.adsabs.harvard.edu/abs/2006SSRv..123..485G} {123, 485}

\bibitem[\protect\citeauthoryear{{Garrison-Kimmel} et~al.,}{{Garrison-Kimmel}
  et~al.}{2019}]{GarrisonKimmel2019}
{Garrison-Kimmel} S.,  et~al., 2019, \mn@doi [\mnras] {10.1093/mnras/stz2507},
  \href {https://ui.adsabs.harvard.edu/abs/2019MNRAS.489.4574G} {489, 4574}

\bibitem[\protect\citeauthoryear{{Gerke}, {Kochanek}  \& {Stanek}}{{Gerke}
  et~al.}{2015}]{Gerke2015}
{Gerke} J.~R.,  {Kochanek} C.~S.,   {Stanek} K.~Z.,  2015, \mn@doi [\mnras]
  {10.1093/mnras/stv776}, \href
  {https://ui.adsabs.harvard.edu/abs/2015MNRAS.450.3289G} {450, 3289}

\bibitem[\protect\citeauthoryear{{Graur} \& {Maoz}}{{Graur} \&
  {Maoz}}{2013}]{Graur2013}
{Graur} O.,  {Maoz} D.,  2013, \mn@doi [\mnras] {10.1093/mnras/sts718}, \href
  {https://ui.adsabs.harvard.edu/abs/2013MNRAS.430.1746G} {430, 1746}

\bibitem[\protect\citeauthoryear{{Graur} et~al.,}{{Graur}
  et~al.}{2014}]{Graur2014}
{Graur} O.,  et~al., 2014, \mn@doi [\apj] {10.1088/0004-637X/783/1/28}, \href
  {https://ui.adsabs.harvard.edu/abs/2014ApJ...783...28G} {783, 28}

\bibitem[\protect\citeauthoryear{{Greggio}}{{Greggio}}{2005}]{Greggio2005}
{Greggio} L.,  2005, \mn@doi [\aap] {10.1051/0004-6361:20052926}, \href
  {https://ui.adsabs.harvard.edu/abs/2005A&A...441.1055G} {441, 1055}

\bibitem[\protect\citeauthoryear{{Griffith}, {Johnson}  \&
  {Weinberg}}{{Griffith} et~al.}{2019}]{Griffith2019}
{Griffith} E.,  {Johnson} J.~A.,   {Weinberg} D.~H.,  2019, \mn@doi [\apj]
  {10.3847/1538-4357/ab4b5d}, \href
  {https://ui.adsabs.harvard.edu/abs/2019ApJ...886...84G} {886, 84}

\bibitem[\protect\citeauthoryear{{Griffith}, {Sukhbold}, {Weinberg}, {Johnson},
  {Johnson}  \& {Vincenzo}}{{Griffith} et~al.}{2021}]{Griffith2021}
{Griffith} E.~J.,  {Sukhbold} T.,  {Weinberg} D.~H.,  {Johnson} J.~A.,
  {Johnson} J.~W.,   {Vincenzo} F.,  2021, \mn@doi [\apj]
  {10.3847/1538-4357/ac1bac}, \href
  {https://ui.adsabs.harvard.edu/abs/2021ApJ...921...73G} {921, 73}

\bibitem[\protect\citeauthoryear{{Griffith}, {Weinberg}, {Buder}, {Johnson},
  {Johnson}  \& {Vincenzo}}{{Griffith} et~al.}{2022}]{Griffith2022}
{Griffith} E.~J.,  {Weinberg} D.~H.,  {Buder} S.,  {Johnson} J.~A.,  {Johnson}
  J.~W.,   {Vincenzo} F.,  2022, \mn@doi [\apj] {10.3847/1538-4357/ac5826},
  \href {https://ui.adsabs.harvard.edu/abs/2022ApJ...931...23G} {931, 23}

\bibitem[\protect\citeauthoryear{{Han} et~al.,}{{Han} et~al.}{2022}]{Han2022}
{Han} J.~J.,  et~al., 2022, arXiv e-prints, \href
  {https://ui.adsabs.harvard.edu/abs/2022arXiv220804327H} {p. arXiv:2208.04327}

\bibitem[\protect\citeauthoryear{{Harris} et~al.,}{{Harris}
  et~al.}{2020}]{Harris2020}
{Harris} C.~R.,  et~al., 2020, \mn@doi [\nat] {10.1038/s41586-020-2649-2},
  \href {https://ui.adsabs.harvard.edu/abs/2020Natur.585..357H} {585, 357}

\bibitem[\protect\citeauthoryear{{Hasselquist} et~al.,}{{Hasselquist}
  et~al.}{2021}]{Hasselquist2021}
{Hasselquist} S.,  et~al., 2021, \mn@doi [\apj] {10.3847/1538-4357/ac25f9},
  \href {https://ui.adsabs.harvard.edu/abs/2021ApJ...923..172H} {923, 172}

\bibitem[\protect\citeauthoryear{{Haywood}, {Di Matteo}, {Lehnert}, {Snaith},
  {Khoperskov}  \& {G{\'o}mez}}{{Haywood} et~al.}{2018}]{Haywood2018}
{Haywood} M.,  {Di Matteo} P.,  {Lehnert} M.~D.,  {Snaith} O.,  {Khoperskov}
  S.,   {G{\'o}mez} A.,  2018, \mn@doi [\apj] {10.3847/1538-4357/aad235}, \href
  {https://ui.adsabs.harvard.edu/abs/2018ApJ...863..113H} {863, 113}

\bibitem[\protect\citeauthoryear{{Helmi}, {White}, {de Zeeuw}  \&
  {Zhao}}{{Helmi} et~al.}{1999}]{Helmi1999}
{Helmi} A.,  {White} S. D.~M.,  {de Zeeuw} P.~T.,   {Zhao} H.,  1999, \mn@doi
  [\nat] {10.1038/46980}, \href
  {https://ui.adsabs.harvard.edu/abs/1999Natur.402...53H} {402, 53}

\bibitem[\protect\citeauthoryear{{Helmi}, {Babusiaux}, {Koppelman}, {Massari},
  {Veljanoski}  \& {Brown}}{{Helmi} et~al.}{2018}]{Helmi2018}
{Helmi} A.,  {Babusiaux} C.,  {Koppelman} H.~H.,  {Massari} D.,  {Veljanoski}
  J.,   {Brown} A. G.~A.,  2018, \mn@doi [\nat] {10.1038/s41586-018-0625-x},
  \href {https://ui.adsabs.harvard.edu/abs/2018Natur.563...85H} {563, 85}

\bibitem[\protect\citeauthoryear{{Holland-Ashford}, {Lopez}  \&
  {Auchettl}}{{Holland-Ashford} et~al.}{2020}]{HollandAshford2020}
{Holland-Ashford} T.,  {Lopez} L.~A.,   {Auchettl} K.,  2020, \mn@doi [\apj]
  {10.3847/1538-4357/ab64e4}, \href
  {https://ui.adsabs.harvard.edu/abs/2020ApJ...889..144H} {889, 144}

\bibitem[\protect\citeauthoryear{{Hopkins} \& {Beacom}}{{Hopkins} \&
  {Beacom}}{2006}]{Hopkins2006}
{Hopkins} A.~M.,  {Beacom} J.~F.,  2006, \mn@doi [\apj] {10.1086/506610}, \href
  {https://ui.adsabs.harvard.edu/abs/2006ApJ...651..142H} {651, 142}

\bibitem[\protect\citeauthoryear{{Hopkins}, {Kere{\v{s}}}, {O{\~n}orbe},
  {Faucher-Gigu{\`e}re}, {Quataert}, {Murray}  \& {Bullock}}{{Hopkins}
  et~al.}{2014}]{Hopkins2014}
{Hopkins} P.~F.,  {Kere{\v{s}}} D.,  {O{\~n}orbe} J.,  {Faucher-Gigu{\`e}re}
  C.-A.,  {Quataert} E.,  {Murray} N.,   {Bullock} J.~S.,  2014, \mn@doi
  [\mnras] {10.1093/mnras/stu1738}, \href
  {https://ui.adsabs.harvard.edu/abs/2014MNRAS.445..581H} {445, 581}

\bibitem[\protect\citeauthoryear{{Hu} et~al.,}{{Hu} et~al.}{2022}]{Hu2022}
{Hu} C.-Y.,  et~al., 2022, arXiv e-prints, \href
  {https://ui.adsabs.harvard.edu/abs/2022arXiv220810528H} {p. arXiv:2208.10528}

\bibitem[\protect\citeauthoryear{{Hudson} et~al.,}{{Hudson}
  et~al.}{2015}]{Hudson2015}
{Hudson} M.~J.,  et~al., 2015, \mn@doi [\mnras] {10.1093/mnras/stu2367}, \href
  {https://ui.adsabs.harvard.edu/abs/2015MNRAS.447..298H} {447, 298}

\bibitem[\protect\citeauthoryear{{Hunter}}{{Hunter}}{2007}]{Hunter2007}
{Hunter} J.~D.,  2007, \mn@doi [Computing in Science and Engineering]
  {10.1109/MCSE.2007.55}, \href
  {https://ui.adsabs.harvard.edu/abs/2007CSE.....9...90H} {9, 90}

\bibitem[\protect\citeauthoryear{{Hurley}, {Pols}  \& {Tout}}{{Hurley}
  et~al.}{2000}]{Hurley2000}
{Hurley} J.~R.,  {Pols} O.~R.,   {Tout} C.~A.,  2000, \mn@doi [\mnras]
  {10.1046/j.1365-8711.2000.03426.x}, \href
  {https://ui.adsabs.harvard.edu/abs/2000MNRAS.315..543H} {315, 543}

\bibitem[\protect\citeauthoryear{{Iwamoto}, {Brachwitz}, {Nomoto}, {Kishimoto},
  {Umeda}, {Hix}  \& {Thielemann}}{{Iwamoto} et~al.}{1999}]{Iwamoto1999}
{Iwamoto} K.,  {Brachwitz} F.,  {Nomoto} K.,  {Kishimoto} N.,  {Umeda} H.,
  {Hix} W.~R.,   {Thielemann} F.-K.,  1999, \mn@doi [\apjs] {10.1086/313278},
  \href {https://ui.adsabs.harvard.edu/abs/1999ApJS..125..439I} {125, 439}

\bibitem[\protect\citeauthoryear{{Johnson}}{{Johnson}}{2019}]{Johnson2019}
{Johnson} J.~A.,  2019, \mn@doi [Science] {10.1126/science.aau9540}, \href
  {https://ui.adsabs.harvard.edu/abs/2019Sci...363..474J} {363, 474}

\bibitem[\protect\citeauthoryear{{Johnson} \& {Weinberg}}{{Johnson} \&
  {Weinberg}}{2020}]{Johnson2020}
{Johnson} J.~W.,  {Weinberg} D.~H.,  2020, \mn@doi [\mnras]
  {10.1093/mnras/staa2431}, \href
  {https://ui.adsabs.harvard.edu/abs/2020MNRAS.498.1364J} {498, 1364}

\bibitem[\protect\citeauthoryear{{Johnson} et~al.,}{{Johnson}
  et~al.}{2021}]{Johnson2021}
{Johnson} J.~W.,  et~al., 2021, \mn@doi [\mnras] {10.1093/mnras/stab2718},
  \href {https://ui.adsabs.harvard.edu/abs/2021MNRAS.508.4484J} {508, 4484}

\bibitem[\protect\citeauthoryear{{Kalirai}, {Hansen}, {Kelson}, {Reitzel},
  {Rich}  \& {Richer}}{{Kalirai} et~al.}{2008}]{Kalirai2008}
{Kalirai} J.~S.,  {Hansen} B. M.~S.,  {Kelson} D.~D.,  {Reitzel} D.~B.,  {Rich}
  R.~M.,   {Richer} H.~B.,  2008, \mn@doi [\apj] {10.1086/527028}, \href
  {https://ui.adsabs.harvard.edu/abs/2008ApJ...676..594K} {676, 594}

\bibitem[\protect\citeauthoryear{{Kennicutt}}{{Kennicutt}}{1998}]{Kennicutt1998}
{Kennicutt} Robert~C. J.,  1998, \mn@doi [\apj] {10.1086/305588}, \href
  {https://ui.adsabs.harvard.edu/abs/1998ApJ...498..541K} {498, 541}

\bibitem[\protect\citeauthoryear{{Kennicutt} \& {Evans}}{{Kennicutt} \&
  {Evans}}{2012}]{Kennicutt2012}
{Kennicutt} R.~C.,  {Evans} N.~J.,  2012, \mn@doi [\araa]
  {10.1146/annurev-astro-081811-125610}, \href
  {https://ui.adsabs.harvard.edu/abs/2012ARA&A..50..531K} {50, 531}

\bibitem[\protect\citeauthoryear{{Kennicutt} \& {de los Reyes}}{{Kennicutt} \&
  {de los Reyes}}{2021}]{Kennicutt2021}
{Kennicutt} Robert~C. J.,  {de los Reyes} M. A.~C.,  2021, \mn@doi [\apj]
  {10.3847/1538-4357/abd3a2}, \href
  {https://ui.adsabs.harvard.edu/abs/2021ApJ...908...61K} {908, 61}

\bibitem[\protect\citeauthoryear{{Kirby}, {Lanfranchi}, {Simon}, {Cohen}  \&
  {Guhathakurta}}{{Kirby} et~al.}{2011}]{Kirby2011}
{Kirby} E.~N.,  {Lanfranchi} G.~A.,  {Simon} J.~D.,  {Cohen} J.~G.,
  {Guhathakurta} P.,  2011, \mn@doi [\apj] {10.1088/0004-637X/727/2/78}, \href
  {https://ui.adsabs.harvard.edu/abs/2011ApJ...727...78K} {727, 78}

\bibitem[\protect\citeauthoryear{{Kirby}, {Cohen}, {Guhathakurta}, {Cheng},
  {Bullock}  \& {Gallazzi}}{{Kirby} et~al.}{2013}]{Kirby2013}
{Kirby} E.~N.,  {Cohen} J.~G.,  {Guhathakurta} P.,  {Cheng} L.,  {Bullock}
  J.~S.,   {Gallazzi} A.,  2013, \mn@doi [\apj] {10.1088/0004-637X/779/2/102},
  \href {https://ui.adsabs.harvard.edu/abs/2013ApJ...779..102K} {779, 102}

\bibitem[\protect\citeauthoryear{{Kirby}, {Gilbert}, {Escala}, {Wojno},
  {Guhathakurta}, {Majewski}  \& {Beaton}}{{Kirby} et~al.}{2020}]{Kirby2020}
{Kirby} E.~N.,  {Gilbert} K.~M.,  {Escala} I.,  {Wojno} J.,  {Guhathakurta} P.,
   {Majewski} S.~R.,   {Beaton} R.~L.,  2020, \mn@doi [\aj]
  {10.3847/1538-3881/ab5f0f}, \href
  {https://ui.adsabs.harvard.edu/abs/2020AJ....159...46K} {159, 46}

\bibitem[\protect\citeauthoryear{{Kobayashi}, {Karakas}  \&
  {Lugaro}}{{Kobayashi} et~al.}{2020}]{Kobayashi2020}
{Kobayashi} C.,  {Karakas} A.~I.,   {Lugaro} M.,  2020, \mn@doi [\apj]
  {10.3847/1538-4357/abae65}, \href
  {https://ui.adsabs.harvard.edu/abs/2020ApJ...900..179K} {900, 179}

\bibitem[\protect\citeauthoryear{{Kroupa}}{{Kroupa}}{2001}]{Kroupa2001}
{Kroupa} P.,  2001, \mn@doi [\mnras] {10.1046/j.1365-8711.2001.04022.x}, \href
  {https://ui.adsabs.harvard.edu/abs/2001MNRAS.322..231K} {322, 231}

\bibitem[\protect\citeauthoryear{{Kruijssen}, {Pfeffer}, {Reina-Campos},
  {Crain}  \& {Bastian}}{{Kruijssen} et~al.}{2019}]{Kruijssen2019}
{Kruijssen} J.~M.~D.,  {Pfeffer} J.~L.,  {Reina-Campos} M.,  {Crain} R.~A.,
  {Bastian} N.,  2019, \mn@doi [\mnras] {10.1093/mnras/sty1609}, \href
  {https://ui.adsabs.harvard.edu/abs/2019MNRAS.486.3180K} {486, 3180}

\bibitem[\protect\citeauthoryear{{Kruijssen} et~al.,}{{Kruijssen}
  et~al.}{2020}]{Kruijssen2020}
{Kruijssen} J.~M.~D.,  et~al., 2020, \mn@doi [\mnras] {10.1093/mnras/staa2452},
  \href {https://ui.adsabs.harvard.edu/abs/2020MNRAS.498.2472K} {498, 2472}

\bibitem[\protect\citeauthoryear{{Krumholz}, {Burkhart}, {Forbes}  \&
  {Crocker}}{{Krumholz} et~al.}{2018}]{Krumholz2018}
{Krumholz} M.~R.,  {Burkhart} B.,  {Forbes} J.~C.,   {Crocker} R.~M.,  2018,
  \mn@doi [\mnras] {10.1093/mnras/sty852}, \href
  {https://ui.adsabs.harvard.edu/abs/2018MNRAS.477.2716K} {477, 2716}

\bibitem[\protect\citeauthoryear{{Larson}}{{Larson}}{1972}]{Larson1972}
{Larson} R.~B.,  1972, \mn@doi [Nature Physical Science]
  {10.1038/physci236007a0}, \href
  {https://ui.adsabs.harvard.edu/abs/1972NPhS..236....7L} {236, 7}

\bibitem[\protect\citeauthoryear{{Larson}}{{Larson}}{1974}]{Larson1974}
{Larson} R.~B.,  1974, \mn@doi [\mnras] {10.1093/mnras/166.3.585}, \href
  {https://ui.adsabs.harvard.edu/abs/1974MNRAS.166..585L} {166, 585}

\bibitem[\protect\citeauthoryear{{Leroy}, {Walter}, {Brinks}, {Bigiel}, {de
  Blok}, {Madore}  \& {Thornley}}{{Leroy} et~al.}{2008}]{Leroy2008}
{Leroy} A.~K.,  {Walter} F.,  {Brinks} E.,  {Bigiel} F.,  {de Blok} W.~J.~G.,
  {Madore} B.,   {Thornley} M.~D.,  2008, \mn@doi [\aj]
  {10.1088/0004-6256/136/6/2782}, \href
  {https://ui.adsabs.harvard.edu/abs/2008AJ....136.2782L} {136, 2782}

\bibitem[\protect\citeauthoryear{{Li}, {Vogelsberger}, {Marinacci}, {Sales}  \&
  {Torrey}}{{Li} et~al.}{2020}]{Li2020}
{Li} H.,  {Vogelsberger} M.,  {Marinacci} F.,  {Sales} L.~V.,   {Torrey} P.,
  2020, \mn@doi [\mnras] {10.1093/mnras/staa3122}, \href
  {https://ui.adsabs.harvard.edu/abs/2020MNRAS.499.5862L} {499, 5862}

\bibitem[\protect\citeauthoryear{{Lian}, {Thomas}, {Maraston}, {Goddard},
  {Comparat}, {Gonzalez-Perez}  \& {Ventura}}{{Lian} et~al.}{2018}]{Lian2018}
{Lian} J.,  {Thomas} D.,  {Maraston} C.,  {Goddard} D.,  {Comparat} J.,
  {Gonzalez-Perez} V.,   {Ventura} P.,  2018, \mn@doi [\mnras]
  {10.1093/mnras/stx2829}, \href
  {https://ui.adsabs.harvard.edu/abs/2018MNRAS.474.1143L} {474, 1143}

\bibitem[\protect\citeauthoryear{{Lian} et~al.,}{{Lian}
  et~al.}{2020}]{Lian2020}
{Lian} J.,  et~al., 2020, \mn@doi [\mnras] {10.1093/mnras/staa867}, \href
  {https://ui.adsabs.harvard.edu/abs/2020MNRAS.494.2561L} {494, 2561}

\bibitem[\protect\citeauthoryear{{Lianou}, {Grebel}  \& {Koch}}{{Lianou}
  et~al.}{2011}]{Lianou2011}
{Lianou} S.,  {Grebel} E.~K.,   {Koch} A.,  2011, \mn@doi [\aap]
  {10.1051/0004-6361/201116998}, \href
  {https://ui.adsabs.harvard.edu/abs/2011A&A...531A.152L} {531, A152}

\bibitem[\protect\citeauthoryear{{Limongi} \& {Chieffi}}{{Limongi} \&
  {Chieffi}}{2018}]{Limongi2018}
{Limongi} M.,  {Chieffi} A.,  2018, \mn@doi [\apjs] {10.3847/1538-4365/aacb24},
  \href {https://ui.adsabs.harvard.edu/abs/2018ApJS..237...13L} {237, 13}

\bibitem[\protect\citeauthoryear{{Linsky} et~al.,}{{Linsky}
  et~al.}{2006}]{Linsky2006}
{Linsky} J.~L.,  et~al., 2006, \mn@doi [\apj] {10.1086/505556}, \href
  {https://ui.adsabs.harvard.edu/abs/2006ApJ...647.1106L} {647, 1106}

\bibitem[\protect\citeauthoryear{{Liu}, {Yong}, {Asplund}, {Ram{\'\i}rez}  \&
  {Mel{\'e}ndez}}{{Liu} et~al.}{2016a}]{Liu2016a}
{Liu} F.,  {Yong} D.,  {Asplund} M.,  {Ram{\'\i}rez} I.,   {Mel{\'e}ndez} J.,
  2016a, \mn@doi [\mnras] {10.1093/mnras/stw247}, \href
  {https://ui.adsabs.harvard.edu/abs/2016MNRAS.457.3934L} {457, 3934}

\bibitem[\protect\citeauthoryear{{Liu}, {Asplund}, {Yong}, {Mel{\'e}ndez},
  {Ram{\'\i}rez}, {Karakas}, {Carlos}  \& {Marino}}{{Liu}
  et~al.}{2016b}]{Liu2016b}
{Liu} F.,  {Asplund} M.,  {Yong} D.,  {Mel{\'e}ndez} J.,  {Ram{\'\i}rez} I.,
  {Karakas} A.~I.,  {Carlos} M.,   {Marino} A.~F.,  2016b, \mn@doi [\mnras]
  {10.1093/mnras/stw2045}, \href
  {https://ui.adsabs.harvard.edu/abs/2016MNRAS.463..696L} {463, 696}

\bibitem[\protect\citeauthoryear{{Liu}, {Asplund}, {Yong}, {Feltzing},
  {Dotter}, {Mel{\'e}ndez}  \& {Ram{\'\i}rez}}{{Liu} et~al.}{2019}]{Liu2019}
{Liu} F.,  {Asplund} M.,  {Yong} D.,  {Feltzing} S.,  {Dotter} A.,
  {Mel{\'e}ndez} J.,   {Ram{\'\i}rez} I.,  2019, \mn@doi [\aap]
  {10.1051/0004-6361/201935306}, \href
  {https://ui.adsabs.harvard.edu/abs/2019A&A...627A.117L} {627, A117}

\bibitem[\protect\citeauthoryear{{Lopez}, {Mathur}, {Nguyen}, {Thompson}  \&
  {Olivier}}{{Lopez} et~al.}{2020}]{Lopez2020}
{Lopez} L.~A.,  {Mathur} S.,  {Nguyen} D.~D.,  {Thompson} T.~A.,   {Olivier}
  G.~M.,  2020, \mn@doi [\apj] {10.3847/1538-4357/abc010}, \href
  {https://ui.adsabs.harvard.edu/abs/2020ApJ...904..152L} {904, 152}

\bibitem[\protect\citeauthoryear{{Lopez}, {Lopez}, {Nguyen}, {Thompson},
  {Mathur}, {Bolatto}, {Vulic}  \& {Sardone}}{{Lopez} et~al.}{2022}]{Lopez2022}
{Lopez} S.,  {Lopez} L.~A.,  {Nguyen} D.~D.,  {Thompson} T.~A.,  {Mathur} S.,
  {Bolatto} A.~D.,  {Vulic} N.,   {Sardone} A.,  2022, arXiv e-prints, \href
  {https://ui.adsabs.harvard.edu/abs/2022arXiv220909260L} {p. arXiv:2209.09260}

\bibitem[\protect\citeauthoryear{{Mackereth} et~al.,}{{Mackereth}
  et~al.}{2019}]{Mackereth2019}
{Mackereth} J.~T.,  et~al., 2019, \mn@doi [\mnras] {10.1093/mnras/sty2955},
  \href {https://ui.adsabs.harvard.edu/abs/2019MNRAS.482.3426M} {482, 3426}

\bibitem[\protect\citeauthoryear{{Madau} \& {Dickinson}}{{Madau} \&
  {Dickinson}}{2014}]{Madau2014}
{Madau} P.,  {Dickinson} M.,  2014, \mn@doi [\araa]
  {10.1146/annurev-astro-081811-125615}, \href
  {https://ui.adsabs.harvard.edu/abs/2014ARA&A..52..415M} {52, 415}

\bibitem[\protect\citeauthoryear{{Madau} \& {Fragos}}{{Madau} \&
  {Fragos}}{2017}]{Madau2017}
{Madau} P.,  {Fragos} T.,  2017, \mn@doi [\apj] {10.3847/1538-4357/aa6af9},
  \href {https://ui.adsabs.harvard.edu/abs/2017ApJ...840...39M} {840, 39}

\bibitem[\protect\citeauthoryear{{Maeder} \& {Meynet}}{{Maeder} \&
  {Meynet}}{1989}]{Maeder1989}
{Maeder} A.,  {Meynet} G.,  1989, \aap, \href
  {https://ui.adsabs.harvard.edu/abs/1989A&A...210..155M} {210, 155}

\bibitem[\protect\citeauthoryear{{Majewski} et~al.,}{{Majewski}
  et~al.}{2017}]{Majewski2017}
{Majewski} S.~R.,  et~al., 2017, \mn@doi [\aj] {10.3847/1538-3881/aa784d},
  \href {https://ui.adsabs.harvard.edu/abs/2017AJ....154...94M} {154, 94}

\bibitem[\protect\citeauthoryear{{Malhan}, {Yuan}, {Ibata}, {Arentsen},
  {Bellazzini}  \& {Martin}}{{Malhan} et~al.}{2021}]{Malhan2021}
{Malhan} K.,  {Yuan} Z.,  {Ibata} R.~A.,  {Arentsen} A.,  {Bellazzini} M.,
  {Martin} N.~F.,  2021, \mn@doi [\apj] {10.3847/1538-4357/ac1675}, \href
  {https://ui.adsabs.harvard.edu/abs/2021ApJ...920...51M} {920, 51}

\bibitem[\protect\citeauthoryear{{Malhan} et~al.,}{{Malhan}
  et~al.}{2022}]{Malhan2022}
{Malhan} K.,  et~al., 2022, \mn@doi [\apj] {10.3847/1538-4357/ac4d2a}, \href
  {https://ui.adsabs.harvard.edu/abs/2022ApJ...926..107M} {926, 107}

\bibitem[\protect\citeauthoryear{{Maoz} \& {Mannucci}}{{Maoz} \&
  {Mannucci}}{2012}]{Maoz2012a}
{Maoz} D.,  {Mannucci} F.,  2012, \mn@doi [\pasa] {10.1071/AS11052}, \href
  {https://ui.adsabs.harvard.edu/abs/2012PASA...29..447M} {29, 447}

\bibitem[\protect\citeauthoryear{{Maoz}, {Mannucci}  \& {Brandt}}{{Maoz}
  et~al.}{2012}]{Maoz2012b}
{Maoz} D.,  {Mannucci} F.,   {Brandt} T.~D.,  2012, \mn@doi [\mnras]
  {10.1111/j.1365-2966.2012.21871.x}, \href
  {https://ui.adsabs.harvard.edu/abs/2012MNRAS.426.3282M} {426, 3282}

\bibitem[\protect\citeauthoryear{{Martell} et~al.,}{{Martell}
  et~al.}{2017}]{Martell2017}
{Martell} S.~L.,  et~al., 2017, \mn@doi [\mnras] {10.1093/mnras/stw2835}, \href
  {https://ui.adsabs.harvard.edu/abs/2017MNRAS.465.3203M} {465, 3203}

\bibitem[\protect\citeauthoryear{{Martin} et~al.,}{{Martin}
  et~al.}{2022}]{Martin2022}
{Martin} N.~F.,  et~al., 2022, \mn@doi [\nat] {10.1038/s41586-021-04162-2},
  \href {https://ui.adsabs.harvard.edu/abs/2022Natur.601...45M} {601, 45}

\bibitem[\protect\citeauthoryear{{Massari}, {Koppelman}  \& {Helmi}}{{Massari}
  et~al.}{2019}]{Massari2019}
{Massari} D.,  {Koppelman} H.~H.,   {Helmi} A.,  2019, \mn@doi [\aap]
  {10.1051/0004-6361/201936135}, \href
  {https://ui.adsabs.harvard.edu/abs/2019A&A...630L...4M} {630, L4}

\bibitem[\protect\citeauthoryear{{Matteucci}}{{Matteucci}}{2012}]{Matteucci2012}
{Matteucci} F.,  2012, {Chemical Evolution of Galaxies},
  \mn@doi{10.1007/978-3-642-22491-1.
}

\bibitem[\protect\citeauthoryear{{Matteucci}}{{Matteucci}}{2021}]{Matteucci2021}
{Matteucci} F.,  2021, \mn@doi [\aapr] {10.1007/s00159-021-00133-8}, \href
  {https://ui.adsabs.harvard.edu/abs/2021A&ARv..29....5M} {29, 5}

\bibitem[\protect\citeauthoryear{{Mel{\'e}ndez}, {Asplund}, {Gustafsson}  \&
  {Yong}}{{Mel{\'e}ndez} et~al.}{2009}]{Melendez2009}
{Mel{\'e}ndez} J.,  {Asplund} M.,  {Gustafsson} B.,   {Yong} D.,  2009, \mn@doi
  [\apjl] {10.1088/0004-637X/704/1/L66}, \href
  {https://ui.adsabs.harvard.edu/abs/2009ApJ...704L..66M} {704, L66}

\bibitem[\protect\citeauthoryear{{Melioli}, {Brighenti}, {D'Ercole}  \& {de
  Gouveia Dal Pino}}{{Melioli} et~al.}{2008}]{Melioli2008}
{Melioli} C.,  {Brighenti} F.,  {D'Ercole} A.,   {de Gouveia Dal Pino} E.~M.,
  2008, \mn@doi [\mnras] {10.1111/j.1365-2966.2008.13446.x}, \href
  {https://ui.adsabs.harvard.edu/abs/2008MNRAS.388..573M} {388, 573}

\bibitem[\protect\citeauthoryear{{Melioli}, {Brighenti}, {D'Ercole}  \& {de
  Gouveia Dal Pino}}{{Melioli} et~al.}{2009}]{Melioli2009}
{Melioli} C.,  {Brighenti} F.,  {D'Ercole} A.,   {de Gouveia Dal Pino} E.~M.,
  2009, \mn@doi [\mnras] {10.1111/j.1365-2966.2009.14725.x}, \href
  {https://ui.adsabs.harvard.edu/abs/2009MNRAS.399.1089M} {399, 1089}

\bibitem[\protect\citeauthoryear{{Miller} \& {Scalo}}{{Miller} \&
  {Scalo}}{1979}]{Miller1979}
{Miller} G.~E.,  {Scalo} J.~M.,  1979, \mn@doi [\apjs] {10.1086/190629}, \href
  {https://ui.adsabs.harvard.edu/abs/1979ApJS...41..513M} {41, 513}

\bibitem[\protect\citeauthoryear{{Minchev}, {Chiappini}  \& {Martig}}{{Minchev}
  et~al.}{2013}]{Minchev2013}
{Minchev} I.,  {Chiappini} C.,   {Martig} M.,  2013, \mn@doi [\aap]
  {10.1051/0004-6361/201220189}, \href
  {https://ui.adsabs.harvard.edu/abs/2013A&A...558A...9M} {558, A9}

\bibitem[\protect\citeauthoryear{{Minchev}, {Chiappini}  \& {Martig}}{{Minchev}
  et~al.}{2014}]{Minchev2014}
{Minchev} I.,  {Chiappini} C.,   {Martig} M.,  2014, \mn@doi [\aap]
  {10.1051/0004-6361/201423487}, \href
  {https://ui.adsabs.harvard.edu/abs/2014A&A...572A..92M} {572, A92}

\bibitem[\protect\citeauthoryear{{Minchev}, {Steinmetz}, {Chiappini}, {Martig},
  {Anders}, {Matijevic}  \& {de Jong}}{{Minchev} et~al.}{2017}]{Minchev2017}
{Minchev} I.,  {Steinmetz} M.,  {Chiappini} C.,  {Martig} M.,  {Anders} F.,
  {Matijevic} G.,   {de Jong} R.~S.,  2017, \mn@doi [\apj]
  {10.3847/1538-4357/834/1/27}, \href
  {https://ui.adsabs.harvard.edu/abs/2017ApJ...834...27M} {834, 27}

\bibitem[\protect\citeauthoryear{{Monelli} et~al.,}{{Monelli}
  et~al.}{2010a}]{Monelli2010a}
{Monelli} M.,  et~al., 2010a, \mn@doi [\apj] {10.1088/0004-637X/720/2/1225},
  \href {https://ui.adsabs.harvard.edu/abs/2010ApJ...720.1225M} {720, 1225}

\bibitem[\protect\citeauthoryear{{Monelli} et~al.,}{{Monelli}
  et~al.}{2010b}]{Monelli2010b}
{Monelli} M.,  et~al., 2010b, \mn@doi [\apj] {10.1088/0004-637X/722/2/1864},
  \href {https://ui.adsabs.harvard.edu/abs/2010ApJ...722.1864M} {722, 1864}

\bibitem[\protect\citeauthoryear{{Montalb{\'a}n} et~al.,}{{Montalb{\'a}n}
  et~al.}{2021}]{Montalban2021}
{Montalb{\'a}n} J.,  et~al., 2021, \mn@doi [Nature Astronomy]
  {10.1038/s41550-021-01347-7}, \href
  {https://ui.adsabs.harvard.edu/abs/2021NatAs...5..640M} {5, 640}

\bibitem[\protect\citeauthoryear{{Muratov}, {Kere{\v{s}}},
  {Faucher-Gigu{\`e}re}, {Hopkins}, {Quataert}  \& {Murray}}{{Muratov}
  et~al.}{2015}]{Muratov2015}
{Muratov} A.~L.,  {Kere{\v{s}}} D.,  {Faucher-Gigu{\`e}re} C.-A.,  {Hopkins}
  P.~F.,  {Quataert} E.,   {Murray} N.,  2015, \mn@doi [\mnras]
  {10.1093/mnras/stv2126}, \href
  {https://ui.adsabs.harvard.edu/abs/2015MNRAS.454.2691M} {454, 2691}

\bibitem[\protect\citeauthoryear{{Myeong}, {Evans}, {Belokurov}, {Sanders}  \&
  {Koposov}}{{Myeong} et~al.}{2018}]{Myeong2018}
{Myeong} G.~C.,  {Evans} N.~W.,  {Belokurov} V.,  {Sanders} J.~L.,   {Koposov}
  S.~E.,  2018, \mn@doi [\apjl] {10.3847/2041-8213/aad7f7}, \href
  {https://ui.adsabs.harvard.edu/abs/2018ApJ...863L..28M} {863, L28}

\bibitem[\protect\citeauthoryear{{Naidu}, {Conroy}, {Bonaca}, {Johnson},
  {Ting}, {Caldwell}, {Zaritsky}  \& {Cargile}}{{Naidu}
  et~al.}{2020}]{Naidu2020}
{Naidu} R.~P.,  {Conroy} C.,  {Bonaca} A.,  {Johnson} B.~D.,  {Ting} Y.-S.,
  {Caldwell} N.,  {Zaritsky} D.,   {Cargile} P.~A.,  2020, \mn@doi [\apj]
  {10.3847/1538-4357/abaef4}, \href
  {https://ui.adsabs.harvard.edu/abs/2020ApJ...901...48N} {901, 48}

\bibitem[\protect\citeauthoryear{{Naidu} et~al.,}{{Naidu}
  et~al.}{2021}]{Naidu2021}
{Naidu} R.~P.,  et~al., 2021, \mn@doi [\apj] {10.3847/1538-4357/ac2d2d}, \href
  {https://ui.adsabs.harvard.edu/abs/2021ApJ...923...92N} {923, 92}

\bibitem[\protect\citeauthoryear{{Naidu} et~al.,}{{Naidu}
  et~al.}{2022}]{Naidu2022}
{Naidu} R.~P.,  et~al., 2022, arXiv e-prints, \href
  {https://ui.adsabs.harvard.edu/abs/2022arXiv220409057N} {p. arXiv:2204.09057}

\bibitem[\protect\citeauthoryear{{Nomoto}, {Kobayashi}  \& {Tominaga}}{{Nomoto}
  et~al.}{2013}]{Nomoto2013}
{Nomoto} K.,  {Kobayashi} C.,   {Tominaga} N.,  2013, \mn@doi [\araa]
  {10.1146/annurev-astro-082812-140956}, \href
  {https://ui.adsabs.harvard.edu/abs/2013ARA&A..51..457N} {51, 457}

\bibitem[\protect\citeauthoryear{{O'Connor} \& {Ott}}{{O'Connor} \&
  {Ott}}{2011}]{OConnor2011}
{O'Connor} E.,  {Ott} C.~D.,  2011, \mn@doi [\apj]
  {10.1088/0004-637X/730/2/70}, \href
  {https://ui.adsabs.harvard.edu/abs/2011ApJ...730...70O} {730, 70}

\bibitem[\protect\citeauthoryear{{Pagel}}{{Pagel}}{2009}]{Pagel2009}
{Pagel} B. E.~J.,  2009, {Nucleosynthesis and Chemical Evolution of Galaxies}

\bibitem[\protect\citeauthoryear{{Peeples} \& {Shankar}}{{Peeples} \&
  {Shankar}}{2011}]{Peeples2011}
{Peeples} M.~S.,  {Shankar} F.,  2011, \mn@doi [\mnras]
  {10.1111/j.1365-2966.2011.19456.x}, \href
  {https://ui.adsabs.harvard.edu/abs/2011MNRAS.417.2962P} {417, 2962}

\bibitem[\protect\citeauthoryear{{Pejcha} \& {Thompson}}{{Pejcha} \&
  {Thompson}}{2015}]{Pejcha2015}
{Pejcha} O.,  {Thompson} T.~A.,  2015, \mn@doi [\apj]
  {10.1088/0004-637X/801/2/90}, \href
  {https://ui.adsabs.harvard.edu/abs/2015ApJ...801...90P} {801, 90}

\bibitem[\protect\citeauthoryear{{Phillips}, {Wheeler}, {Boylan-Kolchin},
  {Bullock}, {Cooper}  \& {Tollerud}}{{Phillips} et~al.}{2014}]{Phillips2014}
{Phillips} J.~I.,  {Wheeler} C.,  {Boylan-Kolchin} M.,  {Bullock} J.~S.,
  {Cooper} M.~C.,   {Tollerud} E.~J.,  2014, \mn@doi [\mnras]
  {10.1093/mnras/stt2023}, \href
  {https://ui.adsabs.harvard.edu/abs/2014MNRAS.437.1930P} {437, 1930}

\bibitem[\protect\citeauthoryear{{Phillips}, {Wheeler}, {Cooper},
  {Boylan-Kolchin}, {Bullock}  \& {Tollerud}}{{Phillips}
  et~al.}{2015}]{Phillips2015}
{Phillips} J.~I.,  {Wheeler} C.,  {Cooper} M.~C.,  {Boylan-Kolchin} M.,
  {Bullock} J.~S.,   {Tollerud} E.,  2015, \mn@doi [\mnras]
  {10.1093/mnras/stu2192}, \href
  {https://ui.adsabs.harvard.edu/abs/2015MNRAS.447..698P} {447, 698}

\bibitem[\protect\citeauthoryear{Press, Teukolsky, Vetterling  \&
  Flannery}{Press et~al.}{2007}]{Press2007}
Press W.~H.,  Teukolsky S.~A.,  Vetterling W.~T.,   Flannery B.~P.,  2007,
  Numerical Recipes 3rd Edition: The Art of Scientific Computing, 3 edn.
Cambridge University Press, \url
  {http://www.amazon.com/Numerical-Recipes-3rd-Scientific-Computing/dp/0521880688/ref=sr_1_1?ie=UTF8&s=books&qid=1280322496&sr=8-1}

\bibitem[\protect\citeauthoryear{{Prodanovi{\'c}}, {Steigman}  \&
  {Fields}}{{Prodanovi{\'c}} et~al.}{2010}]{Prodanovic2010}
{Prodanovi{\'c}} T.,  {Steigman} G.,   {Fields} B.~D.,  2010, \mn@doi [\mnras]
  {10.1111/j.1365-2966.2010.16734.x}, \href
  {https://ui.adsabs.harvard.edu/abs/2010MNRAS.406.1108P} {406, 1108}

\bibitem[\protect\citeauthoryear{{Rocha}, {Peter}  \& {Bullock}}{{Rocha}
  et~al.}{2012}]{Rocha2012}
{Rocha} M.,  {Peter} A. H.~G.,   {Bullock} J.,  2012, \mn@doi [\mnras]
  {10.1111/j.1365-2966.2012.21432.x}, \href
  {https://ui.adsabs.harvard.edu/abs/2012MNRAS.425..231R} {425, 231}

\bibitem[\protect\citeauthoryear{{Roederer} \& {Gnedin}}{{Roederer} \&
  {Gnedin}}{2019}]{Roederer2019}
{Roederer} I.~U.,  {Gnedin} O.~Y.,  2019, \mn@doi [\apj]
  {10.3847/1538-4357/ab365c}, \href
  {https://ui.adsabs.harvard.edu/abs/2019ApJ...883...84R} {883, 84}

\bibitem[\protect\citeauthoryear{{Salpeter}}{{Salpeter}}{1955}]{Salpeter1955}
{Salpeter} E.~E.,  1955, \mn@doi [\apj] {10.1086/145971}, \href
  {https://ui.adsabs.harvard.edu/abs/1955ApJ...121..161S} {121, 161}

\bibitem[\protect\citeauthoryear{{Schleicher} \& {Beck}}{{Schleicher} \&
  {Beck}}{2016}]{Schleicher2016}
{Schleicher} D. R.~G.,  {Beck} R.,  2016, \mn@doi [\aap]
  {10.1051/0004-6361/201628843}, \href
  {https://ui.adsabs.harvard.edu/abs/2016A&A...593A..77S} {593, A77}

\bibitem[\protect\citeauthoryear{{Shank}, {Komater}, {Beers}, {Placco}  \&
  {Huang}}{{Shank} et~al.}{2022}]{Shank2022}
{Shank} D.,  {Komater} D.,  {Beers} T.~C.,  {Placco} V.~M.,   {Huang} Y.,
  2022, \mn@doi [\apjs] {10.3847/1538-4365/ac680c}, \href
  {https://ui.adsabs.harvard.edu/abs/2022ApJS..261...19S} {261, 19}

\bibitem[\protect\citeauthoryear{{Skrutskie} et~al.,}{{Skrutskie}
  et~al.}{2006}]{Skrutskie2006}
{Skrutskie} M.~F.,  et~al., 2006, \mn@doi [\aj] {10.1086/498708}, \href
  {https://ui.adsabs.harvard.edu/abs/2006AJ....131.1163S} {131, 1163}

\bibitem[\protect\citeauthoryear{{Slater} \& {Bell}}{{Slater} \&
  {Bell}}{2013}]{Slater2013}
{Slater} C.~T.,  {Bell} E.~F.,  2013, \mn@doi [\apj]
  {10.1088/0004-637X/773/1/17}, \href
  {https://ui.adsabs.harvard.edu/abs/2013ApJ...773...17S} {773, 17}

\bibitem[\protect\citeauthoryear{{Slater} \& {Bell}}{{Slater} \&
  {Bell}}{2014}]{Slater2014}
{Slater} C.~T.,  {Bell} E.~F.,  2014, \mn@doi [\apj]
  {10.1088/0004-637X/792/2/141}, \href
  {https://ui.adsabs.harvard.edu/abs/2014ApJ...792..141S} {792, 141}

\bibitem[\protect\citeauthoryear{{Soderblom}}{{Soderblom}}{2010}]{Soderblom2010}
{Soderblom} D.~R.,  2010, \mn@doi [\araa]
  {10.1146/annurev-astro-081309-130806}, \href
  {https://ui.adsabs.harvard.edu/abs/2010ARA&A..48..581S} {48, 581}

\bibitem[\protect\citeauthoryear{{Sohn}, {Besla}, {van der Marel},
  {Boylan-Kolchin}, {Majewski}  \& {Bullock}}{{Sohn} et~al.}{2013}]{Sohn2013}
{Sohn} S.~T.,  {Besla} G.,  {van der Marel} R.~P.,  {Boylan-Kolchin} M.,
  {Majewski} S.~R.,   {Bullock} J.~S.,  2013, \mn@doi [\apj]
  {10.1088/0004-637X/768/2/139}, \href
  {https://ui.adsabs.harvard.edu/abs/2013ApJ...768..139S} {768, 139}

\bibitem[\protect\citeauthoryear{{Somerville} \& {Dav{\'e}}}{{Somerville} \&
  {Dav{\'e}}}{2015}]{Somerville2015a}
{Somerville} R.~S.,  {Dav{\'e}} R.,  2015, \mn@doi [\araa]
  {10.1146/annurev-astro-082812-140951}, \href
  {https://ui.adsabs.harvard.edu/abs/2015ARA&A..53...51S} {53, 51}

\bibitem[\protect\citeauthoryear{{Souto} et~al.,}{{Souto}
  et~al.}{2019}]{Souto2019}
{Souto} D.,  et~al., 2019, \mn@doi [\apj] {10.3847/1538-4357/ab0b43}, \href
  {https://ui.adsabs.harvard.edu/abs/2019ApJ...874...97S} {874, 97}

\bibitem[\protect\citeauthoryear{{Spergel} et~al.,}{{Spergel}
  et~al.}{2013}]{Spergel2013}
{Spergel} D.,  et~al., 2013, arXiv e-prints, \href
  {https://ui.adsabs.harvard.edu/abs/2013arXiv1305.5422S} {p. arXiv:1305.5422}

\bibitem[\protect\citeauthoryear{{Spergel} et~al.,}{{Spergel}
  et~al.}{2015}]{Spergel2015}
{Spergel} D.,  et~al., 2015, arXiv e-prints, \href
  {https://ui.adsabs.harvard.edu/abs/2015arXiv150303757S} {p. arXiv:1503.03757}

\bibitem[\protect\citeauthoryear{{Spina}, {Mel{\'e}ndez}, {Casey}, {Karakas}
  \& {Tucci-Maia}}{{Spina} et~al.}{2018}]{Spina2018}
{Spina} L.,  {Mel{\'e}ndez} J.,  {Casey} A.~R.,  {Karakas} A.~I.,
  {Tucci-Maia} M.,  2018, \mn@doi [\apj] {10.3847/1538-4357/aad190}, \href
  {https://ui.adsabs.harvard.edu/abs/2018ApJ...863..179S} {863, 179}

\bibitem[\protect\citeauthoryear{{Spitoni}, {Recchi}  \& {Matteucci}}{{Spitoni}
  et~al.}{2008}]{Spitoni2008}
{Spitoni} E.,  {Recchi} S.,   {Matteucci} F.,  2008, \mn@doi [\aap]
  {10.1051/0004-6361:200809403}, \href
  {https://ui.adsabs.harvard.edu/abs/2008A&A...484..743S} {484, 743}

\bibitem[\protect\citeauthoryear{{Spitoni}, {Matteucci}, {Recchi}, {Cescutti}
  \& {Pipino}}{{Spitoni} et~al.}{2009}]{Spitoni2009}
{Spitoni} E.,  {Matteucci} F.,  {Recchi} S.,  {Cescutti} G.,   {Pipino} A.,
  2009, \mn@doi [\aap] {10.1051/0004-6361/200911768}, \href
  {https://ui.adsabs.harvard.edu/abs/2009A&A...504...87S} {504, 87}

\bibitem[\protect\citeauthoryear{{Spitoni}, {Silva Aguirre}, {Matteucci},
  {Calura}  \& {Grisoni}}{{Spitoni} et~al.}{2019}]{Spitoni2019}
{Spitoni} E.,  {Silva Aguirre} V.,  {Matteucci} F.,  {Calura} F.,   {Grisoni}
  V.,  2019, \mn@doi [\aap] {10.1051/0004-6361/201834188}, \href
  {https://ui.adsabs.harvard.edu/abs/2019A&A...623A..60S} {623, A60}

\bibitem[\protect\citeauthoryear{{Spitoni}, {Verma}, {Silva Aguirre}  \&
  {Calura}}{{Spitoni} et~al.}{2020}]{Spitoni2020}
{Spitoni} E.,  {Verma} K.,  {Silva Aguirre} V.,   {Calura} F.,  2020, \mn@doi
  [\aap] {10.1051/0004-6361/201937275}, \href
  {https://ui.adsabs.harvard.edu/abs/2020A&A...635A..58S} {635, A58}

\bibitem[\protect\citeauthoryear{{Spitoni} et~al.,}{{Spitoni}
  et~al.}{2021}]{Spitoni2021}
{Spitoni} E.,  et~al., 2021, \mn@doi [\aap] {10.1051/0004-6361/202039864},
  \href {https://ui.adsabs.harvard.edu/abs/2021A&A...647A..73S} {647, A73}

\bibitem[\protect\citeauthoryear{{Steyrleithner}, {Hensler}  \&
  {Boselli}}{{Steyrleithner} et~al.}{2020}]{Steyrleithner2020}
{Steyrleithner} P.,  {Hensler} G.,   {Boselli} A.,  2020, \mn@doi [\mnras]
  {10.1093/mnras/staa775}, \href
  {https://ui.adsabs.harvard.edu/abs/2020MNRAS.494.1114S} {494, 1114}

\bibitem[\protect\citeauthoryear{{Stilp}, {Dalcanton}, {Skillman}, {Warren},
  {Ott}  \& {Koribalski}}{{Stilp} et~al.}{2013}]{Stilp2013}
{Stilp} A.~M.,  {Dalcanton} J.~J.,  {Skillman} E.,  {Warren} S.~R.,  {Ott} J.,
   {Koribalski} B.,  2013, \mn@doi [\apj] {10.1088/0004-637X/773/2/88}, \href
  {https://ui.adsabs.harvard.edu/abs/2013ApJ...773...88S} {773, 88}

\bibitem[\protect\citeauthoryear{{Strolger}, {Rodney}, {Pacifici}, {Narayan}
  \& {Graur}}{{Strolger} et~al.}{2020}]{Strolger2020}
{Strolger} L.-G.,  {Rodney} S.~A.,  {Pacifici} C.,  {Narayan} G.,   {Graur} O.,
   2020, \mn@doi [\apj] {10.3847/1538-4357/ab6a97}, \href
  {https://ui.adsabs.harvard.edu/abs/2020ApJ...890..140S} {890, 140}

\bibitem[\protect\citeauthoryear{{Stryker}}{{Stryker}}{1993}]{Stryker1993}
{Stryker} L.~L.,  1993, \mn@doi [\pasp] {10.1086/133286}, \href
  {https://ui.adsabs.harvard.edu/abs/1993PASP..105.1081S} {105, 1081}

\bibitem[\protect\citeauthoryear{{Sukhbold}, {Ertl}, {Woosley}, {Brown}  \&
  {Janka}}{{Sukhbold} et~al.}{2016}]{Sukhbold2016}
{Sukhbold} T.,  {Ertl} T.,  {Woosley} S.~E.,  {Brown} J.~M.,   {Janka} H.~T.,
  2016, \mn@doi [\apj] {10.3847/0004-637X/821/1/38}, \href
  {https://ui.adsabs.harvard.edu/abs/2016ApJ...821...38S} {821, 38}

\bibitem[\protect\citeauthoryear{{Szentgyorgyi} et~al.,}{{Szentgyorgyi}
  et~al.}{2011}]{Szentgyorgyi2011}
{Szentgyorgyi} A.,  et~al., 2011, \mn@doi [\pasp] {10.1086/662209}, \href
  {https://ui.adsabs.harvard.edu/abs/2011PASP..123.1188S} {123, 1188}

\bibitem[\protect\citeauthoryear{{Tacconi} et~al.,}{{Tacconi}
  et~al.}{2018}]{Tacconi2018}
{Tacconi} L.~J.,  et~al., 2018, \mn@doi [\apj] {10.3847/1538-4357/aaa4b4},
  \href {https://ui.adsabs.harvard.edu/abs/2018ApJ...853..179T} {853, 179}

\bibitem[\protect\citeauthoryear{{Tinsley}}{{Tinsley}}{1980}]{Tinsley1980}
{Tinsley} B.~M.,  1980, \mn@doi [\fcp] {10.48550/arXiv.2203.02041}, \href
  {https://ui.adsabs.harvard.edu/abs/1980FCPh....5..287T} {5, 287}

\bibitem[\protect\citeauthoryear{{Tremonti} et~al.,}{{Tremonti}
  et~al.}{2004}]{Tremonti2004}
{Tremonti} C.~A.,  et~al., 2004, \mn@doi [\apj] {10.1086/423264}, \href
  {https://ui.adsabs.harvard.edu/abs/2004ApJ...613..898T} {613, 898}

\bibitem[\protect\citeauthoryear{{Veilleux}, {Maiolino}, {Bolatto}  \&
  {Aalto}}{{Veilleux} et~al.}{2020}]{Veilleux2020}
{Veilleux} S.,  {Maiolino} R.,  {Bolatto} A.~D.,   {Aalto} S.,  2020, \mn@doi
  [\aapr] {10.1007/s00159-019-0121-9}, \href
  {https://ui.adsabs.harvard.edu/abs/2020A&ARv..28....2V} {28, 2}

\bibitem[\protect\citeauthoryear{{Vincenzo}, {Spitoni}, {Calura}, {Matteucci},
  {Silva Aguirre}, {Miglio}  \& {Cescutti}}{{Vincenzo}
  et~al.}{2019}]{Vincenzo2019}
{Vincenzo} F.,  {Spitoni} E.,  {Calura} F.,  {Matteucci} F.,  {Silva Aguirre}
  V.,  {Miglio} A.,   {Cescutti} G.,  2019, \mn@doi [\mnras]
  {10.1093/mnrasl/slz070}, \href
  {https://ui.adsabs.harvard.edu/abs/2019MNRAS.487L..47V} {487, L47}

\bibitem[\protect\citeauthoryear{{Wan} et~al.,}{{Wan} et~al.}{2020}]{Wan2020}
{Wan} Z.,  et~al., 2020, \mn@doi [\nat] {10.1038/s41586-020-2483-6}, \href
  {https://ui.adsabs.harvard.edu/abs/2020Natur.583..768W} {583, 768}

\bibitem[\protect\citeauthoryear{{Weinberg}}{{Weinberg}}{2017}]{Weinberg2017b}
{Weinberg} D.~H.,  2017, \mn@doi [\apj] {10.3847/1538-4357/aa96b2}, \href
  {https://ui.adsabs.harvard.edu/abs/2017ApJ...851...25W} {851, 25}

\bibitem[\protect\citeauthoryear{{Weinberg}, {Andrews}  \&
  {Freudenburg}}{{Weinberg} et~al.}{2017}]{Weinberg2017}
{Weinberg} D.~H.,  {Andrews} B.~H.,   {Freudenburg} J.,  2017, \mn@doi [\apj]
  {10.3847/1538-4357/837/2/183}, \href
  {https://ui.adsabs.harvard.edu/abs/2017ApJ...837..183W} {837, 183}

\bibitem[\protect\citeauthoryear{{Weinberg} et~al.,}{{Weinberg}
  et~al.}{2019}]{Weinberg2019}
{Weinberg} D.~H.,  et~al., 2019, \mn@doi [\apj] {10.3847/1538-4357/ab07c7},
  \href {https://ui.adsabs.harvard.edu/abs/2019ApJ...874..102W} {874, 102}

\bibitem[\protect\citeauthoryear{{Weinberg} et~al.,}{{Weinberg}
  et~al.}{2022}]{Weinberg2022}
{Weinberg} D.~H.,  et~al., 2022, \mn@doi [\apjs] {10.3847/1538-4365/ac6028},
  \href {https://ui.adsabs.harvard.edu/abs/2022ApJS..260...32W} {260, 32}

\bibitem[\protect\citeauthoryear{{Weisz} et~al.,}{{Weisz}
  et~al.}{2014a}]{Weisz2014a}
{Weisz} D.~R.,  et~al., 2014a, \mn@doi [\apj] {10.1088/0004-637X/789/1/24},
  \href {https://ui.adsabs.harvard.edu/abs/2014ApJ...789...24W} {789, 24}

\bibitem[\protect\citeauthoryear{{Weisz}, {Dolphin}, {Skillman}, {Holtzman},
  {Gilbert}, {Dalcanton}  \& {Williams}}{{Weisz} et~al.}{2014b}]{Weisz2014b}
{Weisz} D.~R.,  {Dolphin} A.~E.,  {Skillman} E.~D.,  {Holtzman} J.,  {Gilbert}
  K.~M.,  {Dalcanton} J.~J.,   {Williams} B.~F.,  2014b, \mn@doi [\apj]
  {10.1088/0004-637X/789/2/147}, \href
  {https://ui.adsabs.harvard.edu/abs/2014ApJ...789..147W} {789, 147}

\bibitem[\protect\citeauthoryear{{Weisz}, {Dolphin}, {Skillman}, {Holtzman},
  {Gilbert}, {Dalcanton}  \& {Williams}}{{Weisz} et~al.}{2015}]{Weisz2015}
{Weisz} D.~R.,  {Dolphin} A.~E.,  {Skillman} E.~D.,  {Holtzman} J.,  {Gilbert}
  K.~M.,  {Dalcanton} J.~J.,   {Williams} B.~F.,  2015, \mn@doi [\apj]
  {10.1088/0004-637X/804/2/136}, \href
  {https://ui.adsabs.harvard.edu/abs/2015ApJ...804..136W} {804, 136}

\bibitem[\protect\citeauthoryear{{Wheeler}, {Phillips}, {Cooper},
  {Boylan-Kolchin}  \& {Bullock}}{{Wheeler} et~al.}{2014}]{Wheeler2014}
{Wheeler} C.,  {Phillips} J.~I.,  {Cooper} M.~C.,  {Boylan-Kolchin} M.,
  {Bullock} J.~S.,  2014, \mn@doi [\mnras] {10.1093/mnras/stu965}, \href
  {https://ui.adsabs.harvard.edu/abs/2014MNRAS.442.1396W} {442, 1396}

\bibitem[\protect\citeauthoryear{{Whitten} et~al.,}{{Whitten}
  et~al.}{2021}]{Whitten2011}
{Whitten} D.~D.,  et~al., 2021, \mn@doi [\apj] {10.3847/1538-4357/abee7e},
  \href {https://ui.adsabs.harvard.edu/abs/2021ApJ...912..147W} {912, 147}

\bibitem[\protect\citeauthoryear{{Woosley} \& {Weaver}}{{Woosley} \&
  {Weaver}}{1995}]{Woosley1995}
{Woosley} S.~E.,  {Weaver} T.~A.,  1995, \mn@doi [\apjs] {10.1086/192237},
  \href {https://ui.adsabs.harvard.edu/abs/1995ApJS..101..181W} {101, 181}

\bibitem[\protect\citeauthoryear{{Wright} et~al.,}{{Wright}
  et~al.}{2010}]{Wright2010}
{Wright} E.~L.,  et~al., 2010, \mn@doi [\aj] {10.1088/0004-6256/140/6/1868},
  \href {https://ui.adsabs.harvard.edu/abs/2010AJ....140.1868W} {140, 1868}

\bibitem[\protect\citeauthoryear{{Xiang} \& {Rix}}{{Xiang} \&
  {Rix}}{2022}]{Xiang2022}
{Xiang} M.,  {Rix} H.-W.,  2022, \mn@doi [\nat] {10.1038/s41586-022-04496-5},
  \href {https://ui.adsabs.harvard.edu/abs/2022Natur.603..599X} {603, 599}

\bibitem[\protect\citeauthoryear{{York} et~al.,}{{York}
  et~al.}{2000}]{York2000}
{York} D.~G.,  et~al., 2000, \mn@doi [\aj] {10.1086/301513}, \href
  {https://ui.adsabs.harvard.edu/abs/2000AJ....120.1579Y} {120, 1579}

\bibitem[\protect\citeauthoryear{{Yuan}, {Chang}, {Beers}  \& {Huang}}{{Yuan}
  et~al.}{2020}]{Yuan2020}
{Yuan} Z.,  {Chang} J.,  {Beers} T.~C.,   {Huang} Y.,  2020, \mn@doi [\apjl]
  {10.3847/2041-8213/aba49f}, \href
  {https://ui.adsabs.harvard.edu/abs/2020ApJ...898L..37Y} {898, L37}

\bibitem[\protect\citeauthoryear{{Zahid}, {Kewley}  \& {Bresolin}}{{Zahid}
  et~al.}{2011}]{Zahid2011}
{Zahid} H.~J.,  {Kewley} L.~J.,   {Bresolin} F.,  2011, \mn@doi [\apj]
  {10.1088/0004-637X/730/2/137}, \href
  {https://ui.adsabs.harvard.edu/abs/2011ApJ...730..137Z} {730, 137}

\bibitem[\protect\citeauthoryear{{Zahid}, {Dima}, {Kudritzki}, {Kewley},
  {Geller}, {Hwang}, {Silverman}  \& {Kashino}}{{Zahid}
  et~al.}{2014}]{Zahid2014}
{Zahid} H.~J.,  {Dima} G.~I.,  {Kudritzki} R.-P.,  {Kewley} L.~J.,  {Geller}
  M.~J.,  {Hwang} H.~S.,  {Silverman} J.~D.,   {Kashino} D.,  2014, \mn@doi
  [\apj] {10.1088/0004-637X/791/2/130}, \href
  {https://ui.adsabs.harvard.edu/abs/2014ApJ...791..130Z} {791, 130}

\bibitem[\protect\citeauthoryear{{Zepeda} et~al.,}{{Zepeda}
  et~al.}{2022}]{Zepeda2022}
{Zepeda} J.,  et~al., 2022, arXiv e-prints, \href
  {https://ui.adsabs.harvard.edu/abs/2022arXiv220912224Z} {p. arXiv:2209.12224}

\makeatother
\end{thebibliography}

\begin{appendices}

\renewcommand\theequation{\thesection\arabic{equation}}
\renewcommand\thefigure{\thesection\arabic{figure}}
\setcounter{equation}{0}
\setcounter{figure}{0}

\section{Derivation of the Likelihood Function}
\label{sec:likelihood}

Here we provide a detailed derivation of our likelihood function (Eq.
\ref{eq:likelihood}).
In its most general form, the problem at hand is to treat some set of data as a
stochasitc sample from an evolutionary track in some observed space.
This assumption implies that all of the data would fall perfectly on some
infinitely thin line or curve in the absence of measurement uncertainties.
We make no assumptions about the underlying model that computes the track, so
this approach should be universally applicable to one-zone GCE models of any
parametrization.
Evolutionary tracks also arise in the context of, e.g., stellar streams and
isochrones, indicating that our likelihood function should be easily extensible
to these models as well.
We however phrase our discussion here under the assumption that the observed
quantities are the abundances and ages of stars and that the underlying
framework is a one-zone GCE model (see discussion in~\S~\ref{sec:onezone}).
\par
First, we define the key variables:
\begin{itemize}

	\item[\textbf{1.}] $\script{D} = \{\script{D}_1, \script{D}_2,
	\script{D}_3, ..., \script{D}_N\}$ is the data containing~$N$ individual
	stars with measurement uncertainties described by the covariance matrices
	of each datum~$C = \{C_1, C_2, C_3, ..., C_N\}$.
	The quantities associated with each star are not necessarily the same --
	that is, only some of the stars may have age measurements, or the
	abundances of some nuclear species may not be reliably measured for the
	whole sample.

	\item[\textbf{2.}] \script{M}~is the evolutionary track in chemical and age
	space.
	Although~\script{M} is a smooth and continuous curve in principle, in
	practice it is approximated in a piece-wise linear form computed by some
	numerical code.
	It can therefore also be expressed as a discrete set of~$K$ points
	$\script{M} = \{\script{M}_1, \script{M}_2, \script{M}_3, ...,
	\script{M}_K\}$ in the observed space connected by line segments.
	We demonstrate below that under this numerical approximation, the
	likelihood function for the continuous piece-wise linear track can be
	expressed as a summation over the discretely sampled points.

	\item[\textbf{3.}] $\{\theta\}$ is a chosen set of one-zone model
	parameters.
	These values impact the detailed form of the track~\script{M}~and otherwise
	affect the inferred best-fit values only if there is an assumed prior
	$L(\{\theta\})$ (see equation~\ref{eq:bayes}).

\end{itemize}
\par
Given the track~\script{M}, the likelihood~$L(\script{D} | \{\theta\})$ of
observing the data can be expressed as the line integral of the differential
likelihood along~\script{M}:
\begin{equation}
L(\script{D} | \{\theta\}) = \int_\script{M} dL =
\int_\script{M} L(\script{D} | \script{M}) P(\script{M} | \{\theta\})
d\script{M},
\end{equation}
where~$P(\script{M} | \{\theta\})$ describes the probability that a singular
datum will be drawn from the model at a given point along the track.
The defining characteristic of the IPPP is that~$P(\script{M} | \{\theta\})$
follows a Poisson distribution~\citep{Press2007}:
\begin{equation}
P(\script{M}_j | \{\theta\}) = e^{-N_\lambda}
\prod_i^N \lambda (\script{M}_j | \{\theta\}),
\end{equation}
where for notational convenience below we leave the expression written as a
product over the~$N$ stars in the sample as opposed to~$\lambda^N$.
$\lambda$ is the~\textit{intensity function} describing the expected
number of stars at a specific point along the track~$\script{M}_j$.
$N_\lambda$ denotes the expected~\textit{total} number of stars in the sample
and can be expressed as the line integral of the intensity function along the
track:
\begin{equation}
N_\lambda = \int_\script{M} \lambda(\script{M} | \{\theta\}) d\script{M}.
\end{equation}
$\lambda$ describes the predicted~\textit{observed} distribution of stars in
chemical space and should therefore incorporate any selection effects in the
data.
It can be expressed as the product of the selection function~\script{S}~(see
discussion in~\S~\ref{sec:fitting}) and the~\textit{intrinsic}
distribution~$\Lambda$ according to
\begin{equation}
\lambda(\script{M}_j | \{\theta\}) = \script{S}(\script{M}_j | \{\theta\})
\Lambda(\script{M}_j | \{\theta\}).
\end{equation}
Plugging the Poisson distribution into our expression for the likelihood
function, we obtain
\begin{subequations}\begin{align}
L(\script{D} | \{\theta\}) &= \int_\script{M}
\left(\prod_i^N L(\script{D}_i | \script{M})\right)
\left(e^{-N_\lambda} \prod_i^N \lambda(\script{M} | \{\theta\})\right)
d\script{M}
\\
&= e^{-N_\lambda} \prod_i^N \int_\script{M} L(\script{D}_i | \script{M})
\lambda(\script{M} | \{\theta\}) d\script{M},
\end{align}\end{subequations}
where we have exploited the conditional independence of each datum, allowing us
to substitute~$L(\script{D} | \script{M}) = \prod L(\script{D}_i |
\script{M})$.
We have also dropped the subscript~$j$ in~$\lambda(\script{M}_j | \{\theta\})$
because we are computing the line integral along the track~\script{M}, so a
specific location~$\script{M}_j$ is implicit.
\par
Now taking the logarithm of the likelihood function produces the following
expression for~$\ln L$:
\begin{equation}
\ln L(\script{D} | \{\theta\}) = -N_\lambda + \sum_i^N \ln \left(
\int_\script{M} L(\script{D}_i | \script{M})\lambda(\script{M} | \{\theta\})
d\script{M}
\right).
\label{eq:lnL_with_integral}
\end{equation}
The next step is to assess the likelihood~$L(\script{D}_i | \script{M})$ of
observing each datum given the predicted track.
The line integral within the summation indicates that the most general solution
is to marginalize the likelihood over the entire evolutionary track.
In fact, we find in our tests against mock samples that this marginalizaion is necessary to
ensure that the inferred best-fit parameters are accurate (see discussion
in~\S~\ref{sec:mocks:recovered}).
This requirement arises due to observational uncertainties -- there is no way
of knowing~\textit{a priori} which point on the track any individual datum is
truly associated with.
If this information were available,~$L(\script{D}_i | \script{M})$ would reduce
to a delta function at the known point~$\script{M}_j$.

\begin{figure}
\centering
\includegraphics[scale = 0.6]{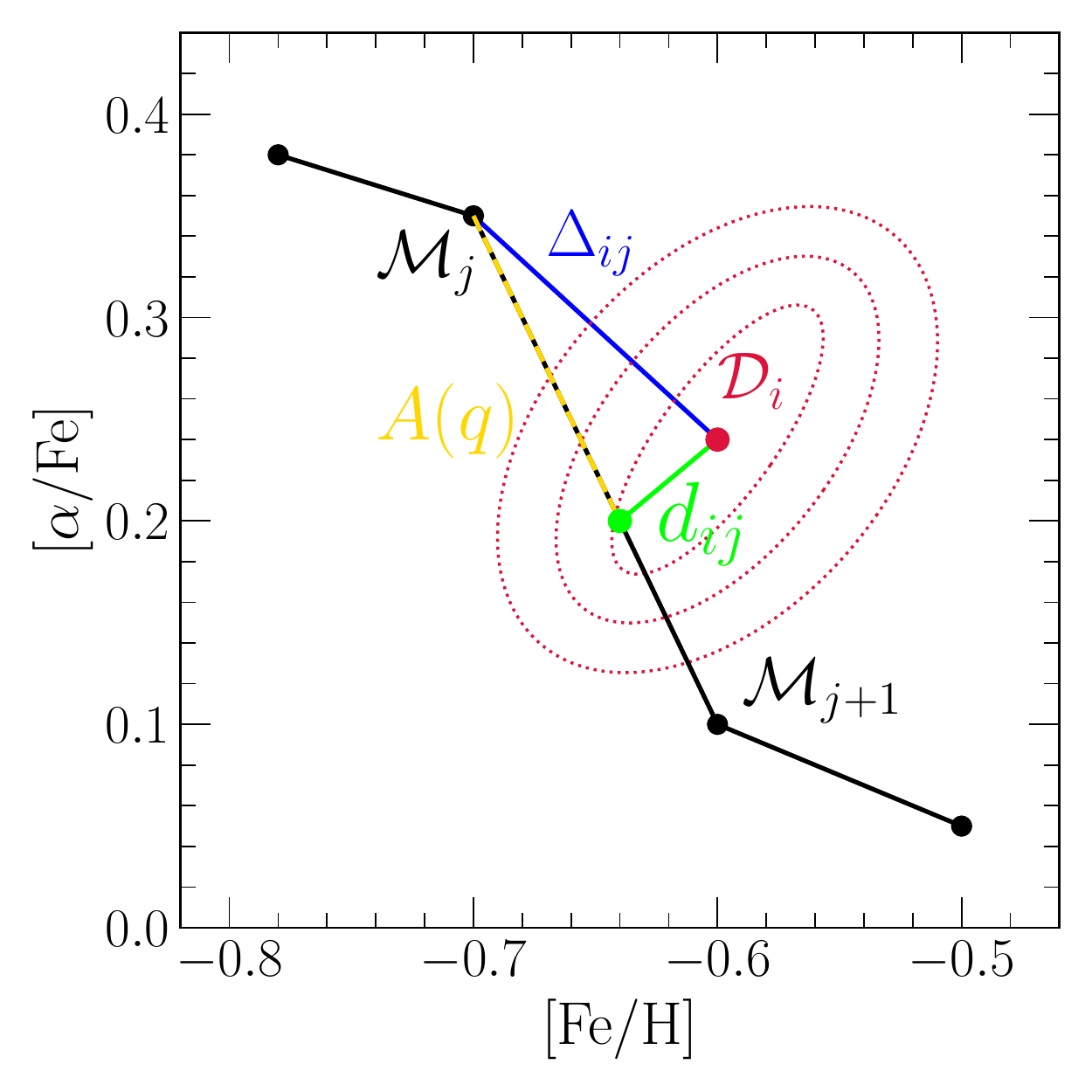}
\caption{
A schematic of our derivation and the quantities involved.
In practice, the evolutionary track~\script{M} is computed by some numerical
code as a piece-wise linear approximation -- here we exaggerate the spacing
between points for illustrative purposes.\tabularnewline
When the spacing~$\Delta\script{M}_j$ between the points~$\script{M}_j$ and
$\script{M}_{j + 1}$ is large compared to the observation uncertainties
associated with the datum~$\script{D}_i$ (shown by the dotted red contours),
the finite length of the line segment becomes an important correction.
Additional vector quantities that appear in our derivation are also noted.
}
\label{fig:marginalization}
\end{figure}

In practice, the track may be complicated in shape and is generally not known
as a smooth and continuous function, instead in some piece-wise linear
approximation computed by a numerical code.
We visualize a hypothetical track and datum in Fig.~\ref{fig:marginalization}
where we have deliberately exaggerated the spacing between two adjacent
points~$\script{M}_j$ and~$\script{M}_{j + 1}$ for illustrative purposes.
In principle, the likelihood of observing some datum~$\script{D}_i$ varies
along the line segment~$\Delta\script{M}_j$ connecting the two points.
To properly take this variation into account, we must integrate along the
length of the line segment:
\begin{equation}
L(\script{D}_i | \script{M}_j) = \int_0^1 L(\script{D}_i | \script{M}_j, q) dq,
\label{eq:l_int_q}
\end{equation}
where~$q$ is a dimensionless parameter defined to be 0 at the point
$\script{M}_j$ and 1 at the point~$\script{M}_{j + 1}$ according to
\begin{equation}
A(q) = \script{M}_j + q(\script{M}_{j + 1} - \script{M}_j)
= \script{M}_j + q\Delta\script{M}_j.
\end{equation}
If the errors associated with the observed datum~$\script{D}_i$ are accurately
described by a multivariate Gaussian, then the likelihood of observing
$\script{D}_i$ given a point along this line segment can be expressed in terms
of its covariance matrix~$C_i$ as
\begin{subequations}\begin{align}
L(\script{D}_i | \script{M}_j, q) &=
\frac{1}{\sqrt{2\pi \det{(C_i)}}}
\exp\left(\frac{-1}{2} d_{ij}(q) C_i^{-1} d_{ij}^T(q)\right)
\label{eq:l_di_mj_q}
\\
d_{ij} &= \script{D}_i - A(q)
\\
&= \script{D}_i - \script{M}_j - q(\script{M}_{j + 1} - \script{M}_j)
\\
&= \Delta_{ij} - q\Delta \script{M}_j,
\label{eq:delta_minus_qmj}
\end{align}\end{subequations}
where~$d_{ij}$ is the vector difference between~$\script{D}_i$ and the
point along the track~$A(q)$ in the observed space.
For notational convenience, we have introduced the variable
$\Delta_{ij} = \script{D}_i - \script{M}_j$ as the vector difference between
the~$i$th datum and the~$j$th point sampled on the track.
We clarify our notation that the subscripts~$i$ and~$ij$ in equation
\refp{eq:l_di_mj_q} above do not refer to rows and columns of matrices, but
rather to the~$i$th datum and the~$j$th point on the model track.
If a multivariate Gaussian is not an accurate description of the measurement
uncertainties in any one datum, then equation~\refp{eq:l_di_mj_q} must
be replaced with some alternative characterization of the likelihood of
observation, such a kernel density estimate evaluated at the point~$A(q)$.
We however continue our derivation under the assumption of multivariate
Gaussian uncertainties.
\par
Before evaluating equation~\refp{eq:l_int_q}, we first compute the square
$d_{ij}(q)C_i^{-1}d_{ij}^T(q)$ and isolate the terms that depend on~$q$:
\begin{subequations}\begin{align}
\begin{split}
d_{ij}(q) C_i^{-1} d_{ij}(q)^T &= \Delta_{ij} C_i^{-1} \Delta_{ij}^T -
2q\Delta_{ij}C_i^{-1}\Delta \script{M}_j^T +
\\
&\qquad q^2\Delta \script{M}_j C_i^{-1} \Delta \script{M}_j^T
\end{split}
\label{eq:chisquared}
\\
&= \Delta_{ij} C_i^{-1} \Delta_{ij}^T - 2bq + aq^2,
\end{align}\end{subequations}
where we have introduced the substitutions
$a = \Delta\script{M}_j C_i^{-1} \Delta\script{M}_j^T$ and
$b = \Delta_{ij} C_i^{-1} \Delta\script{M}_j^T$.
Plugging this expression into the exponential in equation~\refp{eq:l_di_mj_q}
and integrating from~$q = 0$ to~$1$ according to equation~\refp{eq:l_int_q}
yields the following expression for~$L(\script{D}_i | \script{M}_j)$:
\begin{subequations}\begin{align}
\begin{split}
L(\script{D}_i | \script{M}_j) &= \frac{1}{\sqrt{2\pi \det{(C_i)}}}
\exp\left(\frac{-1}{2}\Delta_{ij} C_i^{-1} \Delta_{ij}^T\right)
\\
&\qquad \int_0^1 \exp\left(\frac{-1}{2}(aq^2 - 2bq)\right)dq
\end{split}
\\
\begin{split}
&= \frac{1}{\sqrt{2\pi\det{(C_i)}}}
\exp\left(\frac{-1}{2}\Delta_{ij}C_i^{-1}\Delta_{ij}^T\right)
\sqrt{\frac{\pi}{2a}}
\\
&\qquad \exp\left(\frac{b^2}{2a}\right)
\left[\erf\left(\frac{a - b}{\sqrt{2a}}\right) - \erf\left(\frac{b}{\sqrt{2a}}
\right)\right].
\end{split}
\end{align}\end{subequations}
For notational convenience, we introduce the corrective term~$\beta_{ij}$ given
by
\begin{equation}
\beta_{ij} = \sqrt{\frac{\pi}{2a}} \exp\left(\frac{b^2}{2a}\right)
\left[\erf\left(\frac{a - b}{\sqrt{2a}}\right) - \erf\left(\frac{b}{\sqrt{2a}}
\right)\right],
\label{eq:corrective_beta}
\end{equation}
such that~$L(\script{D}_i | \script{M}_j)$ can be expressed as
\begin{equation}
L(\script{D}_i | \script{M}_j) = \frac{\beta_{ij}}{\sqrt{2\pi \det{(C_i)}}}
\exp\left(\frac{-1}{2}\Delta_{ij} C_i^{-1} \Delta_{ij}^T\right).
\end{equation}
With this expression for the likelihood~$L(\script{D}_i | \script{M}_j)$ of
observing the datum~$\script{D}_i$ marginalized over the length of the line
segment~$\Delta\script{M}_j$,~$L(\script{D}_i | \script{M})$ can now be
written a summation over each individual line segment.
As mentioned above, the numerical piece-wise linear approximation of the smooth
and continuous form reduces to a summation over the individual points
$\script{M} = \{\script{M}_1, \script{M}_2, \script{M}_3, ..., \script{M}_K\}$
at which the track is sampled:
\begin{equation}\begin{split}
&\ln L(\script{D} | \{\theta\}) = -N_\lambda
- \sum_i^N \ln \left(\sqrt{2\pi \det{(C_i)}}\right) +
\\
&\qquad \sum_i^N \ln \left(
\sum_j^K \beta_{ij}
\exp\left(\frac{-1}{2}\Delta_{ij}C_i^{-1}\Delta_{ij}^T\right)
\lambda(\script{M}_j | \{\theta\})
\right).
\end{split}\end{equation}
Although we have exaggerated the spacing between points for illustrative
purposes, Fig.~\ref{fig:marginalization} indicates that
$q\Delta\script{M}_j \ll \Delta_{ij}$ in the opposing case in which
$\Delta\script{M}_j$ is small compared to the measurement uncertainties.
As a consequence,~$\beta_{ij} \approx 1$ and this corrective term can be safely
neglected.
In some cases, however, computing the evolutionary track~\script{M}~may be
computationally expensive, making it potentially advantageous to reduce the
the number of computed points~$K$ in exchange for a slightly more complicated
likelihood calculation.
\par
As discussed above, the intensity function~$\lambda$ quantifies the observed
density of points, incorporating any selection effects present in the data into
the predicted intrinsic density~$\Lambda$.
In a one-zone GCE model,~$\Lambda$ is given by the SFR at the point
$\script{M}_j$ (to incorporate the effects dying stars or stars at a given
evolutionary stage, one can modify the selection function~\script{S}).
This multiplicative factor on the likelihood~$L$ can be incorporated by simply
letting the pair-wise component of the datum~$\script{D}_i$ and the point along
the track~$\script{M}_j$ take on a weight
$w_j \equiv \script{S}(\script{M}_j | \{\theta\}) \dot{M}_\star(\script{M}_j |
\{\theta\})$ determined by the survey selection function~\script{S}~and the
SFR~$\dot{M}_\star$ at the point~$\script{M}_j$.
The predicted number of instances~$N_\lambda$, originally expressed as the
line integral of~$\lambda$, can now be expressed as the sum of the
weights~$w_j$.
The following likelihood function then arises:
\begin{equation}
\ln L(\script{D} | \{\theta\}) \propto
\sum_i^N \ln \left(\sum_j^K
\beta_{ij} w_j \exp\left(\frac{-1}{2}\Delta_{ij} C_i^{-1} \Delta_{ij}^T\right)
\right) - \sum_j^K w_j,
\label{eq:lnL_minus_weights}
\end{equation}
where we have omitted the term~$\sum \ln \left(\sqrt{2\pi \det{(C_i)}}\right)$
because it is a constant that can safely be neglected in the interest of
optimization.
This likelihood function considers each pair-wise combination of the data and
model, weighting the likelihood according to the predicted density of
observations and penalizing models by the sum of their weights.
This term can also be described as a reward for models that explain the
observations in as few predicted instances as possible.
\par
In many one-zone GCE models, however, the normalization of the SFH is
irrelevant to the evolution of the abundances.
Because the metallicity is given by the metal mass~\textit{relative} to the ISM
mass, the normalization often cancels.
Because the SFH determines the weights, it is essential in these cases to
ensure that the sum of the weights has no impact on the inferred likelihood.
To this end, we consider a density~$\rho$ with some unknown overall
normalization defined relative to the intensity function according to
\begin{subequations}\begin{align}
\lambda(\script{M} | \{\theta\}) &= N_\lambda \rho(\script{M} | \{\theta\})
\\
\int_\script{M} \rho(\script{M} | \{\theta\}) d\script{M} &= 1.
\end{align}\end{subequations}
Plugging~$\rho$ into equation~\refp{eq:lnL_with_integral} and pulling
$N_\lambda$ out of the natural logarithm yields the following expression:
\begin{equation}\begin{split}
\ln L(\script{D} | \{\theta\}) &= -N_\lambda + N \ln N_\lambda +
\sum_i^N \ln \left(\sqrt{2\pi \det{(C_i)}}\right) +
\\
&\qquad \sum_i^N \ln \left(
\int_\script{M} L(\script{D}_i | \script{M}) \rho(\script{M} | \{\theta\})
d\script{M} \right).
\end{split}\end{equation}
With~$\rho$ in place of~$\lambda$ and the extra term~$N \ln N_\lambda$,
reducing this equation proceeds in the exact same manner as above, resulting
in the following likelihood function:
\begin{equation}\begin{split}
\ln L(\script{D} | \{\theta\}) &= -N_\lambda + N \ln N_\lambda +
\sum_i^N \ln \left(\sqrt{2\pi \det{(C_i)}}\right) +
\\
&\qquad \sum_i^N \ln \left(
\sum_j^K \beta_{ij} w_j
\exp\left(\frac{-1}{2}\Delta_{ij} C_i^{-1} \Delta_{ij}^T\right)\right).
\end{split}\end{equation}
For notational convenience, we have left the normalization of the weights
written as~$N_\lambda$.
In the interest of optimizing the likelihood function, we take the partial
derivative of~$\ln L$ with respect to~$N_\lambda$ and find that it is equal to
zero when~$N_\lambda = N$.
Because~$\rho$ is by definition un-normalized, we can simply choose this
overall scale (this is also the ``most correct'' scale in the sense that the
number of stars in the sample is exactly as predicted).
The first two terms in the above expression for~$\ln L$ then become
$-N + N \ln N$, a constant for a given sample which can safely be neglected
for optimization along with the term incorporating the determinants of the
covariance matrices.
We arrive at the following expression for the likelihood function in cases
where the normalization of the SFH does not impact the evolution of the
abundances:
\begin{subequations}\begin{align}
\ln L(\script{D} | \{\theta\}) &\propto \sum_i^N \ln \left( \sum_j^K \beta_{ij}
w_j \exp\left(\frac{-1}{2}\Delta_{ij} C_i^{-1} \Delta_{ij}^T\right) \right)
\label{eq:lnL_fracweight}
\\
\sum_j^K w_j &= 1,
\label{eq:lnL_fracweightsum}
\end{align}\end{subequations}
where the second expression arises from the requirement that the line integral
of the un-normalized density~$\rho$ along the track equal 1.
\par
In summary, when inferring best-fit parameters for one-zone GCE models in which
the normalization of the SFH is irrelevant to the evolution of the abundances,
authors should adopt equations~\refp{eq:lnL_fracweight} and
\refp{eq:lnL_fracweightsum}.
If the model is instead parametrized in such a manner that the normalization
does indeed impact the abundance evolution, then authors should adopt
equation~\refp{eq:lnL_minus_weights}.
Such models can arise, e.g., when the mass-loading factor~$\eta$ grows with the
stellar mass to mimic the deepending of the potential
well~\citep[e.g.,][]{Conroy2022}.
In either case, the corrective term~$\beta_{ij}$ given by equation
\refp{eq:corrective_beta} is approximately 1 and can be safely neglected when
the track is densely sampled relative to the observational uncertainties.
In the present paper, our GCE models are parametrized in such a manner that the
normalization of the SFH does~\textit{not} impact the enrichment history, and
we adopt equations~\refp{eq:lnL_fracweight} and~\refp{eq:lnL_fracweightsum}
accordingly.

\renewcommand\theequation{\thesection\arabic{equation}}
\renewcommand\thefigure{\thesection\arabic{figure}}
\setcounter{equation}{0}
\setcounter{figure}{0}

\section{The Yield-Outflow Degeneracy}
\label{sec:degeneracy}

\begin{figure*}
\centering
\includegraphics[scale = 0.38]{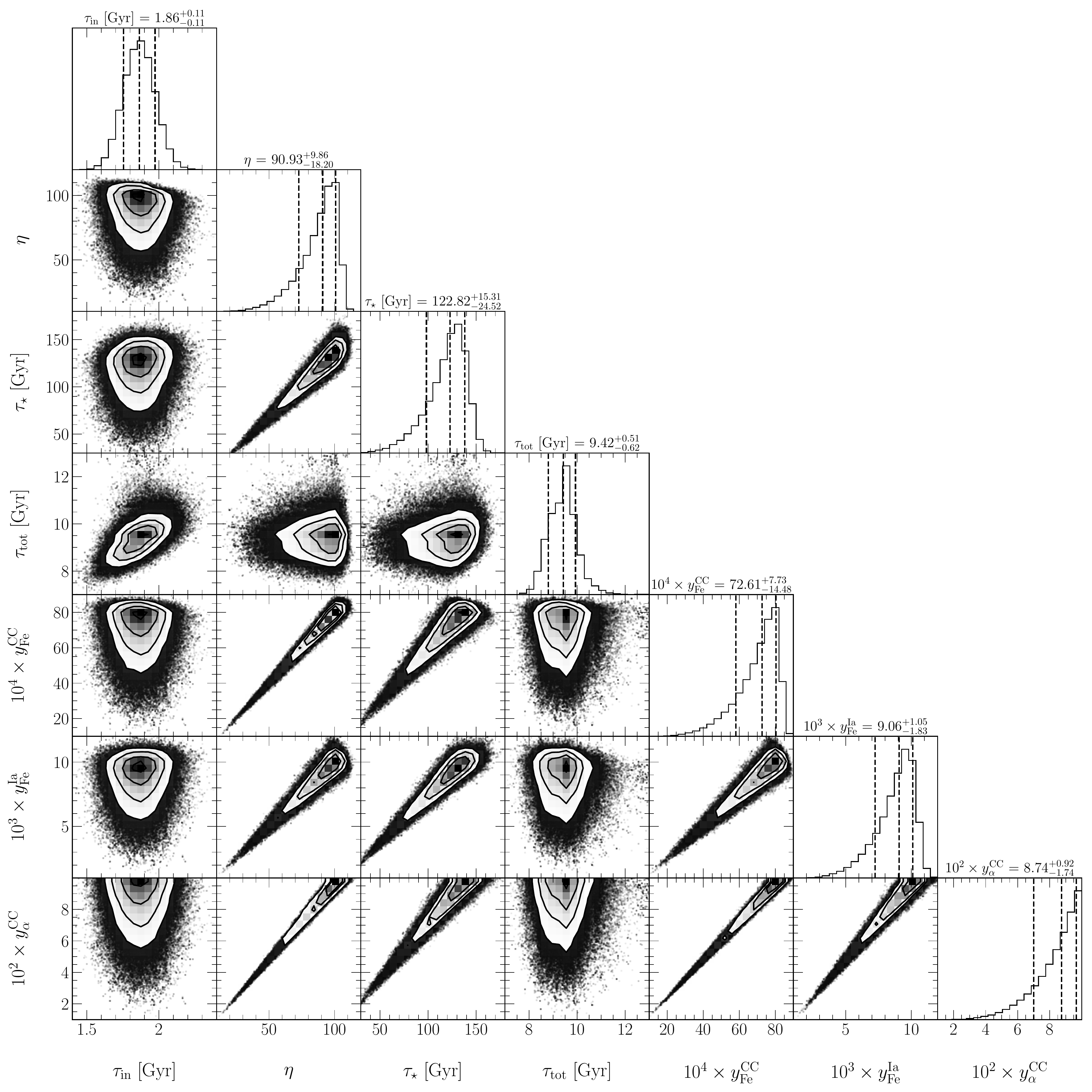}
\caption{
The same as Fig.~\ref{fig:fiducial_mock_corner}, but with the alpha element
yield from massive stars~$\yacc$ as an additional free parameter.
Motivated both by theoretical models of O nucleosynthesis in massive stars and
the convenience for scaling parameters up or down, we have adopted
$\yacc = 0.01$ in this paper to set the scale of this degeneracy.
Here we include a prior that enforces~$\yacc < 0.1$, without which the
likelihood distribution extends to arbitrarily high values.
}
\label{fig:degeneracy}
\end{figure*}

Under the instantaneous recycling approximation, early work in GCE demonstrated
that galaxies with ongoing accretion of metal-poor gas reached an equilibrium
metal abundance in which the newly produced metal mass is balanced by losses to
star formation and, if present, outflows (e.g.,~\citealp{Larson1972}, and more
recently~\citealp{Weinberg2017}).
These ``open-box'' models offered a simple solution to the ``closed-box''
models suffering from the so-called ``G-dwarf problem'' whereby the frequency
of super-solar metallicity stars was extremely over-predicted (see the review
in, e.g.,~\citealp{Tinsley1980}).
These results were corroborated by~\citet{Dalcanton2007} who argued that
metal-enriched outflows are the only mechanism that can significantly reduce
effective yields from SNe.
\par
Recent theoretical explorations of SN explosions propose that many massive
stars collapse directly to black holes at the ends of their lives as opposed to
exploding as CCSNe (\citealp{OConnor2011, Pejcha2015, Ertl2016, Sukhbold2016};
see also discussion in~\citealp{Griffith2021}).
This scenario is supported by the observation of a~$\sim$25~\msun~red
supergiant in NGC 6946 (the ``Fireworks Galaxy'') that disappeared from view
after a brief outburst in 2009, indicative of a failed SN
(\citealp*{Gerke2015};~\citealp{Adams2017, Basinger2021}).
These results add to the theoretical uncertainties in stellar evolution and
nuclear reaction networks which significantly impact predicted nucleosynthetic
yields.
Observationally, it is feasible to constrain relative but not absolute yields.
For example, the ``two-process model'' (\citealp{Weinberg2019, Weinberg2022};
\citealp*{Griffith2019};~\citealp{Griffith2022}) quantifies the median trends
in abundance ratios relative to Mg along the high- and low-alpha sequences to
disentangle the relative contributions of prompt and delayed nucleosynthetic
sources of various elements.
Yield ratios can also be derived from individual SN remnants as in, e.g.,
\citet*{HollandAshford2020}.
However, these investigations cannot constrain the absolute yields of
individual elements.
\par
In GCE models, there are many parametrizations of outflows.
The publicly available GCE codes~\textsc{FlexCE}~\citep{Andrews2017},
\textsc{OMEGA}~\citep{Cote2017} and~\vice~\citep{Johnson2020} assume the form
of equation~\refp{eq:massloading}, implicitly assuming that massive stars are
the dominant source of energy in outflow-driving winds.
Recently,~\citet{delosReyes2022} modelled the evolution of the Sculptor dwarf
spheroidal by letting the outflow rate be linearly proportional to the the
SN rate~$\dot{N}_\text{II} + \dot{N}_\text{Ia}$.
\citet*{Kobayashi2020} constructed a model for the Milky Way in which
outflows develop in the early phases of the evolution, but die out as the
Galaxy grows.
Based on theoretical models suggesting that the re-accretion timescales of
ejected metals are short ($\sim$100 Myr;~\citealp{Melioli2008, Melioli2009,
Spitoni2008, Spitoni2009}), some authors even neglect outflows entirely when
modelling the Milky Way~\citep[e.g.,][]{Minchev2013, Minchev2014, Minchev2017,
Spitoni2019, Spitoni2021}.
Although these models neglecting outflows are able to reproduce many
observables within the Milky Way disc, this argument is at odds with the
empirical result that multi-phase kiloparsec-scale outflows are ubiquitous
around galaxies of a broad range of stellar masses (see, e.g., the recent
review in~\citealt{Veilleux2020}).
Furthermore, measurements of the deuterium abundance (\citealp{Linsky2006};
\citealp*{Prodanovic2010}) and the~$^3$He/$^4$He ratio~\citep{Balser2018} in
the local ISM indicate near-primordial values.
These results indicate that much of the gas in the Galaxy has not been
processed by stars, further suggesting that ambient ISM is readily swept up in
outflows and replaced by unprocessed baryons through
accretion~\citep{Weinberg2017b, Cooke2022}.
\par
Suffice it to say that the community has settled on neither the proper
parametrization nor the importance of mass-loading in GCE models.
As discussed in~\S~\ref{sec:onezone}, the strength of outflows (i.e., the value
of~$\eta$ in this work) is strongly degenerate with the absolute scale of
effective nucleosynthetic yields because they are the primary source and sink
terms in describing enrichment rates (Eq.~\ref{eq:enrichment}).
In this paper, we have applied our fitting method on an assumed scale in which
the oxygen yield from massive stars is fixed at~$\yacc = 0.01$, though if
outflows are to be neglected, the assumption of~$\eta = 0$ fulfills the same
purpose.
While variations in assumptions regarding massive star explodability and the
black hole landscape can lower yields by factors
of~$\sim2 - 3$~\citep{Griffith2021}, values lower by an order of magnitude or
more can be achieved if a significant fraction of SN ejecta is immediately
lost to a hot outflow as proposed by~\citet{Peeples2011}.
Unless star formation is sufficiently slow, this modification is a necessary
for models that assume~$\eta = 0$ as otherwise unphysically high metal
abundances will arise.
There is some observational support for this scenario in that galactic outflows
are observed to be more metal-rich than the ISM of the host galaxy
(\citealp*{Chisholm2018};~\citealp{Cameron2021}), but the metallicities are
not as high as the SN ejecta themselves and cold-phase material is generally
observed in the outflows as well (e.g., in M82,~\citealp{Lopez2020}, and in
NGC 253,~\citealp{Lopez2022}; see also the review in~\citealt{Veilleux2020}).
\par
Motivated by this discourse, we quantify the strength of the yield-outflow
degeneracy by introducing~\yacc~as an additional free parameter in our fit to
our fiducial mock sample described in~\S~\ref{sec:mocks:fiducial}.
We include a prior enforcing~$\yacc < 0.1$; otherwise we find that the MCMC
algorithm allows~$\eta$,~$\tau_\star$ and the SN yields to reach arbitrarily
high values.
Otherwise, we follow the exact same procedure to recover the known evolutionary
parameters of the input model.
Fig.~\ref{fig:degeneracy} shows the resultant posterior distributions.
As expected, there are extremely strong degeneracies in all yields with one
another and with the outflow parameter~$\eta$.
There is an additional degeneracy between the SFE timescale~$\tau_\star$ and
the yields that arises because the position of the ``knee'' in
the~\afe-\feh~plane can be fit with either a high-yield and slow star formation
or a low yield and fast star formation (when we set the overall scale with
$\yacc = 0.01$, we find a degeneracy of the opposite sign; see discussion
in~\S~\ref{sec:mocks:recovered} and in~\citealt{Weinberg2017}).
The strength of these degeneracies is especially striking considering that this
is mock data drawn from an input model with known evolutionary parameters.
In practice, the overall yield scale has factors of~$\sim$$2 - 3$ uncertainty
but not an order of magnitude.
It may therefore be preferable to find best-fit models at a few discrete
values of~\yacc~and understand how other parameters change rather than treat
it as a free parameter.
\par
In detail, this degeneracy arises whenever a parameter influences either the
centroid of the MDF or the position or shape of the evolutionary track in
the~\afe-\feh~diagram.
The infall timescale~$\tau_\text{in}$ and the total duration of star formation
$\tau_\text{tot}$ are unaffected by this degeneracy because they do not
significantly impact these details of the enrichment history (see discussion
in~\S~\ref{sec:mocks:recovered}).
Regardless of the choice of yields and the values of~$\eta$ and~$\tau_\star$,
the shape of the MDF is constrained by a sufficiently large sample, allowing
precise derivations of~$\tau_\text{in}$ and~$\tau_\text{tot}$ with our fitting
method.
Determining the duration of star formation in this manner may open a new
pathway for constraining the early epochs of star formation in both intact
and disrupted dwarf galaxies as well as deriving quenching times for the
now-quiescent systems (see discussion in~\S~\ref{sec:mocks:variations}).

\end{appendices}

\label{lastpage}
\end{document}